\documentclass[12pt]{article}

\usepackage{epsf,epsfig,epstopdf}
\usepackage{graphics}

\usepackage{comment}

\usepackage{url}
\usepackage{amsmath,amssymb}
\usepackage{cite}
\usepackage{a4wide}
\numberwithin{equation}{section}

\def\braket#1{\mathinner{\langle{#1}\rangle}}

\begin{document}

\allowdisplaybreaks
\thispagestyle{empty}

\begin{flushright}
{\small
ITP-UU-12/05\\
SPIN-12/04 \\
FR-PHENO-2012-004\\
arXiv:1202.2260 [hep-ph]\\[0.2cm] 
{\bf July 2, 2012}}
\end{flushright}

\vspace{\baselineskip}

\begin{center}
\vspace{0.5\baselineskip}
\textbf{\Large\boldmath
NLL soft and Coulomb resummation for squark \\[0.2cm]
and gluino production at the LHC
}
\\
\vspace{3\baselineskip}
{\sc P.~Falgari$^a$, C.~Schwinn$^b$, C.~Wever$^a$}\\
\vspace{0.7cm}
{\sl ${}^a$Institute for Theoretical Physics and Spinoza Institute,\\
Utrecht University, 3508 TD Utrecht, The Netherlands\\
\vspace{0.3cm}
${}^b$ Albert-Ludwigs Universit\"at Freiburg, 
Physikalisches Institut, \\
D-79104 Freiburg, Germany }

\vspace*{1.2cm}
\textbf{Abstract}\\
\vspace{1\baselineskip}
\parbox{0.9\textwidth}{ 
We present predictions of the total cross sections for
pair production of squarks and gluinos at the LHC, 
including the stop-antistop production process.
Our calculation supplements full fixed-order NLO predictions with resummation of threshold logarithms and Coulomb 
singularities at next-to-leading logarithmic (NLL) accuracy, including bound-state effects.   The numerical effect of higher-order Coulomb terms can be as big or larger than that of soft-gluon corrections.
For a selection of  benchmark points accessible with data from the 2010-2012 LHC runs, resummation leads to an enhancement of the 
total inclusive squark and gluino production cross section in the $15$-$30 \%$ range. For individual production processes of gluinos, the corrections can be much larger. The theoretical uncertainty in the prediction of the hard-scattering cross sections is typically reduced to the $\pm 10\% $ level.
}
\end{center}

\newpage
\setcounter{page}{1}
\section{Introduction}

Despite the good agreement of the Standard Model (SM) with a wealth of
experimental data, both empirical reasons (e.g. the observation of
dark matter) and theoretical arguments (such as the naturalness
problem and the desire for gauge coupling unification) point to
physics beyond the SM.  One of the most thoroughly studied extensions
of the SM is Supersymmetry (SUSY). In particular the Minimal
Supersymmetric Standard Model (MSSM) with $R$-parity conservation and
superpartner masses at the TeV scale could provide a solution to the
above issues.  The search for SUSY at the TeV scale is therefore a
central part of the physics program of the Large Hadron Collider (LHC)
at CERN.  In the context of the MSSM, and of any other $R$-parity
conserving model, the supersymmetric partners of the SM particles are
produced in pairs, and squarks and gluinos, coupling strongly to
quarks and gluons, have typically the highest production
rates. Experimental searches for SUSY have been performed at LEP, the
Tevatron and the LHC in various final-state signatures,
see~\cite{Bueso:2011aa} for a recent review. For squark- and
gluino-pair production the tightest bounds generically arise from
jets+missing energy signatures, where the two LHC experiments have set
lower bounds on the mass of squarks and gluinos of about $800\,
$GeV-$1\,$TeV \cite{Aad:2011ib,Chatrchyan:2011zy}, depending on the
precise underlying theoretical model assumed. Once upgraded to its
nominal energy of $14\,$TeV the LHC should be sensitive to squark and
gluino masses of up to $3\,$TeV~\cite{Baer:2009dn}.

SUSY searches and, if squarks and gluinos are discovered, the
measurement of their properties rely on a precise theoretical
understanding of the production mechanism and on accurate predictions
of the observables used in the analysis. From the theoretical point of
view, the simplest of such observables is the total production cross
section, on which we will focus in this work. For QCD mediated
processes, such as squark- and gluino-pair production, the Born cross
section is notoriously affected by large theoretical uncertainties,
such that the inclusion of at least the next term in the expansion in
the strong coupling constant $\alpha_s$ is mandatory for a reliable
prediction. Next-to-leading order (NLO) SUSY QCD corrections for
production of squarks and gluinos were computed in
\cite{Beenakker:1996ch} and implemented in the program
\texttt{PROSPINO} \cite{Beenakker:1996ed,Beenakker:1997ut}. The
corrections are large, up to $100\%$ of the tree-level result, and
lead to a significant reduction of the scale dependence of the cross
section.  Electroweak contributions were also investigated
\cite{Bornhauser:2007bf,Hollik:2007wf,Hollik:2008yi,Hollik:2008vm,Mirabella:2009ap,Germer:2010vn}, but
found to be much smaller than the QCD contributions, less than
$5-10\%$ of the Born result.

The size of the ${\cal O}(\alpha_s)$ corrections raises the question of the
magnitude of unknown higher-order QCD corrections, and makes it desirable to
include at least the dominant contributions beyond NLO. It is known that a
non-negligible part of the full NLO corrections arises from the partonic
threshold region, defined by the limit $\beta \equiv \sqrt{1-4 M^2/\hat{s}}
\rightarrow 0$, with $M$ the average mass of the particles produced and
$\hat{s}$ the \emph{partonic} centre-of-mass energy. In the threshold region
the partonic cross section is dominated by soft-gluon emission off the
initial- and final-state coloured particles and by Coulomb interactions of the
two non-relativistic heavy particles, which give rise to singular terms of the
form $\alpha_s \ln^{2,1} \beta$ and $\alpha_s/\beta$, respectively.  These
corrections can be resummed to all orders in $\alpha_s$, thus leading to
improved predictions of the cross section and smaller theoretical
uncertainties. Note that to obtain the total \emph{hadronic} cross section,
the partonic cross section is convoluted with parton luminosity functions. The
convolution scans over regions where $\beta$ is not necessarily small, unless
$M$ is close to the hadronic centre-of-mass energy $s$. Hence, in these
regions the threshold-enhanced terms cannot be expected \emph{a priori} to
give the dominant contribution to the cross section.  However, one often finds
after convoluting with the parton luminosity that the threshold contributions
give a reasonable approximation to the total hadronic cross section (see
Figures~\ref{fig:nlosing} and~\ref{fig:Kfact_stop} below for the case of squark and gluino production),
so resummation is also relevant for improving predictions of the hadronic cross
 section.

Resummation of soft logarithms for squark and gluino production at the
next-to-leading (NLL) logarithmic accuracy have been presented in
\cite{Kulesza:2008jb,Kulesza:2009kq,Beenakker:2009ha,Beenakker:2010nq,Beenakker:2011fu}
using the Mellin-space resummation formalism developed by
\cite{Sterman:1986aj,Catani:1989ne,Kidonakis:1997gm,Bonciani:1998vc}. Recently
the same formalism has been extended to NNLL order~\cite{Czakon:2009zw}
and applied to squark-antisquark production
\cite{Beenakker:2011sf}. These works do not resum Coulomb corrections
to all orders, though the numerically dominant terms are
accounted for at fixed order. All-order resummation of Coulomb
contributions and bound-state effects for squark-gluino and
gluino-gluino production were on the other hand investigated in
\cite{Hagiwara:2009hq,Kauth:2011vg,Kauth:2011bz}, without the
inclusion of soft resummation.  In
\cite{Langenfeld:2009eg,Langenfeld:2010vu} partial NNLL resummation of
soft logarithms has been used to construct approximated NNLO results
for the squark-antisquark production cross section.  Recently a new
formalism for the combined resummation of soft and Coulomb corrections
has been developed \cite{Beneke:2009rj,Beneke:2010da}, and applied to
NLL resummation of squark-antisquark production \cite{Beneke:2010da},
and NNLL resummation of $t \bar{t}$ hadroproduction
\cite{Beneke:2011mq}.  Contrary to the traditional Mellin-space
formalism, in our approach, which is based on soft-collinear effective
theory (SCET) and potential non-relativistic QCD (pNRQCD), resummation
is performed directly in momentum space via renormalization-group
evolution equations~\cite{Becher:2006nr,Becher:2006mr,Becher:2007ty}. The combined soft-Coulomb effects have been found
to be sizeable for the case of squark-antisquark
production~\cite{Beneke:2010da} and lead to a reduction of the scale
uncertainty, as has been observed as well in~\cite{Beenakker:2011sf}.

In this work we extend the results given in \cite{Beneke:2010da} to
the remaining production processes for squarks and gluinos at NLL
accuracy, i.e. squark-squark, squark-gluino and gluino-gluino
production. We also consider separately the production of pairs of
stops, which requires the extension of the formalism presented in
\cite{Beneke:2010da} to particles pair-produced in a $P$-wave state.
The paper is organized as follows: in Section \ref{sec:resum} we give
an overview of squark and gluino production processes, set up the
calculation and briefly review the resummation formalism we employ,
listing the ingredients needed for NLL resummation.  The validity of the formalism for $P$-wave induced processes, which is necessary for
resummation of the stop-antistop cross section, is established in
Appendix \ref{sec:Pwave}. Numerical results for the cross sections 
 are presented in Section
\ref{sec:results}, including predictions for a representative set of the benchmark points proposed in~\cite{AbdusSalam:2011fc} and a comparison to results using the Mellin-space formalism\cite{Beenakker:2009ha,Beenakker:2010nq}. Finally in Section \ref{sec:conclusion} we
present our conclusions and outlook.
Explicit expressions for resummation functions appearing in the NLL cross sections are provided in Appendix~\ref{sec:resfunc}, while Appendix~\ref{sec:bcutdeterm} contains some details on the scales used in the momentum-space resummation and on our method to estimate ambiguities in the  resummation procedure.
 
\section{NLL resummation for squark and gluino production}
\label{sec:resum}
At hadron colliders, the dominant production channels for  squarks $\tilde q$ and gluinos
$\tilde g$ are pair-production processes of the form
\begin{equation}
  \label{eq:had-process}
  N_1 N_2 \to \tilde{s} \tilde{s}'X,
\end{equation}
where $N_{1,2}$ denote the incoming hadrons and $\tilde{s}$, $\tilde{s}'$ the two sparticles.
The total hadronic cross sections for the processes~\eqref{eq:had-process} can be obtained
by convoluting short-distance production cross sections  $ \hat{\sigma}_{p
  p'}(\hat s, \mu_f)$ 
for the partonic processes 
\begin{equation}
  \label{eq:part-process}
  pp'\to \tilde{s} \tilde{s}'X\;, \quad p,p'\in\{q,\bar q,g\},
\end{equation}
with the parton luminosity functions $L_{p p'}(\tau,\mu)$:
\begin{equation}
\label{eq:sigma-had}
\sigma_{N_1 N_2 \to \tilde{s} \tilde{s}'X}(s) = \int_{\tau_0}^1 d \tau \sum_{p, p'=q,\bar{q}, g} L_{p p'}(\tau,\mu_f) \hat{\sigma}_{p p'}(\tau s, \mu_f) \, ,
\end{equation} 
where $\tau_0=4M^2/s$,  with the average sparticle mass
\begin{equation}
  \label{eq:mav}
M=\frac{m_{\tilde{s}}+m_{\tilde{s}'}}{2}.
\end{equation}
The parton luminosity functions  
are defined from the parton density functions (PDFs) as
\begin{equation}
L_{p p'}(\tau,\mu)=\int_0^1 d x_1 d x_2 \delta(x_1 x_2-\tau) f_{p/N_1}(x_1,\mu) f_{p'/N_2}(x_2,\mu)  \, .
\end{equation}

We perform a NLL resummation of threshold logarithms and Coulomb corrections to the partonic cross section,
counting both $\alpha_s/\beta$ and $\alpha_s\ln\beta$ as quantities of order
one, where $\beta=(1-4M^2/\hat s)^{1/2}$ is the heavy-particle velocity.
Our predictions include all corrections to the Born cross section of the schematic form 
\begin{equation}
\label{eq:nll-def}
  \hat{\sigma}_{p p'}^{\text{NLL}} \propto \,\hat\sigma^{(0)}\, 
\sum_{k=0} \left(\frac{\alpha_s}{\beta}\right)^k \,
\exp\Big[\ln\beta\,g_0(\alpha_s\ln\beta)+ g_1(\alpha_s\ln\beta)\Big].
\end{equation}
A resummation at NNLL accuracy in the counting $\alpha_s\ln\beta\sim
1$, $\alpha_s/\beta\sim 1$ , which is beyond the scope of this paper
but has recently been performed for top-pair
production~\cite{Beneke:2011mq}, would include in addition terms of
the form $\exp[\alpha_sg_2(\alpha_s\ln\beta)]$ and corrections of
order $\mathcal{O}(\alpha_s)$ relative to the NLL cross section,
including NLO corrections to the Coulomb potential and other
higher-order potentials, as well as the non-logarithmic one-loop hard
corrections. The recent NNLL calculation of squark-antisquark
production\cite{Beenakker:2011sf} included the corrections of the
$g_2$-type related to soft corrections and the hard
$\mathcal{O}(\alpha_s)$ corrections, but kept only the
$(\alpha_s/\beta)^1$-term in the sum over $k$.  In
Section~\ref{sec:processes} we collect some facts about the production
processes of gluinos and the superpartners of the light quarks at LO
and NLO while the formalism employed for the NLL resummation is
reviewed in Section~\ref{sec:resummation}. The production of
stop pairs is included in~\ref{sec:stop}, while details about the  choice of the soft scale in the momentum-space resummation formalism and our procedure to estimate the remaining theoretical uncertainty  are discussed
in~\ref{sec:scales}.
 
\subsection{LO and NLO results}
\label{sec:processes}

At leading order~\cite{Kane:1982hw,Harrison:1982yi,Dawson:1983fw}, the following partonic channels contribute to the
production of light-flavour squarks 
and gluinos:
\begin{eqnarray}
\label{eq:processes}
 gg, \, q_i \bar{q}_j &\rightarrow& \tilde{q} \bar{\tilde{q}} \,, \nonumber\\
  q_i q_j&\rightarrow& \tilde{q} \tilde{q}, \qquad \bar{q}_i
 \bar{q}_j\rightarrow \bar{\tilde{q}} \bar{\tilde{q}} \,, \nonumber\\
  g q_i &\rightarrow& \tilde{g} \tilde{q},\,\qquad 
  g \bar{q}_i  \rightarrow \tilde{g} \bar{\tilde{q}}  \,, \nonumber\\
  gg, \, q_i \bar{q}_i  & \rightarrow& \tilde{g} \tilde{g}  \,, 
\end{eqnarray}
where $i,\, j=u, \, d, \, s, \, c, \, b$. 
At NLO further partonic processes contribute to the cross section. 
To keep the notation as simple as possible, in (\ref{eq:processes}) we have suppressed the helicity and flavour 
indices of the squarks. It is understood that in the predictions for the cross sections presented below the contributions of the
ten light-flavour squarks ($\tilde{u}_{L/R}, \, \tilde{d}_{L/R}, \, \tilde{c}_{L/R}, \, \tilde{s}_{L/R}, \, \tilde{b}_{L/R}$) are always summed over. 
Furthermore,  the ten scalars are assumed to be degenerate in mass, with the common light-flavour squark 
mass given by $m_{\tilde{q}}$. 
In the following, the charge-conjugate subprocesses for squark-squark and
gluino-squark productions will be included in our results.
As input for the convolution (\ref{eq:sigma-had}) we will use the  MSTW08 set of PDFs~\cite{Martin:2009iq} at the appropriate perturbative order 
(the LO PDFs for Born-level predictions and the NLO PDFs for the NLO and NLL results) and
set the factorization scale  to the average mass of the produced sparticles, $\mu_f=M$. 
We use a set of PDFs with an improved  accuracy at large $x$ provided to us by the MSTW 
collaboration that has also been employed for the NLL results in~\cite{Beenakker:2011fu}.

\begin{figure}[t!]
  \centering
  \includegraphics[width=0.49 \linewidth]{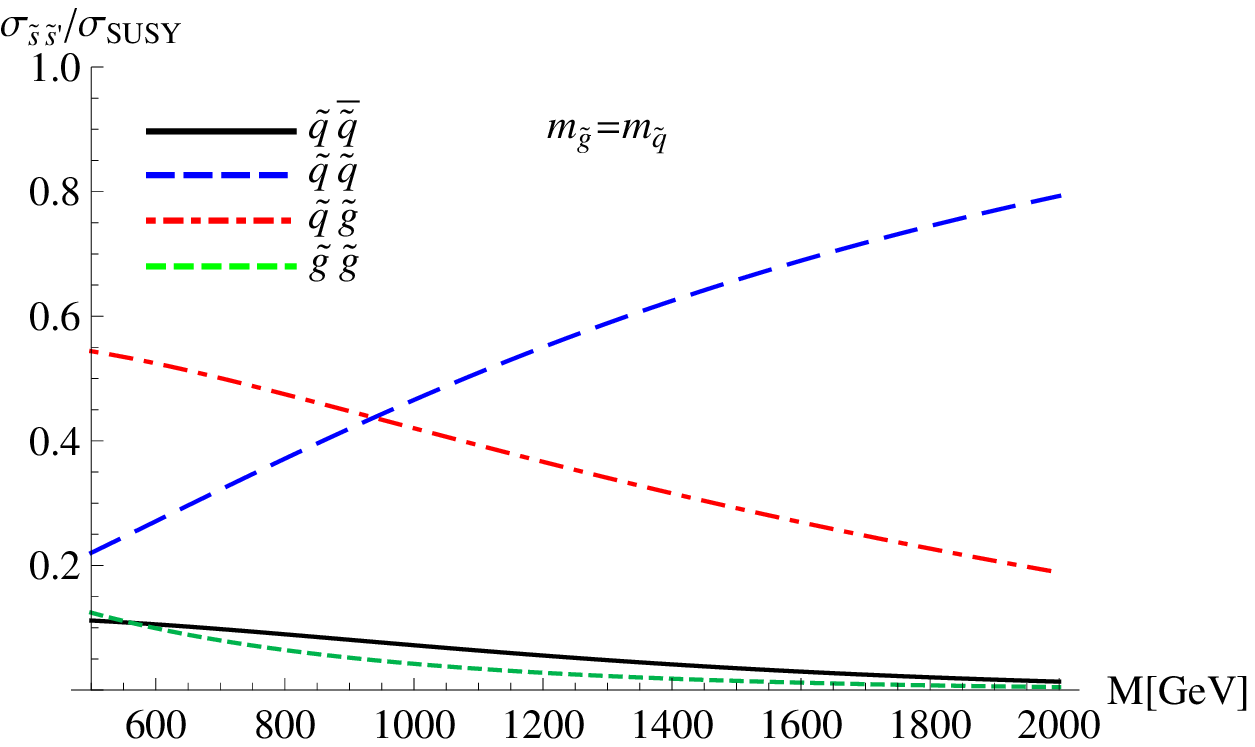}
  \includegraphics[width=0.49 \linewidth]{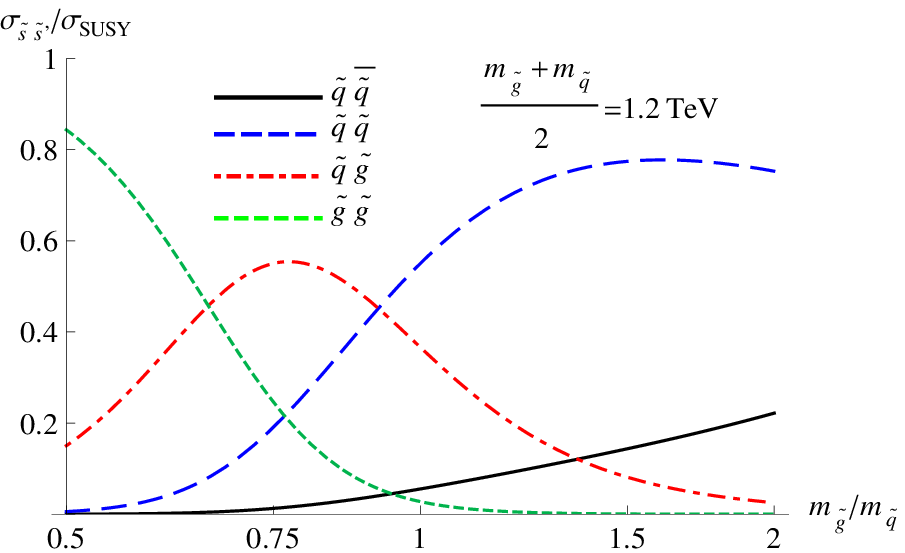}
  \caption{Ratio of the LO production cross sections for the
    processes~\eqref{eq:processes} to the total Born production rate of coloured
    sparticles, $\sigma_{\text{SUSY}}$, for the LHC with $\sqrt s=7$
    TeV. Left:  Mass dependence for a fixed mass ratio $m_{\tilde q}=m_{\tilde
      g}=M$. Right:  Dependence on the ratio $m_{\tilde g}/m_{\tilde q}$ for
    a fixed average mass $(m_{\tilde{q}}+m_{\tilde{g}})/2=1.2$~TeV.
}
  \label{fig:sigmaLO}
\end{figure}

To illustrate the relative magnitude of the various processes depending on the
squark and gluino masses, the ratio of the total hadronic cross section for
the processes~\eqref{eq:processes} to the total inclusive cross section for
squark and gluino production $\sigma_{\text{SUSY}}=\sigma_{PP\to \tilde
  q\bar{\tilde q}+ \tilde q\tilde q+ \tilde g \tilde q+ \tilde g\tilde
  g}$ is shown in Figure~\ref{fig:sigmaLO} for the LHC with $\sqrt s=7$~TeV
centre-of-mass energy.
From the left-hand side plot, showing the relative contributions of the various processes 
as a function of a common squark and gluino mass, it can be seen that
squark-squark and squark-gluino production are by far the dominant channels
over the full mass range considered. 
In the right-hand side plot, the relative contributions are shown
 as a function of the squark-gluino mass ratio for average mass
 $\frac{m_{\tilde g}+m_{\tilde q}}{2}=1.2$~TeV and it is seen that only for gluinos that are significantly
 lighter than squarks, gluino-pair production becomes the dominant production channel.
\begin{figure}[t!]
  \centering
  \includegraphics[width=0.49 \linewidth]{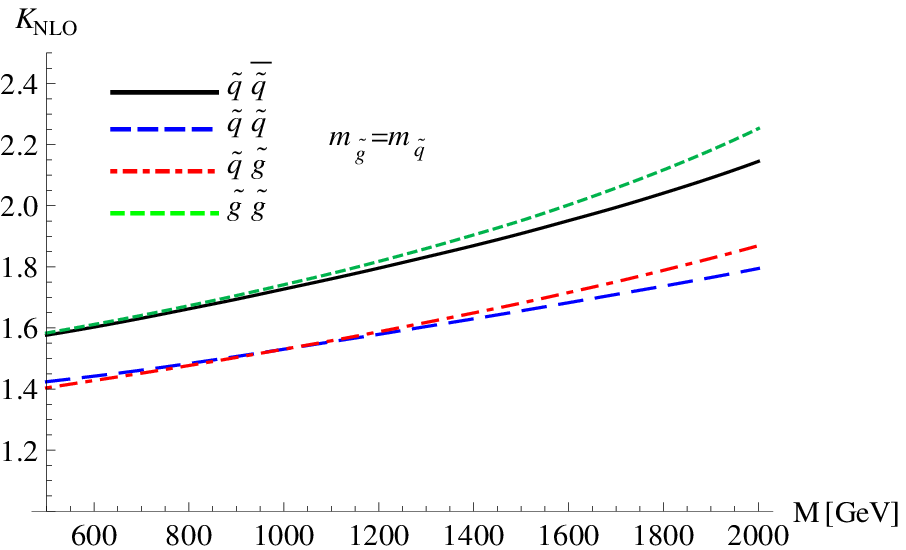}
  \includegraphics[width=0.49 \linewidth]{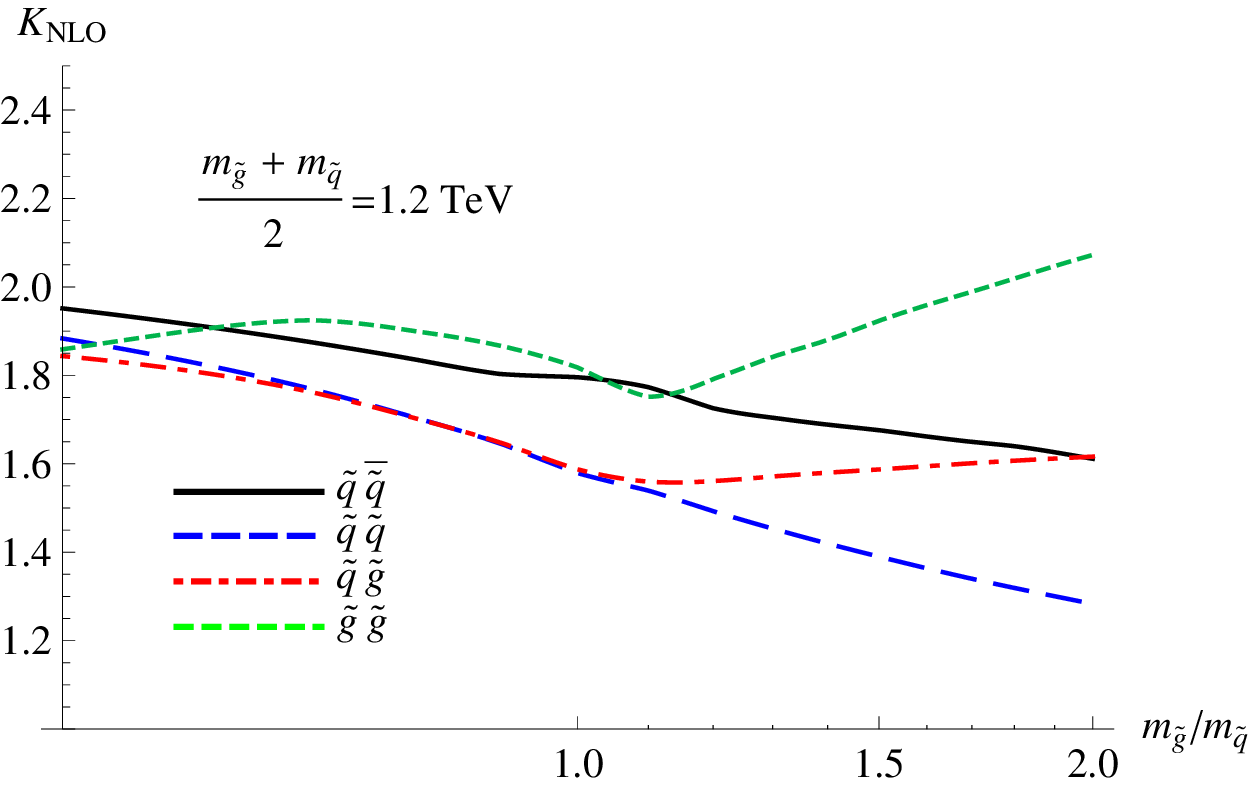}
  \caption{NLO $K$-factor for the processes~\eqref{eq:processes} at the LHC
    with $\sqrt s=7$ TeV. Left: Mass dependence for a fixed mass ratio
    $m_{\tilde q}=m_{\tilde g}=M$. Right: Dependence
    on the ratio $m_{\tilde g}/m_{\tilde q}$ for a fixed average mass
    $(m_{\tilde q}+m_{\tilde g})/2=1.2$~TeV.
}
  \label{fig:KNLO}
\end{figure}
In Figure~\ref{fig:KNLO} we show the $K$-factor
$K_{\text{NLO}}=\sigma_{\text{NLO}}/\sigma_{\text{LO}}$ for the
SUSY-QCD corrections for the various production processes as obtained
from \texttt{PROSPINO} \cite{Beenakker:1996ed}.\footnote{To see the genuine size of the NLO
  corrections, the $K$-factors have been computed using the NLO PDFs  for the Born cross sections.}
 The corrections are positive and enhance the cross
section from $40 \%$ for squark-gluino production with light sparticle
masses up to $100\%$ or larger for squark-antisquark and gluino-pair production at large
sparticle masses.

Since the focus of this work is on higher-order corrections that are enhanced
in the threshold limit $\beta\to 0$, we consider here the corresponding terms
appearing at NLO.
To this end, we decompose the partonic cross section $\hat{\sigma}_{p p'}$ 
into a complete colour basis, 
and parametrize the higher-order corrections as
\begin{equation} \label{eq:partonic}
 \hat{\sigma}_{p p'}= \sum_{R_\alpha} \hat{\sigma}^{(0),R_\alpha}_{p p'} \left\{1+\frac{\alpha_s}{4 \pi} f^{(1),R_\alpha}_{p p'} +... \right\} \, .
\end{equation}
The sum is over the irreducible colour representations
appearing in the decomposition $R\otimes R'=\sum
R_\alpha$, where $R, \, R'$ are the $SU(3)$ representations of the two
final-state sparticles.  The relevant decompositions for squark and gluino production
are given by
\begin{equation}
\label{eq:susy-reps}
  \begin{aligned}
\tilde q\bar{\tilde q}&:&3\otimes \bar 3&=1\oplus 8\,,\\ 
\tilde q \tilde q&:& 3\otimes 3&= \bar 3 \oplus 6\,,\\ 
 \tilde q\tilde g&:&3\otimes 8&=3\oplus \bar 6 \oplus 15\,,\\
  \tilde g\tilde g&:&   8 \otimes 8&=1 \oplus 8_s \oplus 8_a \oplus 10
  \oplus \overline{10} \oplus 27\,.
  \end{aligned}
\end{equation}
 The
explicit basis tensors for the various representations have been constructed
in~\cite{Beneke:2009rj} (see
also~\cite{Kulesza:2009kq,Beenakker:2009ha}), where it has been shown that 
an $s$-channel colour basis based on the decompositions~\eqref{eq:susy-reps}
is advantageous for the all-order summation of soft-gluon corrections.
In Eq. (\ref{eq:partonic}) $\hat{\sigma}^{(0),R_\alpha}_{p p'}$ represents the tree-level cross section for a given process in colour channel $R_\alpha$, while the 
$f^{(1),R_\alpha}_{p p'}$ are colour-specific NLO scaling functions. The colour-separated Born cross sections
for squark and gluino production  are available in \cite{Kulesza:2009kq,Beenakker:2009ha,Beenakker:2010nq}. 
The NLO scaling functions, on the contrary, are only known numerically in their
colour-averaged form \cite{Beenakker:1996ch}.
However, a simple formula is available for the threshold limit of the NLO scaling functions, containing all the threshold-enhanced contributions, 
for arbitrary colour representation $R_\alpha$ \cite{Beneke:2009ye}:
\begin{eqnarray}
\label{eq:NLOapprox}
f^{(1),R_\alpha}_{p p'} &=& -\frac{2 \pi^2 D_{R_\alpha}}{\beta} \sqrt{\frac{2 m_r}{M}}+4 (C_r+C_{r'}) \left[\ln^2 \left(\frac{8 M \beta^2}{\mu_f}\right)+8-\frac{11 \pi^2}{24}\right]\nonumber\\
&&-4 (C_{R_\alpha}+4 (C_r+C_{r'})) \ln \left(\frac{8 M \beta^2}{\mu_f}\right)+12 C_{R_\alpha}+h_{pp'}^{(1),R_\alpha}+{\cal O}(\beta),
\end{eqnarray}
with $M$ the average mass of the two particles produced~\eqref{eq:mav}, while 
$m_r$ denotes the reduced mass, 
$m_r=m_{\tilde{s}} m_{\tilde{s}'}/(m_{\tilde{s}}+m_{\tilde{s}'})$.
$C_r$, $C_{r'}$ and $C_{R_\alpha}$ are the Casimir invariants for the colour representations of the initial-state particles, $p$ and $p'$, and for the 
irreducible representation $R_\alpha$ of the SUSY pair.
The coefficients $D_{R_\alpha}$ of the Coulomb
potential for the production of heavy particles in $SU(3)$ representations $R$ and $R'$ in the colour channel $R_\alpha$ are given in terms of the quadratic Casimir operators for the various representations:
\begin{equation}
\label{eq:dralpha}
  D_{R_\alpha}=\frac{1}{2}(C_{R_\alpha}-C_R-C_{R'})\,,
\end{equation}
where negative values correspond to an attractive Coulomb potential,
positive values to a repulsive one.  The numerical values for the
representations relevant for squark and gluino production can be found
in~\cite{Kats:2009bv,Beneke:2010da} and are collected in
Table~\ref{tab:Coulomb}.  The coefficient $h_{pp'}^{(1),R_\alpha}$ is
the one-loop contribution to the hard matching coefficient appearing
in Eq. (\ref{eq:fact}) below, and represents the only process-specific
quantity in Eq. (\ref{eq:NLOapprox}). It has been obtained recently
for squark-antisquark production and gluino-gluino
production~\cite{Beenakker:2011sf,Kauth:2011vg} but is not known yet
for the remaining production processes. The knowledge of
$h_{pp'}^{(1),R_\alpha}$ is required for NNLL
resummation~\cite{Beenakker:2011sf}, but not at NLL accuracy as
considered here, so the $h_{pp'}^{(1),R_\alpha}$ will be always set to
zero in the following. Using the Born cross sections for the different
colour channels~\cite{Kulesza:2009kq,Beenakker:2009ha}
and~\eqref{eq:NLOapprox} one can reproduce the threshold expansions of
the NLO corrections
in~\cite{Beenakker:1996ch}.\footnote{In~\cite{Beenakker:1996ch} there
  is a typo in the sign of the Coulomb correction for like-flavour
  $\tilde q \tilde q$ production, i.e. the function $f_{qq}^{V+S}$ in
  eq. (54). Also note that~\cite{Beenakker:1996ch} choose to expand
  the cross section of $\tilde q\tilde g$ production in the variable
  $\bar\beta=\sqrt{1-4m_{\tilde q}m_{\tilde g}/(s-(m_{\tilde
      q}-m_{\tilde g})^2)}\approx \beta\sqrt{M/2m_r}$ which leads to
  the appearance of additional $\ln(m_r/M)$ terms. For this process
  they also observe an apparent non-factorization of the
  colour-averaged NLO threshold corrections from the Born cross
  section. This is nevertheless consistent with~\eqref{eq:NLOapprox}
  since the Born $\tilde q\tilde g$ cross section in the
  colour-triplet channel is not proportional to that for the other
  channels~\cite{Beenakker:2009ha}.  }

 \begin{table}[t]
   \centering
   \begin{tabular}{|c|c|c|c|c|c} \hline
    $\tilde q\bar{\tilde q}$ &$D_1=-\frac{4}{3}$&$D_8= 
   \frac{1}{6}$ &&\\\hline
          $\tilde q\tilde q$&  $D_{\bar 3}=-\frac{2}{3}$
       & $D_6=\frac{1}{3}$ &&\\ \hline
       $\tilde g\tilde q$&  $D_{3}=-\frac{3}{2} $
       & $D_{\bar 6}=-\frac{1}{2}\,,$ & $D_{15}=+\frac{1}{2}$&\\\hline
            $\tilde g\tilde g$ & $D_{1}=-3$&$D_8=-\frac{3}{2}$ &$D_{10}=0$ &$D_{27}=1$
     \\\hline
   \end{tabular}
   \caption{Numerical values of the coefficients of the Coulomb potential~\eqref{eq:dralpha} for squark and gluino production processes. 
   Negative values correspond to an attractive potential.}
   \label{tab:Coulomb}
 \end{table}

\begin{figure}[t!] 
  \centering
  \includegraphics[width=0.49 \linewidth]{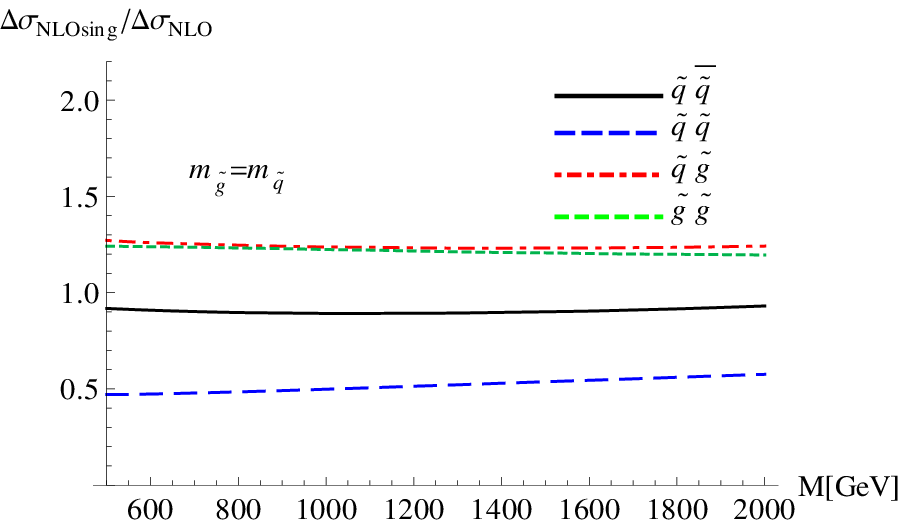}
  \includegraphics[width=0.49 \linewidth]{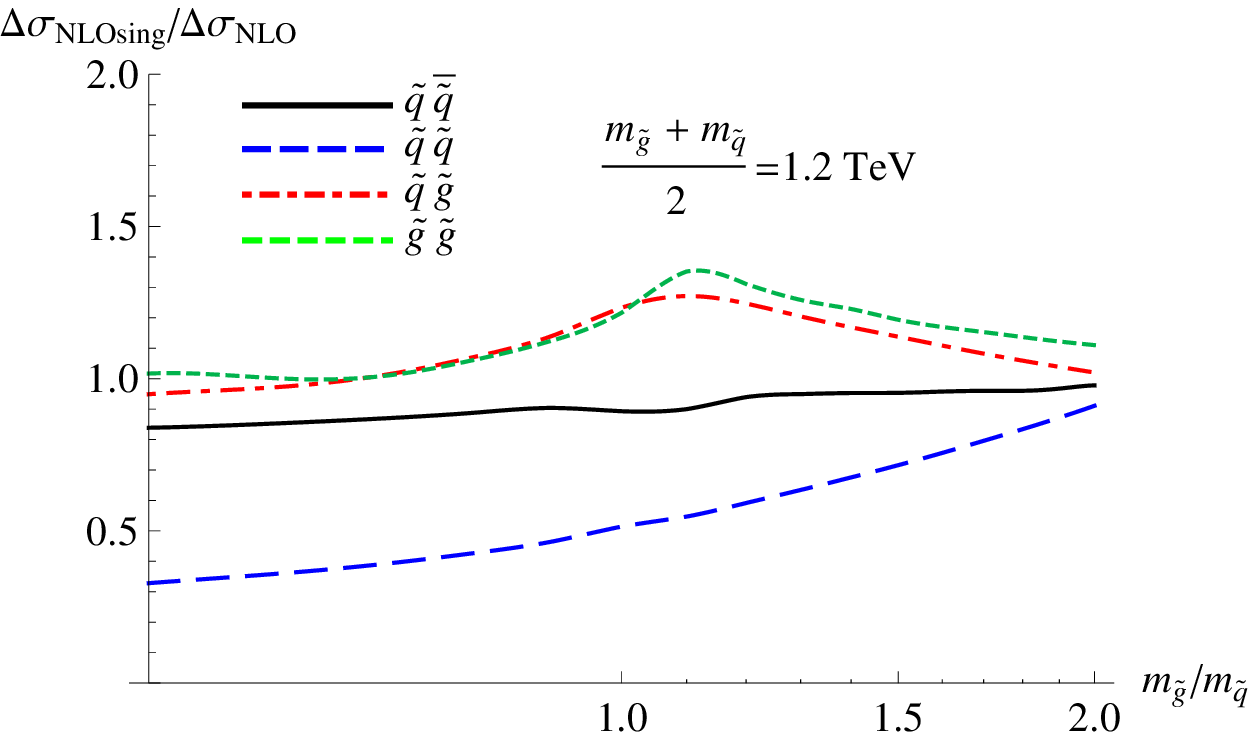}
  \caption{Ratio of the singular NLO contributions obtained from~\eqref{eq:NLOapprox} to the exact NLO corrections for the LHC with $\sqrt s=7$ TeV.  Left: Mass dependence for a fixed mass-ratio
    $m_{\tilde q}=m_{\tilde g}=M$. Right: Dependence
    on the ratio $m_{\tilde g}/m_{\tilde q}$ for a fixed average mass
    ($m_{\tilde{q}}+m_{\tilde{g}})/2=1.2$~TeV.}
  \label{fig:nlosing}
\end{figure}

 In Figure~\ref{fig:nlosing} we study to which extent the full NLO
 corrections as obtained from \texttt{PROSPINO} are approximated by
 the singular NLO corrections, obtained by dropping all constant terms
 from~\eqref{eq:NLOapprox}, including $\ln 2$ terms, and convoluting
 the resulting partonic cross section~\eqref{eq:partonic} with the
 parton luminosities.  For the Born cross sections
 $\hat\sigma^{(0),R_\alpha}_{pp'}$ in~\eqref{eq:partonic} the exact
 expressions, without use of the threshold approximation, have been
 kept, but colour channels with a vanishing threshold limit of the
 Born cross section at leading order in $\beta$ have been dropped.
 For the case of degenerate squark and gluino masses it is seen that
 the difference of the threshold-enhanced contributions to the full
 NLO corrections is at the $10-30\%$ level over the whole mass range
 considered, with the exception of the squark-squark production
 channel where the threshold contributions account for only $40-50\%$
 of the full NLO corrections.  For $m_{\tilde g}>m_{\tilde q}$ the
 singular terms overestimate the corrections for the processes
 involving gluinos, while the agreement for squark-antisquark
 production improves. For $m_{\tilde g}<m_{\tilde q}$ the singular
 terms approximate the full corrections very well for all processes
 apart from squark-squark production.\footnote{Note that one could improve the threshold approximation by including the constant terms in~\eqref{eq:NLOapprox} once the coefficients  $h_{pp'}^{(1),R_\alpha}$ are known. This could be particularly relevant for the squark-squark process where the threshold contributions to the NLO cross section are relatively small due to the smaller colour charges involved and an accidental cancellation of Coulomb corrections between the same-flavour and different-flavour production channels.}
 Comparing to
 Figure~\ref{fig:sigmaLO}, it is seen that the singular terms capture
 the NLO corrections to the dominant processes for larger mass ratios
 (i.e. squark-squark production for $m_{\tilde g}\sim 2 m_{\tilde q}$
 and gluino-pair production for $m_{\tilde g}\sim 0.5 m_{\tilde q}$)
 rather well.  For degenerate squark and gluino masses the quality of
 the threshold approximation for the dominant squark-squark and
 squark-gluino processes is somewhat worse.  In all cases, the
 inclusion of the threshold enhanced NLO corrections in addition to
 the Born terms improves the agreement with the full NLO results.
 This motivates the computation of the higher-order threshold-enhanced
 terms through resummation, as performed in the remainder of this
 work.

\subsection{Soft-gluon and Coulomb resummation}
\label{sec:resummation}
Next, we briefly review the formalism for the combined resummation of soft- and Coulomb-gluon corrections~\cite{Beneke:2009rj,Beneke:2010da} and provide the relevant ingredients for squark and gluino production at NLL accuracy.
We also discuss some features of our implementation  that differ from that used previously for squark-antisquark production in~\cite{Beneke:2010da}.

The combined soft-Coulomb resummation for the production of squarks
and gluinos is based on a factorization of the hard-scattering total
cross section for partonic subprocesses of the
type~\eqref{eq:part-process}. It can be shown that near the partonic
threshold,
\begin{equation}
\label{eq:threshold}
\hat s \sim (m_{\tilde{s}}+m_{\tilde{s}^\prime})^2,
\end{equation}
the partonic cross section factorizes into three
contributions~\cite{Beneke:2010da},   a
 hard function $H$,  a soft function $W$ containing soft gluons to all orders,  and a potential function $J$ summing Coulomb-gluon exchange:
\begin{equation}
\label{eq:fact}
  \hat\sigma_{pp'}(\hat s,\mu)
= \sum_{R_\alpha}H^{R_\alpha}_{pp'}(m_{\tilde q},m_{\tilde g},\mu)
\;\int d \omega\;
J_{R_\alpha}(E-\frac{\omega}{2})\,
W^{R_\alpha}(\omega,\mu)\, .
\end{equation}
Here $E=\sqrt{\hat s}-2 M$ is the energy relative to the production
threshold and the sum is over the colour
representations~\eqref{eq:susy-reps}.  In~\eqref{eq:fact} the
$s$-channel colour basis mentioned above, that  can be shown to diagonalize the soft
function to all orders~\cite{Beneke:2009rj}, is chosen for the
hard-scattering amplitudes. Independent of the
sparticle type, the soft function then depends only on the colour
representations of the initial-state partons and the irreducible
representation $R_\alpha$ of the sparticle pair appearing in the
decompositions~\eqref{eq:susy-reps}, in agreement with the picture
that soft-gluon radiation is only sensitive to the total colour charge
of the slowly moving sparticle pair~\cite{Bonciani:1998vc}.  The
formula~\eqref{eq:fact} has been derived in~\cite{Beneke:2010da} for
$S$-wave dominated production processes up to NNLL accuracy. This covers
all production processes of squarks and gluinos, apart from
quark-antiquark initiated stop-antistop production, that proceeds
through a $P$-wave. The applicability of the formalism to
stop production is discussed in Section~\ref{sec:stop} and
Appendix~\ref{sec:Pwave}.

It can be argued that the natural scale for the evaluation of the hard
function in~\eqref{eq:fact}, leading to well-behaved higher-order
corrections, is of the order of $\mu_h\sim 2M$, while the natural
scale for soft-gluon radiation is of the order of $\mu_s\sim
M\beta^2$. We use the momentum-space resummation formalism
of~\cite{Becher:2006nr,Becher:2006mr,Becher:2007ty} to evolve the soft
and hard functions from their natural scales to the factorization
scale $\mu_f$ used for the evaluation of the parton distribution
functions, commonly taken to be of the order of $\mu_f\sim M$. In this way,
logarithms of $\mu_s/\mu_f\sim\beta^2$ are summed to all orders. 
The precise prescription for the choice of the soft scale adopted in our calculation is discussed in~\ref{sec:scales}.
The exchange of multiple Coulomb gluons can be summed up using  the
method of Coulomb Green's functions in non-relativistic QCD~\cite{Fadin:1987wz,Hoang:2000yr,Beneke:1999qg}.

For resummation at NLL accuracy, the
leading-order hard and soft functions are required as fixed-order
input to the evolution equations.  The leading-order soft function is
trivial, $W^{(0)R_\alpha}(\omega)=\delta(\omega)$.  The leading-order
hard functions are obtained from the threshold limit of the Born cross
section for the colour channel $R_\alpha$~\cite{Beneke:2010da},
\begin{equation}
\label{eq:sigma-hard}
  \hat\sigma_{pp'}^{(0),R_\alpha}(\hat s)
\underset{\hat s\to 4M^2}{\approx}
 \frac{(2m_r)^2}{2\pi}\sqrt{\frac{E}{2m_r}} H^{(0),R_\alpha}_{pp'} \, .
\end{equation}  
Although
the Born-cross sections in the threshold limit appear on the left-hand
side in~\eqref{eq:sigma-hard}, we keep the exact expressions in our
numerical implementation, so the hard functions in practice depend on
$\hat s$. This incorporates some higher-order terms in $\beta$, albeit not
systematically.
\footnote{This
  corresponds to the treatment
  of~\cite{Kulesza:2009kq,Beenakker:2009ha,Beenakker:2010nq}, up to
  the fact that we set the hard function for a given production and
  colour channel to zero if the Born cross section vanishes at
  threshold at leading order in $\beta$, even if the full Born cross section for this channel is
  non-vanishing. This affects the subprocesses $q\bar q\to\tilde
  g\tilde g$ in the singlet channel and $q_iq_i\to \tilde q_i\tilde
  q_i$ in the triplet channel. The numerical effect is, however, negligible.}
 Here our current treatment differs from that used for squark-antisquark production 
 in~\cite{Beneke:2010da} where only the threshold limit of the Born cross section was used to compute the hard function.

For the resummation of Coulomb corrections, we use results for the
non-relativistic Coulomb Green's function obtained for top-quark
production at electron-positron colliders and stop
production~\cite{Fadin:1987wz,Bigi:1991mi}.
For positive values of $E$ and vanishing decay widths of the sparticles, the $S$-wave potential function is given by
the Sommerfeld factor
\begin{equation}
\label{eq:sommerfeld}
J_{R_\alpha}(E)=
\frac{(2 m_r)^2\pi D_{R_\alpha}\alpha_s }{2\pi}
\left(e^{\pi D_{R_\alpha}\alpha_s
\sqrt{\frac{2m_r}{E }}}-1\right)^{-1}
\, ,\qquad E>0,
\end{equation}
with the coefficients $D_{R_\alpha}$ of the Coulomb
potential given in~\eqref{eq:dralpha}.
For an attractive potential,  a 
series of bound states develops below threshold with energies
\begin{equation}
\label{eq:bound}
  E_n=-\frac{2m_{\text{r}}\alpha_s^2 D_{R_\alpha}^2}{4 n^2}\,.
\end{equation}
Their  contribution to the $S$-wave potential function is given by  
\begin{equation}
\label{eq:bound-GF}
J_{R_\alpha}^{\text{bound}}(E)=2
\sum_{n=1}^\infty \delta(E-E_n)
 \left(\frac{2m_{\text{r}} (-D_{R_\alpha})\alpha_s}{2 n}\right)^3
\,,\qquad E<0.
\end{equation}
For sufficiently broad squarks and gluinos with decay widths exceeding
the binding energy of the would-be bound states,
$\Gamma_{\tilde{s}}+\Gamma_{\tilde{s}'}>|E_1-E_2|$, the bound state
poles are smeared out by the finite lifetimes.\footnote{The actual formation of bound
  states is possible if the decay widths of the sparticles are smaller
  than the decay-rate of the bound state. For long-lived gluinos, or
  light stops without allowed two-body decays, higher-order
  corrections to bound-state production have been recently obtained
  in~\cite{Kauth:2009ud,Younkin:2009zn}. Since the bound states decay
  into gluon or photon pairs, this scenario leads to very different
  collider signatures compared to the missing-energy signatures of
  continuum production and is not considered further here.}  We consider here a
situation where the widths of the squarks and gluinos are large enough
to prevent the formation of bound states, but small enough that the
use of a narrow-width approximation is justified, which is
the case for SUSY scenarios with moderate
mass ratios of squarks and gluinos where 
$\Gamma_{\tilde{s}}/M_{\tilde{s}} \sim 1\%$. The
contributions to the total cross section below the nominal production
threshold, $\hat s=2 M$, can then be included by setting the sparticle
widths to zero and including the bound-state
poles~\eqref{eq:bound-GF}. For other observables, the finite width can be taken into account to a first approximation by the replacement
$E\to E+i (\Gamma_{\tilde{s}}+\Gamma_{{\tilde{s}}'})/2$ in the
potential function, see e.g.~\cite{Hagiwara:2009hq,Kauth:2011vg,Kauth:2011bz} for recent studies of the invariant-mass spectrum of gluino-pair and squark-gluino production. The study of finite-width corrections for larger
decay widths (e.g. for gluino masses $m_{\tilde g}\gtrsim 2 m_{\tilde
  q}$)  is left for future work.  In
the numerical results presented in this work, the contributions of the
bound state poles for $E<0$ will always be included in our default
implementation, and are convoluted with the resummed soft function as
described in~\cite{Beneke:2011mq}. Note that in the previous results
for squark-antisquark production~\cite{Beneke:2010da} the bound-state
corrections have been added without soft-gluon resummation.
 
The resummed cross section at NLL accuracy is obtained by inserting the
potential function~\eqref{eq:sommerfeld} and the solutions to the evolution
equations of the hard and soft functions~\cite{Becher:2009kw,Beneke:2009rj} into the
factorization formula~\eqref{eq:fact}. Using the solutions in momentum-space
obtained in~\cite{Becher:2007ty}, the NLL cross section is written as
\begin{align}
\hat\sigma^{\text{NLL}}_{pp'}(\hat s,\mu_f)
 =&\sum_{R_\alpha}H^{(0),R_\alpha}_{pp'}(\mu_h)\,
 U_i(M,\mu_h,\mu_s,\mu_f)
\frac{e^{-2 \gamma_E \eta}}{\Gamma(2 \eta)}\,
\int_0^\infty \!d \omega \,
\frac{ J_{R_{\alpha}}(M \beta^2-\tfrac{\omega}{2})}{\omega} 
\left(\frac{\omega}{2M}\right)^{2 \eta}\,.
\label{eq:resum-NLL}
\end{align}
Here the label $i$ jointly refers to the colour of the initial-state partons and the
representation $R_\alpha$ of the sparticle pair.  The function
$\eta=\frac{2 \alpha_s(C_r+C_{r'})}{\pi}\ln(\mu_s/\mu_f)+\dots$ contains single
logarithms, while the resummation function $U_i$ sums the Sudakov double
logarithms $\alpha_s\log^2\frac{\mu_h}{\mu_f}$ and
$\alpha_s\log^2\frac{\mu_s}{\mu_f}$. Explicit expressions up to NLL accuracy
are given in Appendix \ref{sec:resfunc}. For $\mu_s<\mu_f$ the function $\eta$ is negative and the
factor $\omega^{2\eta-1}$ in the resummed cross section~\eqref{eq:resum-NLL}
has to be understood in the distributional sense, as discussed in detail
in~\cite{Beneke:2011mq}.  We have used the non-relativistic expression $E= M
\beta^2$ in the argument of the potential function, that is valid near the
partonic threshold~\eqref{eq:threshold}.  This follows the default treatment
of top-pair production in~\cite{Beneke:2011mq} and leads to the customary
expansion of the cross section in $\beta$ (see Eq. \ref{eq:NLOapprox}). We
also perform this replacement in the definition of the hard
functions~\eqref{eq:sigma-hard}.  The difference between this default
implementation and the results obtained by consistently keeping the expression
$E=\sqrt{\hat s}-2M$ (as in the previous results for squark-antisquark
production~\cite{Beneke:2010da}) will be used to estimate the effect of
subleading terms in the cross section, as discussed in Section~\ref{sec:scales}.

In order to assess the importance of the Coulomb corrections and to compare to
the results of the Mellin approach~\cite{Beenakker:2009ha} we will also
present results without Coulomb resummation, obtained by inserting the trivial
potential function $J^{(0)}(E)=\frac{(2 m_{r})^2}{2\pi}\sqrt{E/2m_r}$
into~\eqref{eq:resum-NLL}. In this approximation, that we will denote by
NLL${}_{s+h}$, a fully analytical expression for the resummed cross section can be
obtained:
\begin{equation}
\label{eq:NLLs}
\hat\sigma^{\text{NLL}_{s+h}}_{pp'}=
\sum_{R_{\alpha}} \,\hat\sigma^{(0)}_{pp'}(\hat s,\mu_h)
  U_i(M,\mu_h,\mu_s,\mu_f)
\frac{\sqrt{\pi} e^{-2\eta\gamma_E}}{2\Gamma(2\eta+\frac{3}{2})}\,
\beta^{4\eta} .
\end{equation}

Since contributions to the cross section from outside the threshold region can be numerically non-negligible, we match the NLL resummed cross section
 to the fixed-order NLO calculation by subtracting the NLO expansion of the NLL expression and 
 adding back the full NLO corrections:
\begin{equation}
  \hat\sigma^{\text{matched}}_{pp'}(\hat s)
  =\left[\hat\sigma^{\text{NLL}}_{pp'}(\hat s)-
    \hat\sigma^{\text{NLL}(1)}_{pp'}(\hat s)\right]
 + \hat\sigma^{\text{NLO}}_{pp'}(\hat s)\,,
\label{eq:cross_matched}
\end{equation}
where $\hat\sigma^{\text{NLO}}_{pp'}(\hat s)$ is the fixed-order NLO
cross section obtained in standard perturbation theory, as implemented
in \texttt{PROSPINO} \cite{Beenakker:1996ed}, and
$\hat\sigma^{\text{NLL}(1)}_{pp'}$ is the resummed cross section
expanded to NLO, as given in~\cite{Beneke:2010da}.  The total hadronic
cross section at NLL is then obtained by
convoluting~\eqref{eq:cross_matched} with the parton luminosity, as
in~\eqref{eq:sigma-had}.

\subsection{Stop-antistop production}
\label{sec:stop}
\begin{figure}[t!]
\begin{center}
\includegraphics[width=0.49 \linewidth]{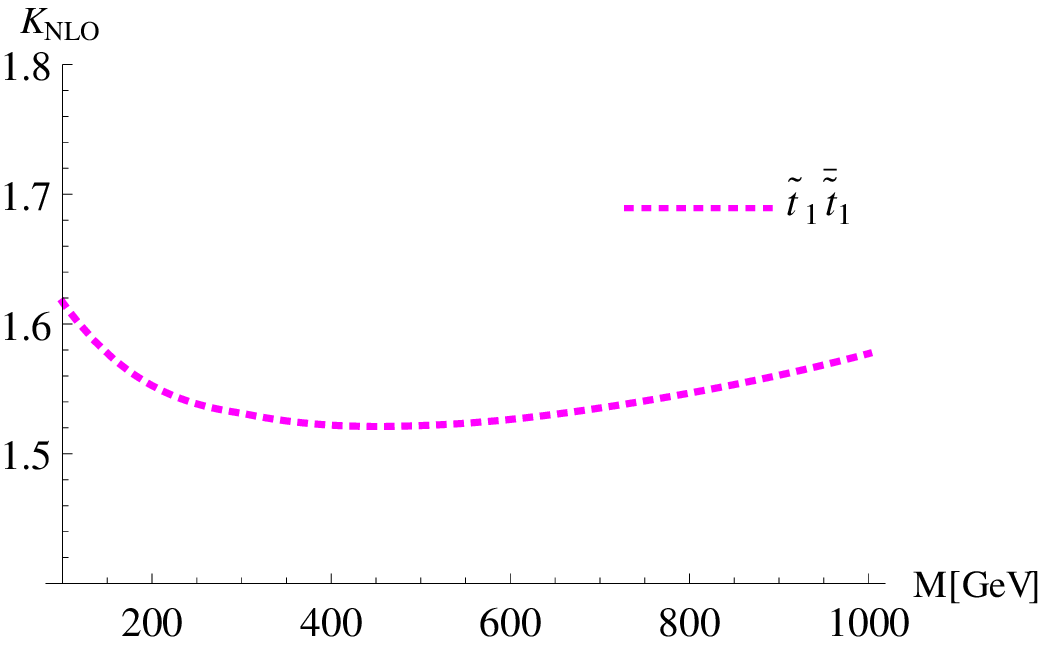}
\includegraphics[width=0.49 \linewidth]{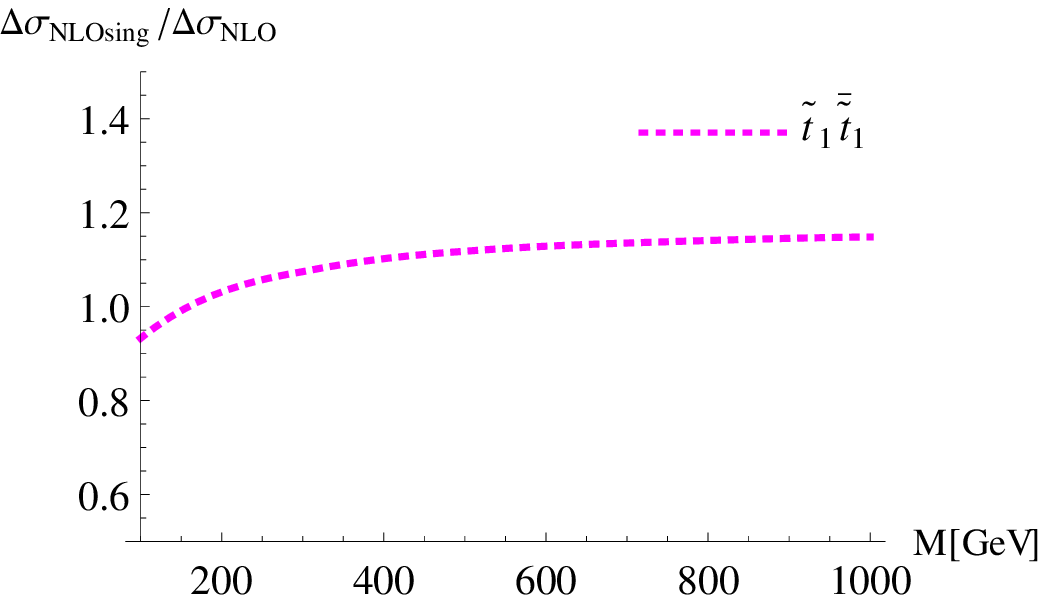}
\end{center}
\caption{Left: NLO $K$-factor for stop-pair production at $\sqrt{s}=7\,$TeV as a function of the stop mass.
Right: ratio of the singular NLO contributions obtained from Eqs. (\ref{eq:NLOapprox}) and (\ref{eq:NLOapprox-P}) to the full NLO cross section for
$PP \rightarrow  \tilde{t}_1 \bar{\tilde{t}}_1 $. } 
\label{fig:Kfact_stop}
\end{figure}

Beside the channels listed in (\ref{eq:processes}), in Section \ref{sec:results} we will also present predictions for stop-pair production:
\begin{figure}[t]
\begin{center}
\includegraphics[width=0.7 \linewidth]{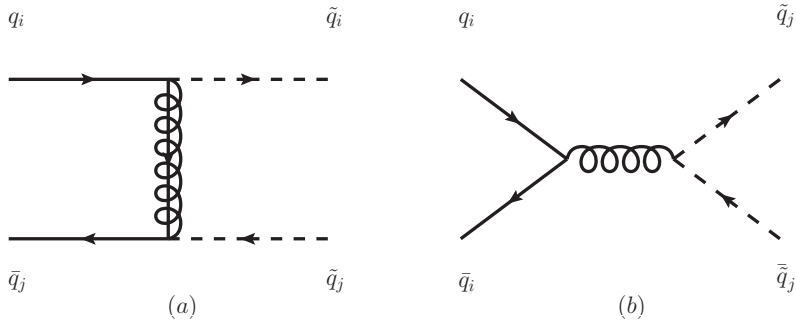}
\end{center}
\caption{Tree-level diagram topologies contributing to $q_k \bar{q}_l \rightarrow \tilde{q}_i \bar{\tilde{q}}_j$.}
\label{fig:squarktree}
\end{figure}
\begin{eqnarray}
\label{eq:stops}
gg,q_i\bar q_i \rightarrow \tilde{t}_j \bar{\tilde{t}}_j \,, 
\end{eqnarray}
where only the initial states appearing at leading order have been shown.
The NLO SUSY-QCD corrections have been computed in~\cite{Beenakker:1997ut} and are implemented in \texttt{PROSPINO} \cite{Beenakker:1996ed}.
Contrary to the light-flavour squark case, in most scenarios the mixing of the two weak eigenstates $\tilde{t}_L$, $\tilde{t}_R$,
and the mass difference of the resulting mass eigenstates, $\tilde{t}_1= \tilde{t}_L \cos \theta_{\tilde{t}} +\tilde{t}_R \sin \theta_{\tilde{t}} $,  
$\tilde{t}_2= -\tilde{t}_L \sin \theta_{\tilde{t}} +\tilde{t}_R \cos \theta_{\tilde{t}} $, is non-negligible. 
Off-diagonal production of the mass eigenstates, e.g. $\tilde
t_1\bar{\tilde {t_2}}$, appears at NLO in SUSY-QCD, and through electroweak
contributions. It is therefore suppressed compared to diagonal production~\cite{Beenakker:1997ut,Bozzi:2005sy}  and
will not be considered here.
It must also be mentioned that because of the absence of a significant top-quark component inside the nucleon the processes
$\tilde{t} \tilde{t}$ and $\tilde{g} \tilde{t}$ first contribute to the cross section at NLO, and are thus numerically suppressed.
The NLO $K$-factor for the process $PP \rightarrow  \tilde{t}_1 \bar{\tilde{t}}_1 $ is shown in Figure \ref{fig:Kfact_stop} for 
a centre-of-mass energy of 7 TeV, and the mass range $M=100-1000\,$GeV. As can be seen, NLO corrections are in the 
$50-60\%$ range. The predictions for the second mass eigenstate differ only in the fixed-order NLO results, and for a given mass the numerical difference between the cross sections for $\tilde t_1\bar{\tilde t}_1$ and  $\tilde t_2\bar{\tilde t}_2$ production is below $2\%$ for the mass range considered in this work.  We therefore omit results for the process $PP \rightarrow  \tilde{t}_2 \bar{\tilde{t}}_2$. 

Contrary to the production of a light squark-antisquark pair, stop-pair production in the $q \bar{q}$ channel cannot be mediated by a 
$t$-channel diagram like in Figure \ref{fig:squarktree}(a), again due to the extreme suppression of top-quark PDFs inside the proton. 
As a result, at LO in QCD a stop-antistop pair is produced in a $P$-wave state
in quark-antiquark collisions.
As shown in Appendix~\ref{sec:Pwave}, the resummation formalism can be extended in a straightforward way to $q\bar{q}$-initiated stop-antistop production at NLL, 
and the only modification of~\eqref{eq:fact} is the replacement of the potential function by that appropriate for $P$-wave processes.
For stable particles, above threshold the result is given
by~\cite{Bigi:1991mi} (see also~\cite{Cassel:2009wt})
\begin{equation}
\label{eq:P-GF}
\begin{aligned}
  J_{R_\alpha}^P (E)&= 2m_{r} E\left(1+\frac{(\alpha_sD_{R_\alpha})^2 2m_{r}}{4E}
 \right)  J_{R_\alpha} (E)\,,\qquad E>0.
\end{aligned}
\end{equation}
The bound-state contributions of the $P$-wave Green's function can be
found in~\cite{Bigi:1991mi} but are not needed here, since only the
repulsive colour-octet channel appears in our application to stop-pair
production. 
By expanding~\eqref{eq:P-GF} in the strong coupling constant one obtains the coefficients of the fixed-order Coulomb corrections. While
the  one-loop Coulomb correction agrees with that for $S$-wave
production~\eqref{eq:NLOapprox}, the second Coulomb
correction  $\frac{(D_{R_\alpha}\alpha_s)^2}{12}
(\pi^2+3)/\beta^2$  differs from the $S$-wave case.
Due to the different normalization of the $P$-wave Green's function, the definition
of the leading-order hard functions for $P$-wave production reads
\begin{equation}
 \hat\sigma_{pp',P}^{(0),R_\alpha}(\hat s)
\underset{\hat s\to 4M^2}{\approx}
 \frac{(2m_r)^4}{2\pi}\sqrt{\left(\frac{E}{2m_r}\right)^3} H^{(0),R_\alpha}_{pp'} \, .
\end{equation}

In addition to the the combined soft/Coulomb resummation, we will
again consider the NLL${}_{s+h}$ approximation where the trivial
$P$-wave potential function $J^{(0)}(E)=\frac{(2m_{r})^4}{2\pi}
(E/2m_r)^{3/2}$ is used in the resummation formula, leading to the
analytical result
\begin{equation}
\label{eq:NLLs-P}
\hat\sigma^{\text{NLL}_{s+h}}_{pp',P}=
\sum_{R_{\alpha}} \,\hat\sigma^{(0)}_{pp'}(\hat s,\mu_h)
  U_i(M,\mu_h,\mu_s,\mu_f)
\frac{3\sqrt{\pi} e^{-2\eta\gamma_E}}{4\Gamma(2\eta+\frac{5}{2})}\,
\beta^{4\eta} .
\end{equation}

In analogy to the $S$-wave result~\eqref{eq:NLOapprox}, one can use the resummation formalism to obtain the threshold-enhanced one-loop scaling functions for $P$-wave production in the colour channel $R_\alpha$ from initial-state partons in the representations $r$ and $r'$:
\begin{multline}
\label{eq:NLOapprox-P}
   f_{pp',P}^{(1)R_\alpha}= - \frac{2 \pi^2 D_{R_\alpha}}{\beta}
   \sqrt{\frac{2 m_r}{M}} +
4\,(C_r+C_{r'}) \bigg[\ln^2\left(\frac{8 E}{\mu_f}\right)
+\frac{104}{9} -\frac{11 \pi^2}{24}\bigg]
\\
- \,4 \,\left(C_{R_\alpha}+ \frac{16}{3}\,(C_r+C_{r'})\right) \,
\ln\left(\frac{8 E}{\mu_f}\right) 
+ \frac{44}{3}  C_{R_\alpha} +h^{(1)}_i (\mu_f) \, .
\end{multline}
In agreement with~\cite{Beenakker:2010nq} one finds that the coefficient of the single logarithms related to initial-state radiation is multiplied by a factor of $\frac{4}{3}$ compared 
to the $S$-wave case while the double logarithm and the logarithms related to final state radiation proportional to $C_{R_\alpha}$ are unchanged. In addition, the constant terms 
are different which is irrelevant at NLL accuracy but has to be taken into account if one aims to extract  the one-loop hard function from a computation of the NLO cross section.
In Figure \ref{fig:Kfact_stop} we study the accuracy of the threshold approximation defined by inserting the NLO singular terms, obtained from Eqs. (\ref{eq:NLOapprox}) and (\ref{eq:NLOapprox-P}) by dropping constant terms, into~\eqref{eq:partonic}. The ratio of the singular NLO corrections to  the full NLO corrections to the hadronic cross section obtained using \texttt{PROSPINO} is shown in Figure \ref{fig:Kfact_stop} (right plot). Analogously to squark-antisquark production, the threshold terms provide an excellent approximation of the full NLO result.

\subsection{Scale choices}
\label{sec:scales}

As explained in Section~\ref{sec:resummation}, the resummed partonic
cross section (\ref{eq:fact}) depends on a number of scales related to
the factorization of hard, soft and Coulomb effects.  The dependence
on these scales would cancel in the exact result, but a residual
dependence remains at a given logarithmic order.  As already pointed
out, our default choice for the factorization and hard scales are
$\mu_f\equiv M$ and $\mu_h \equiv 2 M$, respectively. On the other
hand, the resummation of all NLL effects related to Coulomb exchange
requires that the scale in the potential function $J_{R_\alpha}$ is
chosen of the order of $\sqrt{2 m_r M} \beta$, which is the typical virtuality of
Coulomb gluons. A more detailed analysis shows in fact that for an
attractive Coulomb potential the Coulomb scale freezes when $\beta
\sim |D_{R_\alpha}| \alpha_s$, due to bound-state formation. We thus
choose the scale in $J_{R_\alpha}$ to be
\begin{equation} \label{eq:muC}
\mu_C = \text{Max} \left\{2 \alpha_s(\mu_C) m_r |D_{R_\alpha}|,2 \sqrt{2 m_r M} \beta\right\} \, . 
\end{equation} 
Note that, for a repulsive potential, $D_{R_\alpha}>0$, no bound states arise, so that (\ref{eq:muC}) is not completely
justified. However in this case resummation of Coulomb corrections leads to small effects, and $J_{R_\alpha}$ vanishes for small $\beta$,
so that the precise choice of $\mu_C$ in this limit has a negligible numerical impact on predictions of the cross section.     

The choice of the soft scale $\mu_s$ presents some subtleties. The
exponentiation of all NLL $\ln \beta$-terms in the partonic cross
section would require a choice $\mu_s \sim M \beta^2$. However a
running scale leads to strong oscillations of the cross section for
small $\beta$, due to the prefactor $e^{-2 \gamma_E \eta}/\Gamma(2
\eta)$ in (\ref{eq:resum-NLL}), amplified by the factor
$\omega^{2\eta}$ and terms in the function $U_i$, and eventually hits
the Landau pole of the strong coupling constant $\alpha_s$ when $\beta
\to 0$. To overcome these problems two different approaches have been
used in the literature:
\paragraph{Fixed $\mu_s$:} In \cite{Becher:2006nr,Becher:2006mr,Becher:2007ty} the choice of a fixed soft scale was advocated. 
Such a scale is determined from the minimization of the one-loop \emph{soft} corrections to the hadronic cross section,
\begin{equation}
\label{eq:def-mus}
0 =\mu_s\frac{d}{d \mu_s}
\sum_{p,p'}\int_{\tau_0}^1 d\tau \,L_{pp'}(\tau,\mu_s)
\frac{\hat\sigma^{(1)}_{pp',\text{soft}} (\tau s,\mu_s)}{
\sigma^{(0)}_{N_1N_2}(s,\mu_s)}\,.
\end{equation}
In this approach threshold logarithms are resummed  in an average sense and not locally at the level
of the partonic cross section. However one can argue that, for threshold dominated processes, the choice (\ref{eq:def-mus}) 
preserves the hierarchy between the soft and short-distance scales
and that logarithmic corrections $\log(1-\tau_0)$ to the {\it
  hadronic} cross section are correctly resummed. This was the method
adopted in~\cite{Beneke:2010da} for resummation of the
squark-antisquark production cross section. The explicit value of the
scales determined with the minimization procedure (\ref{eq:def-mus})
are given in Eq. (\ref{eq:muS}) in Appendix \ref{sec:bcutdeterm}.

\paragraph{Running $\mu_s$:} For the NNLL resummation of soft effects in $t \bar t$ production presented in 
\cite{Beneke:2011mq} a different approach was adopted. There a running soft scale, 
\begin{equation} \label{eq:runmus_a}
\mu^{>}_s = k_s M \beta^2 \, ,
\end{equation}
was used in the interval $\beta>\beta_{\text{cut}}$, and replaced by a fixed soft scale
\begin{equation} \label{eq:runmus_b}
\mu^{<}_s = k_s M \beta_{\text{cut}}^2
\end{equation}
below the cutoff. With this scale choice, logarithms of $\beta$ are exponentiated locally in the \emph{partonic} cross section in the large-$\beta$ region, where the use of a fixed 
soft scale cannot be a priori justified.
On the other hand if $\beta_{\text{cut}}$ is not too big, in the lower interval the \emph{hadronic} cross section
is in fact dominated by logarithms of $\beta_{\text{cut}}$, as can be explicitly checked by convoluting the partonic 
cross section with toy parton luminosities~\cite{Becher:2007ty}, so that the use of a fixed scale once again correctly resums the dominant 
logarithms. The precise value of $\beta_{\text{cut}}$ is chosen through the prescription described in \cite{Beneke:2011mq},
which is reviewed in Appendix \ref{sec:bcutdeterm}. The default choice for the prefactor $k_s$ adopted here is $k_s=1$. We have observed that, 
for the SUSY processes considered here, the NLL expression and its NNLO expansion are generally stable against variations of $k_s$ for this choice.\footnote{Our choice of $k_s$ deviates from the one adopted in \cite{Beneke:2011mq}, 
where $k_s=2$. This corresponds to resumming some of the $\ln 2$ terms in the fixed-order cross section alongside the 
threshold-enhanced $\ln \beta$ contributions.}

The two possible choices of the soft scale $\mu_s$ just discussed are one of the ambiguities associated with threshold 
resummation. Others are related to the choice of the hard and Coulomb scales and to power-suppressed terms which 
are not controlled by resummation. Additionally, one has to consider the ambiguity arising from the choice of the factorization
scale $\mu_f$. The latter clearly also applies to the fixed-order NLO result. Thus, to reliably ascertain the residual uncertainty 
of the fixed-order and resummed results we present in Section \ref{sec:results}, we adopt the following procedure: 
\begin{itemize}
\item {\bf Scale uncertainty:} for both the NLO and NLL result the factorization scale $\mu_f$ is varied between half and twice the 
default value, i.e. $M/2<\mu_f<2 M$. For the NLL result, this is done keeping the other scales $\mu_h$, $\mu_C$ $\mu_s$ and 
the parameters $\beta_\text{cut}$ and $k_s$ fixed.
\item {\bf Resummation uncertainty:} both hard and Coulomb scales are varied between half and twice the default values, i.e.
$M<\mu_h<4 M$ and $\mu_C^{(0)}/2<\mu_C<2 \mu_C^{(0)}$, where $\mu_C^{(0)}$ is the solution of the implicit equation 
(\ref{eq:muC}). In addition, for the NLL implementation with a fixed soft scale, $\mu_s$ is varied between half and twice
its default value, while for the running-scale implementation uncertainties related to the choice of $\beta_\text{cut}$ and $k_s$
are estimated according to the procedure given in \cite{Beneke:2011mq} (and reviewed in Appendix \ref{sec:bcutdeterm}).  
Finally, as anticipated below (\ref{eq:resum-NLL}), we take the difference in parametrizing
the resummed cross section in terms of $\beta$ or $\hat{E}$ as a measure of the effect of power-suppressed terms. 
All the scales and the parameters $\beta_\text{cut}$ and $k_s$ are varied one at the time keeping the other fixed to their central values, and the
resulting errors are summed in quadrature.
\item {\bf PDF uncertainty:} we estimate the error  due to uncertainties in the PDFs  
using the $68\%$ confidence level eigenvector set of the MSTW08NLO 
PDFs~\cite{Martin:2009iq}.
\end{itemize}
An additional source of error arises from the uncertainty on the $\alpha_s$-determination. This effect has been found to be of the order of $3\%$ for the NLO cross sections of squark-squark, squark-antisquark and squark-gluino production and up to $8\%$ for gluino-pair production~\cite{Beenakker:2011fu}. We expect a similar uncertainty of the NLL results.

In the following we will often refer to the sum in quadrature of scale and resummation uncertainty as ``total theoretical uncertainty".       
Note that the terminology adopted here differs slightly from the one used for $t \bar{t}$ production in  \cite{Beneke:2011mq} where the errors from variation of the hard and Coulomb scales, and of the soft scale for the fixed-scale
implementation, had been incorporated into the scale uncertainty, while we consider them as resummation ambiguities. 
Additionally, in \cite{Beneke:2011mq} independent and simultaneous variations of the factorization and renormalization scale have been considered, whereas in this work 
we identify the factorization and renormalization scales and vary them as one scale, i.e. $M/2<\mu_f\equiv \mu_r<2 M$. This is the default 
procedure implemented in the numerical code \texttt{PROSPINO} used for the computation of the fixed-order NLO result \cite{Beenakker:1996ed}.

\begin{figure}[p]
\begin{center}
\includegraphics[width=0.49 \linewidth]{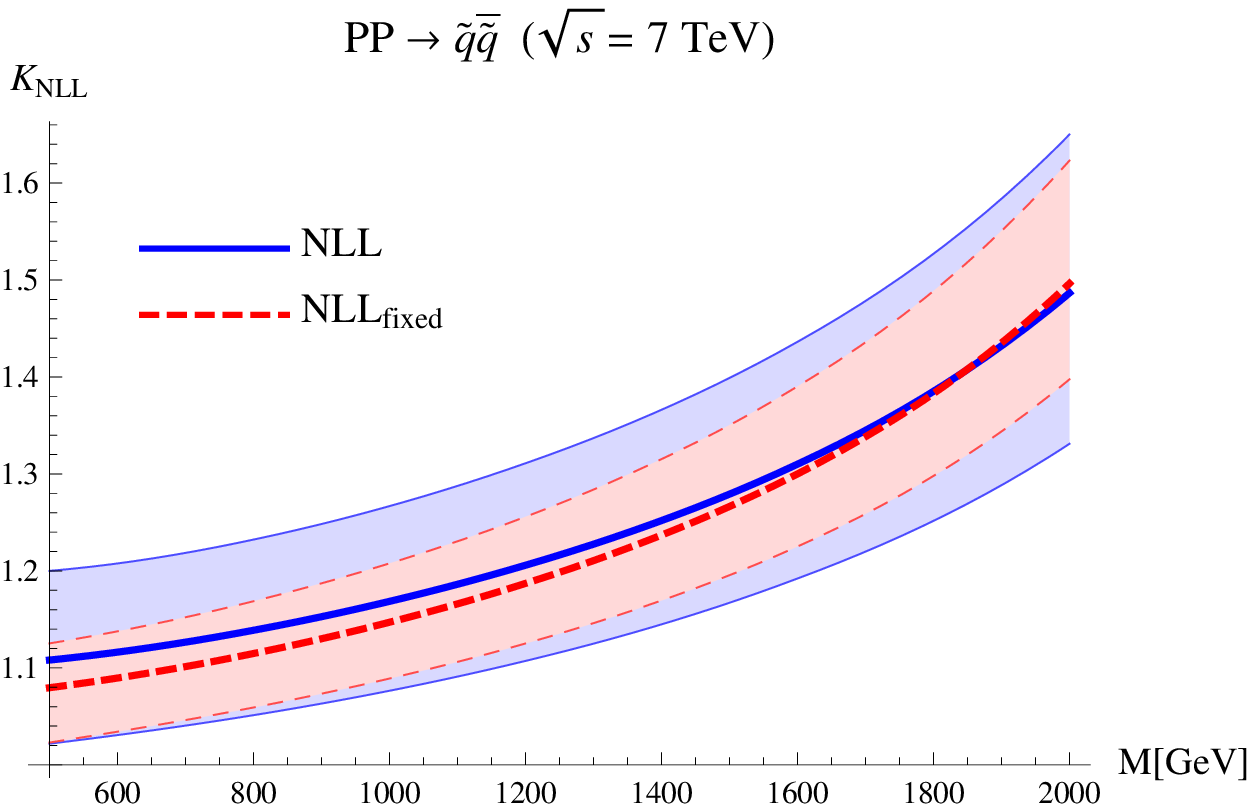}
\includegraphics[width=0.49 \linewidth]{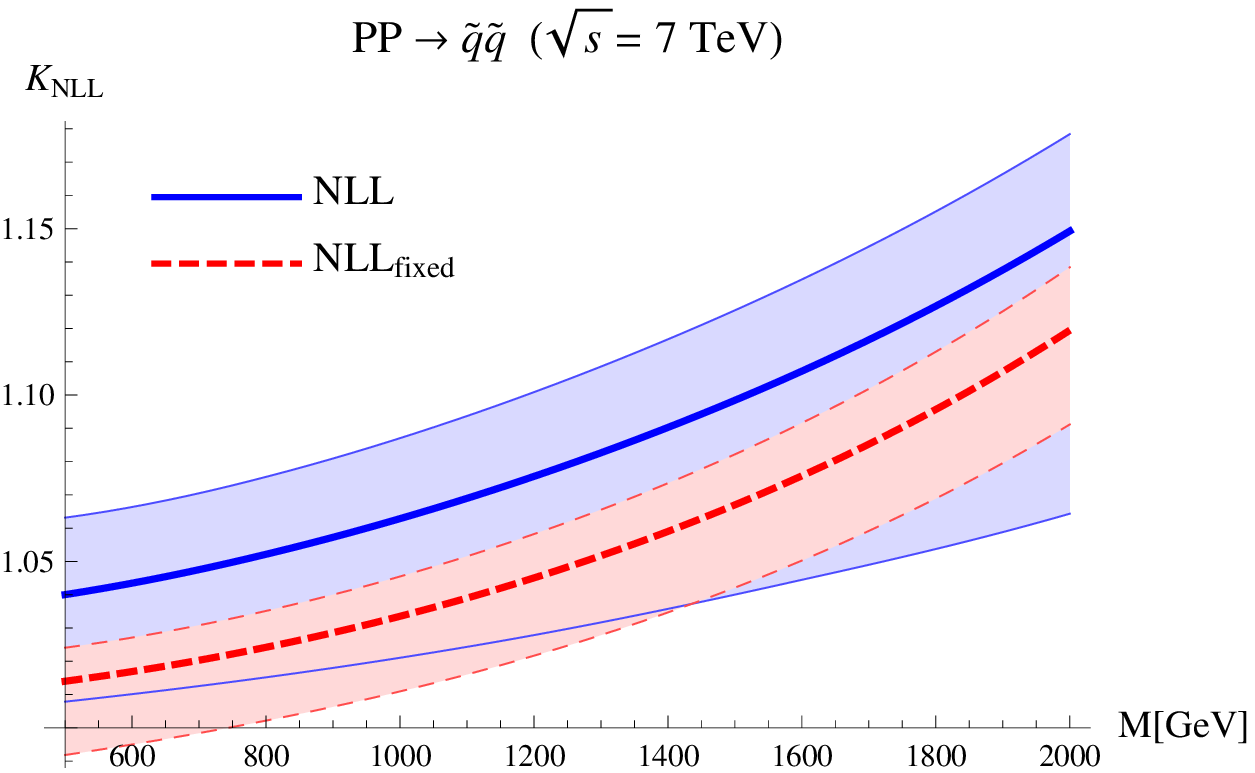}\\
\includegraphics[width=0.49 \linewidth]{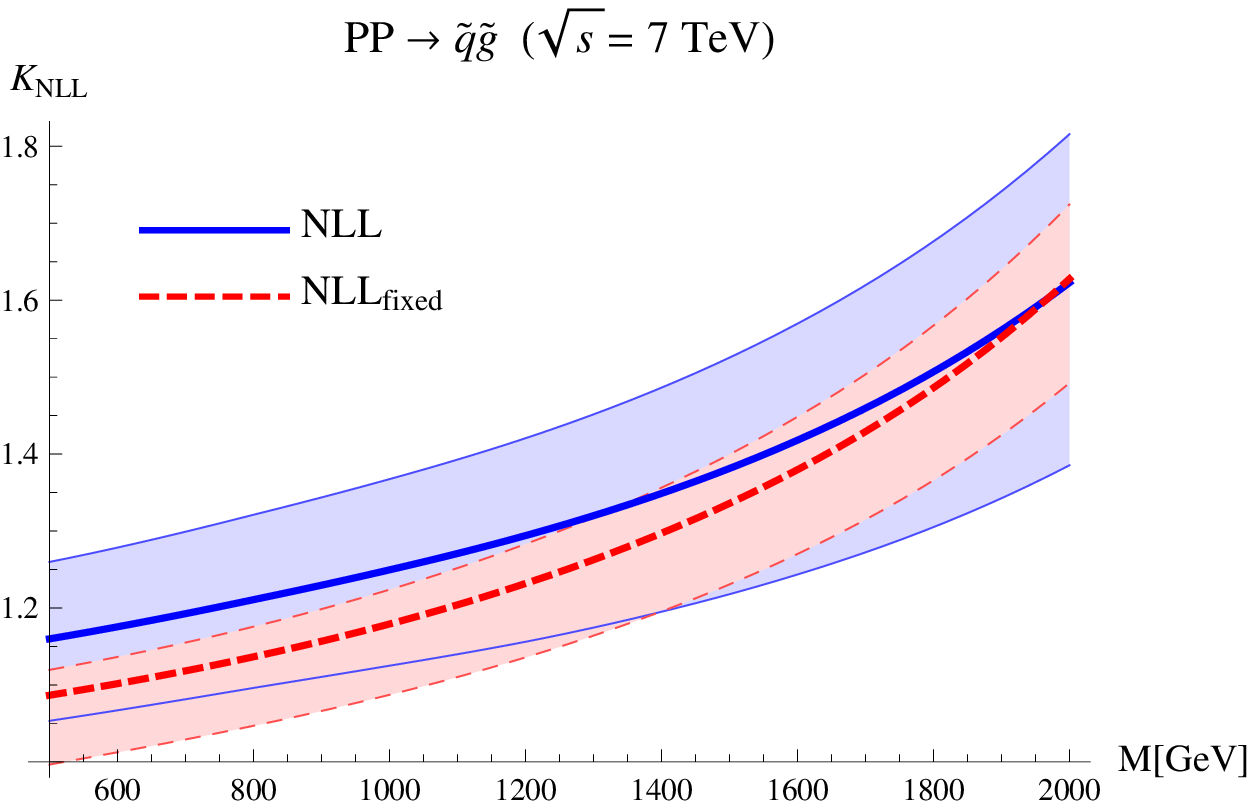}
\includegraphics[width=0.49 \linewidth]{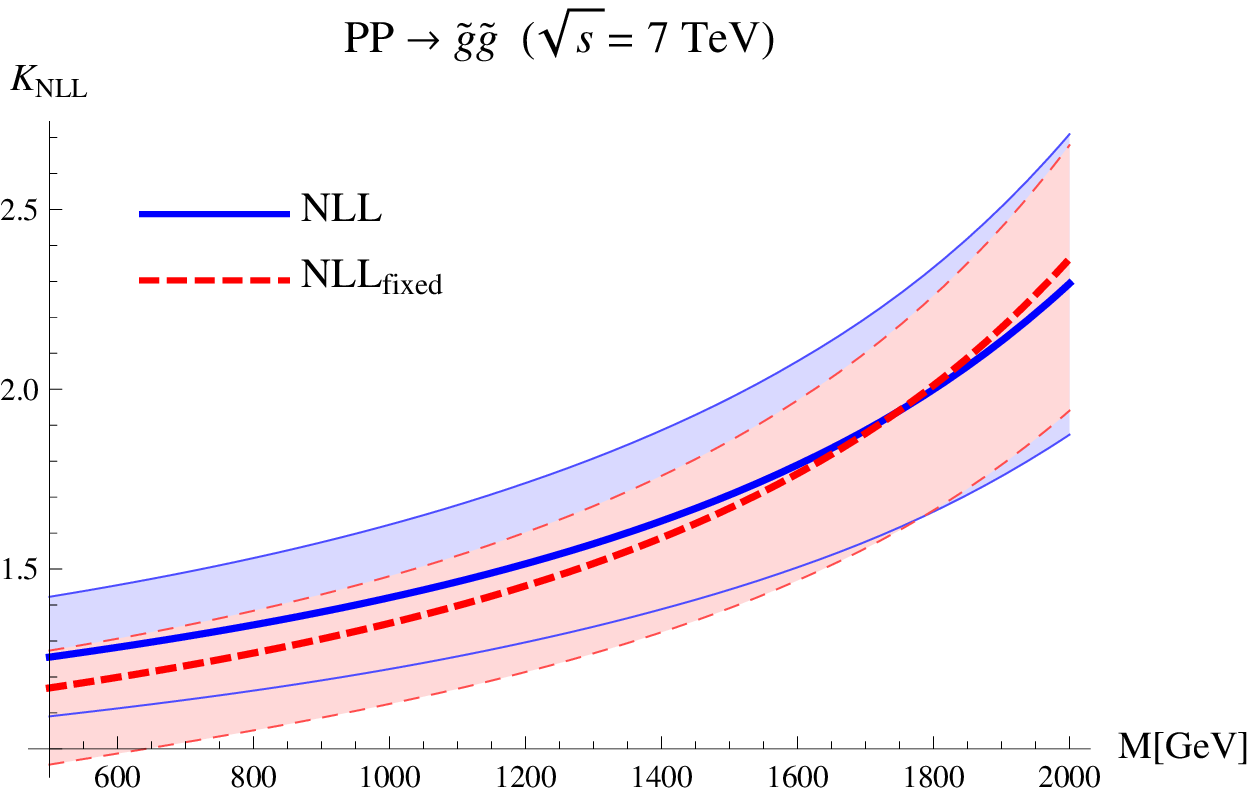}\\
\includegraphics[width=0.49 \linewidth]{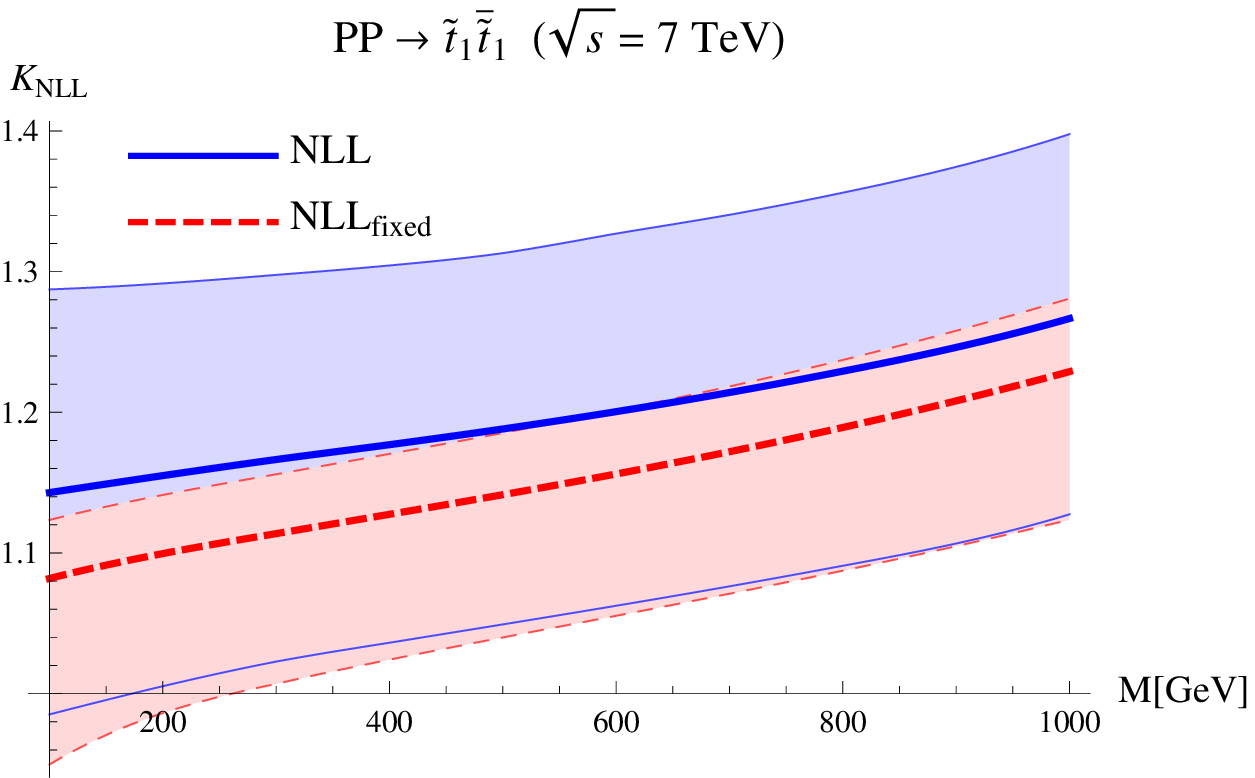}
\includegraphics[width=0.49 \linewidth]{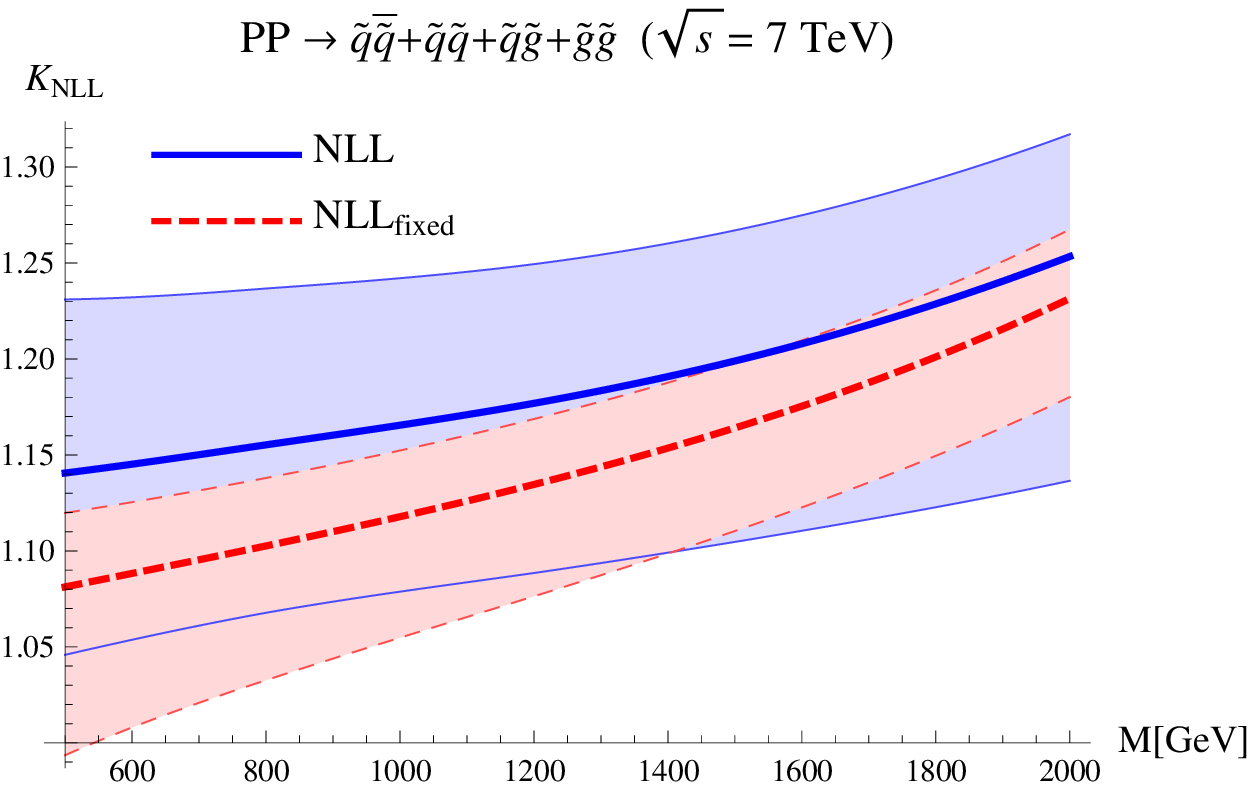}
\end{center}
\caption{Resummation uncertainty for the NLL resummed result with a running soft scale (NLL, solid blue) and a fixed soft scale (NLL$_\text{fixed}$, dashed red) for
 squark-antisquark (top-left), squark-squark (top-right),
  squark-gluino (centre-left), gluino-gluino (centre-right),
  stop-antistop (bottom-left) production and the inclusive gluino and light-flavour squark cross section (bottom-right) at LHC with $\sqrt{s} = 7\,$ TeV.
  The central line 
represents the $K$-factor for the default scale choice, while the band gives the resummation uncertainty associated with the result.  
See text for explanation.}
\label{fig:fix_VS_run}
\end{figure}

It is interesting to study how the choice of a fixed or running soft
scale affects the NLL resummed cross section, especially in view of
the uncertainties just discussed.  In Figure \ref{fig:fix_VS_run} we
plot the NLL $K$-factor, defined in Eq. (\ref{eq:Kfact}), as a
function of a common SUSY mass $M=m_{\tilde g}=m_{\tilde q}$ for the
four processes listed in (\ref{eq:processes}), for a centre-of-mass
energy of $7\,$TeV (the situation for $\sqrt{s} =\,$ 14 TeV is
qualitatively similar). Results for the stop-pair production process
and the total SUSY production rate are also shown.  The thick lines
represent the central values for the two implementations, whereas the
bands (delimited by thinner lines) correspond to the resummation
uncertainty as defined above. The central values are in good agreement
for squark-antisquark and gluino-pair production, and for
squark-gluino production at larger masses. For squark-squark
production the agreement is less satisfactory, especially for smaller
masses. This is consistent with the observation from
Figure~\ref{fig:nlosing} that the NLO corrections for squark-squark
production are not as dominated by the threshold contributions as
those for the other processes.  In all cases, however, the two
different NLL predictions are consistent with each other once the
uncertainty associated with the resummation procedure is taken into
account. It can also be seen that the uncertainty band for the
fixed-scale implementation NLL$_\text{fixed}$ is mostly contained
inside the uncertainty band of the running-scale result, with the
possible exception of the small-mass region. In light of this, in
Section \ref{sec:results} we will take the matched NLO/NLL result,
Eq. (\ref{eq:cross_matched}), with a running soft scale,
Eqs. (\ref{eq:runmus_a}) and (\ref{eq:runmus_b}), as our default and
best prediction.
 
\section{Numerical results}
\label{sec:results}

In this section we present numerical results for the cross sections of
the five SUSY processes introduced in Section~\ref{sec:processes}
and~\ref{sec:stop}.  In Section~\ref{sec:Kfac} we discuss the impact
of the NLL soft and Coulomb corrections on the central value of the
total cross sections and the uncertainties for the production of
light-flavour squarks and gluinos.  In Section~\ref{sec:bench} we
provide predictions for a selection of the benchmark points defined
in~\cite{AbdusSalam:2011fc}. The results for stop-antistop production
are presented in~\ref{sec:stop-nll}.  In order to facilitate the use
of our results, the arXiv submission of this paper includes grids with
predictions for the LHC with $\sqrt{s}=7$ and $8$~TeV, for
light-flavour squark and gluino masses from $200-2000$~GeV and stop
masses from $100-1000$ GeV ($200-2500$~GeV and $100-1200$ GeV,
respectively, for $\sqrt{s}=8$~TeV). We also provide a
\texttt{Mathematica} file containing interpolations of the cross
sections with an accuracy that is typically better than $\sim 1\%$,
and at worst $1-3\%$ for almost degenerate masses close to the edges
of the grid, $m_{\tilde q} \simeq m_{\tilde g} < 400$ GeV and
$m_{\tilde q} \simeq m_{\tilde g} > 1800$ GeV.

\subsection{Squark and gluino production at NLL}
\label{sec:Kfac}

To  illustrate how different classes of corrections contribute to the total cross section, we introduce three different NLL implementations:
\begin{itemize}
\item {\bf NLL}: our default implementation. Contains the full combined soft and Coulomb resummation, Eq. (\ref{eq:resum-NLL}),
including bound-state contributions below threshold, Eq. (\ref{eq:bound-GF}). 
For the soft scale we adopt the running scale given in Eqs. (\ref{eq:runmus_a}), (\ref{eq:runmus_b}). 
\item {\bf NLL$_{\text{no BS}}$}: as above, but without the inclusion of bound-state effects.  
\item {\bf NLL$_{s+h}$}: this implementation includes resummation of soft and hard logarithms only, without Coulomb resummation. This is
obtained using Eqs. (\ref{eq:NLLs}) and (\ref{eq:NLLs-P}). 
\end{itemize}
The three NLL approximations defined above are always matched to the exact NLO results computed with \texttt{PROSPINO}, 
according to (\ref{eq:cross_matched}). As input for the convolution with the parton luminosity functions, Eq. (\ref{eq:sigma-had}), 
we adopt the MSTW08NLO PDF set \cite{Martin:2009iq} and the associated strong coupling constant $\alpha_s(M_Z)=0.1202$.
Unless otherwise specified, the parameter $r$, defined as 
\begin{equation}
r=\frac{m_{\tilde{g}}}{m_{\tilde{q}}} \, ,
\end{equation}
is set to one.

\begin{figure}[p]
\begin{center}
\includegraphics[width=0.49 \linewidth]{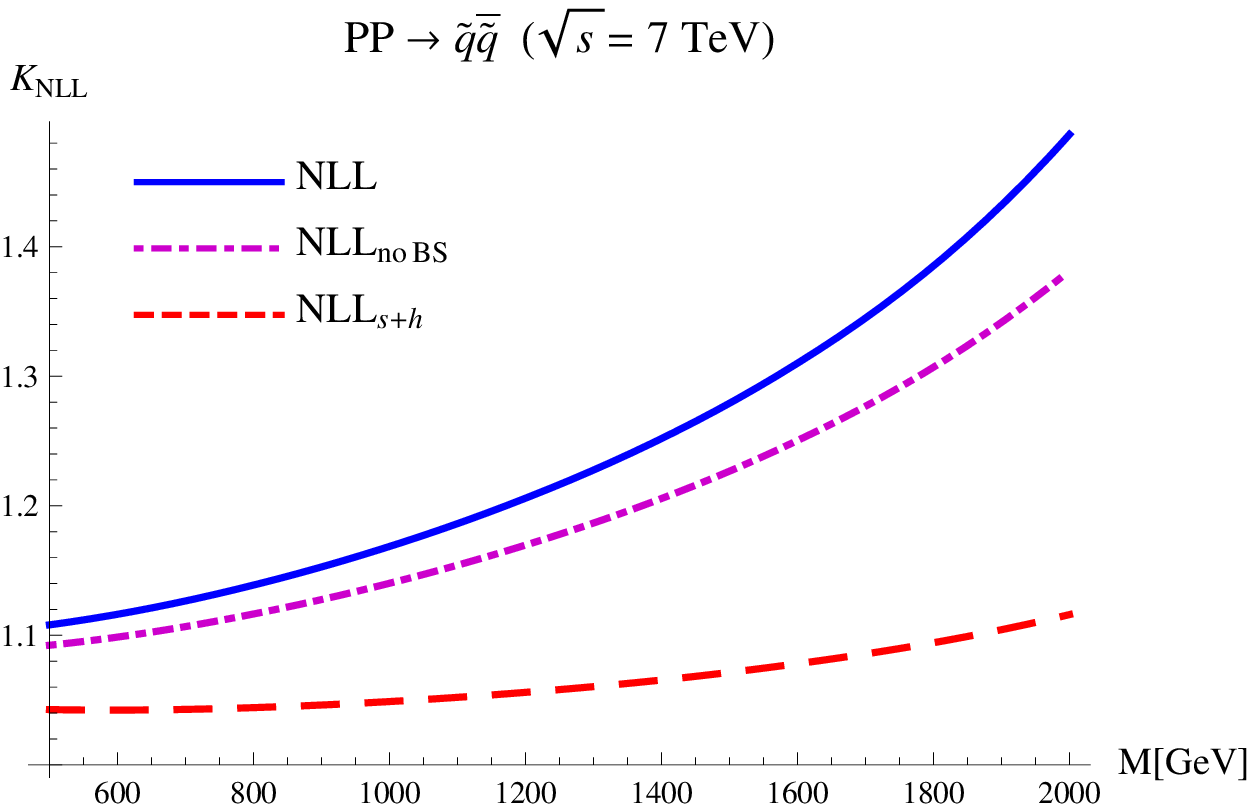}
\includegraphics[width=0.49 \linewidth]{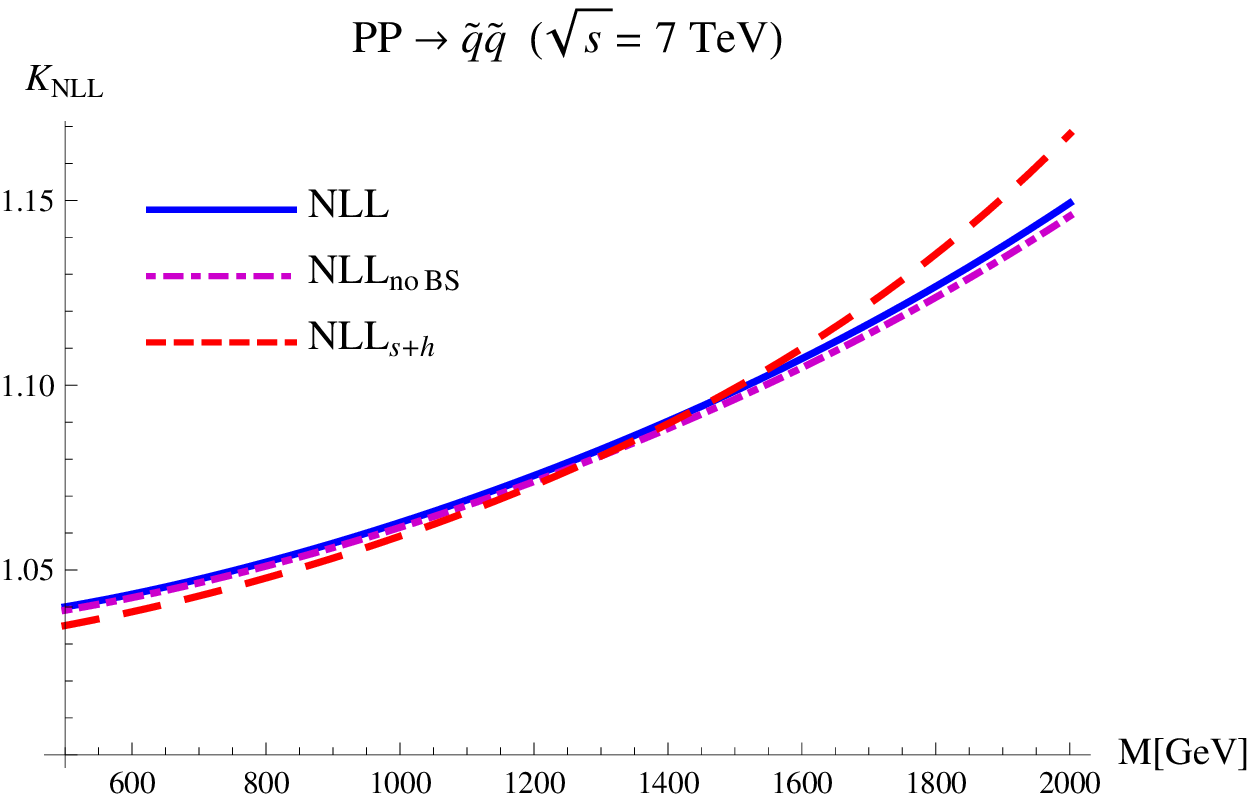}\\
\includegraphics[width=0.49 \linewidth]{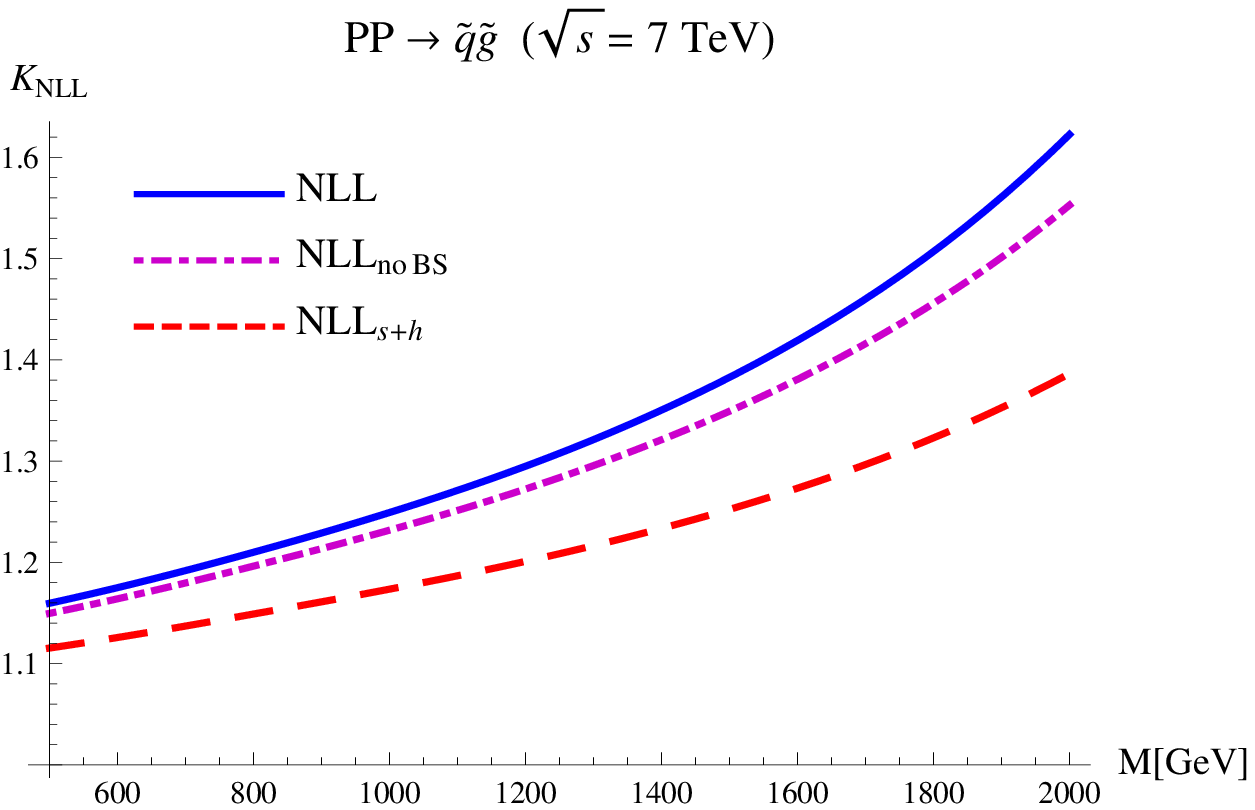}
\includegraphics[width=0.49 \linewidth]{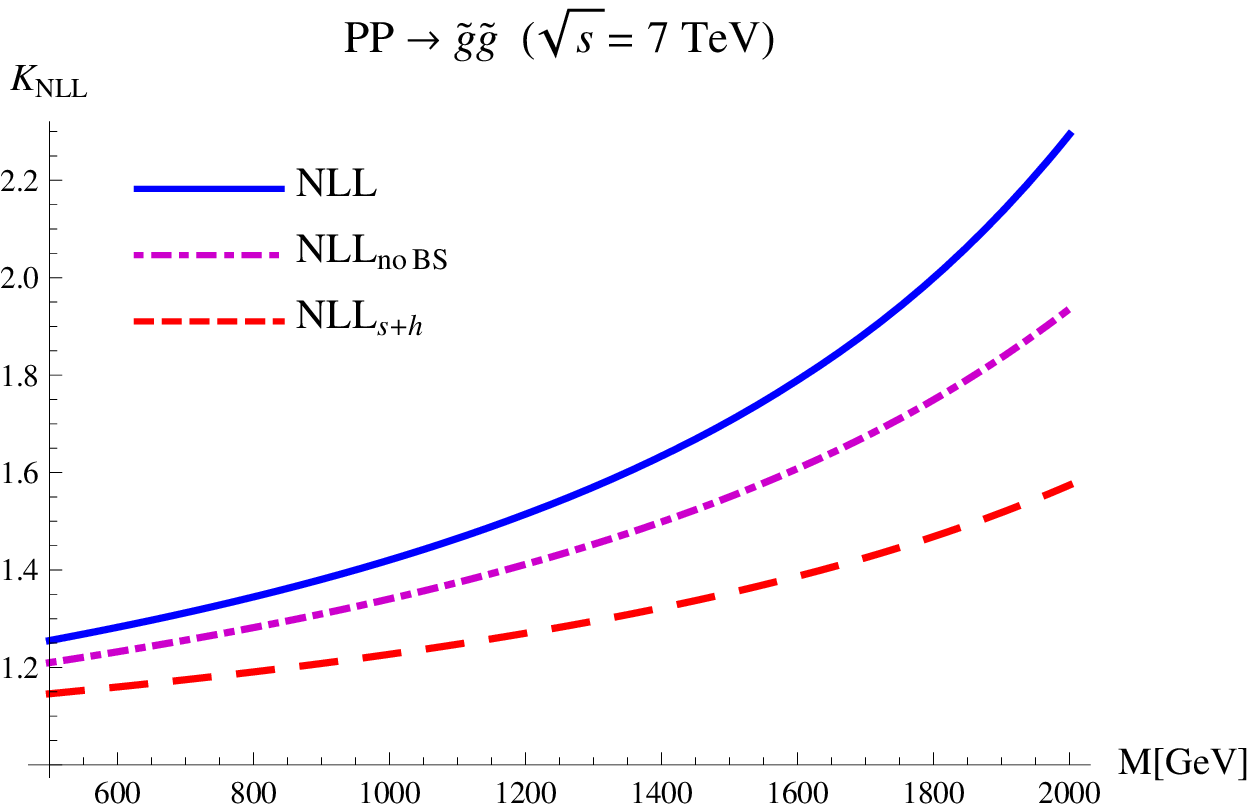}\\
\includegraphics[width=0.49 \linewidth]{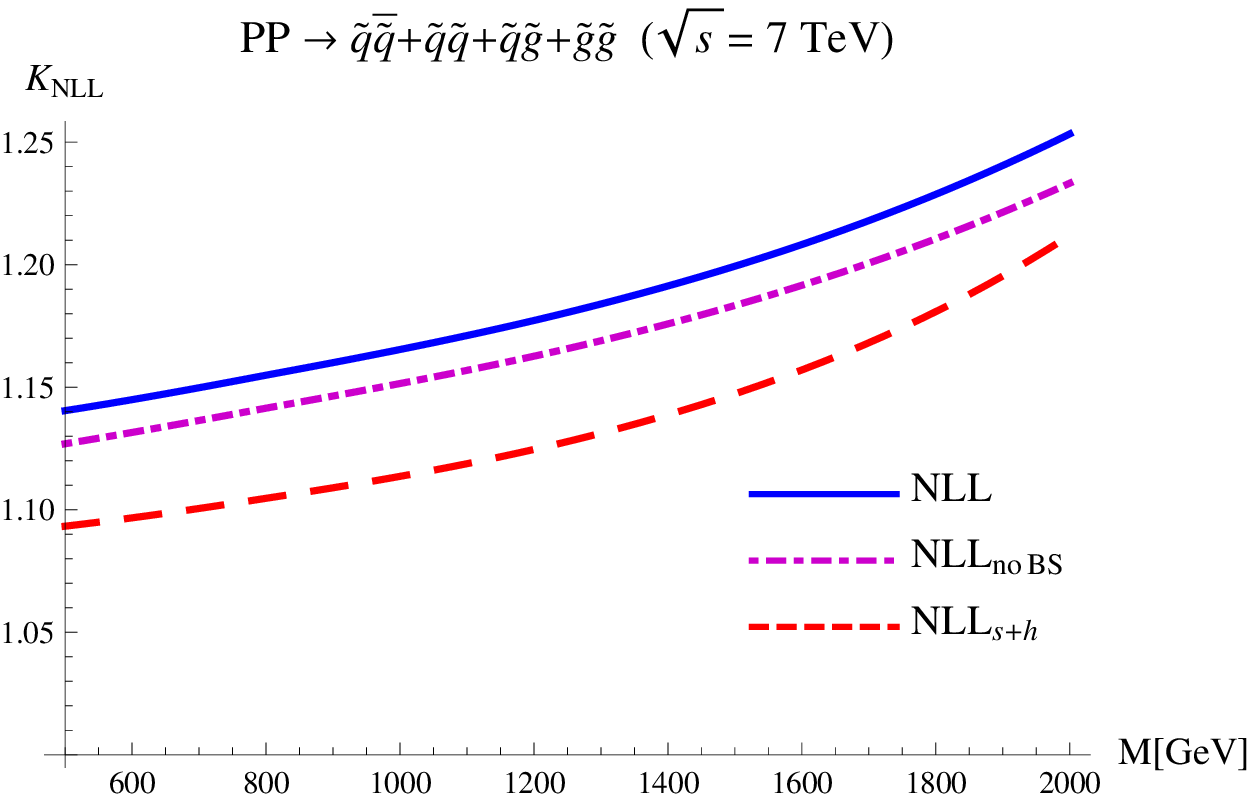}
\end{center}
\caption{NLL $K$-factor for squark-antisquark (top-left), squark-squark (top-right), squark-gluino (centre-left)
and gluino-gluino (centre-right) production at LHC with $\sqrt{s} = 7\,$ TeV, and for the sum of the four processes (bottom). The plots show $K_\text{NLL}$ as a function of $M$ for different NLL approximations: NLL (solid blue), NLL$_\text{no BS}$ (dot-dashed purple) and NLL$_{s+h}$ (dashed red). See the text for explanation.}
\label{fig:K7}
\end{figure}

\begin{figure}[p]
\begin{center}
\includegraphics[width=0.49 \linewidth]{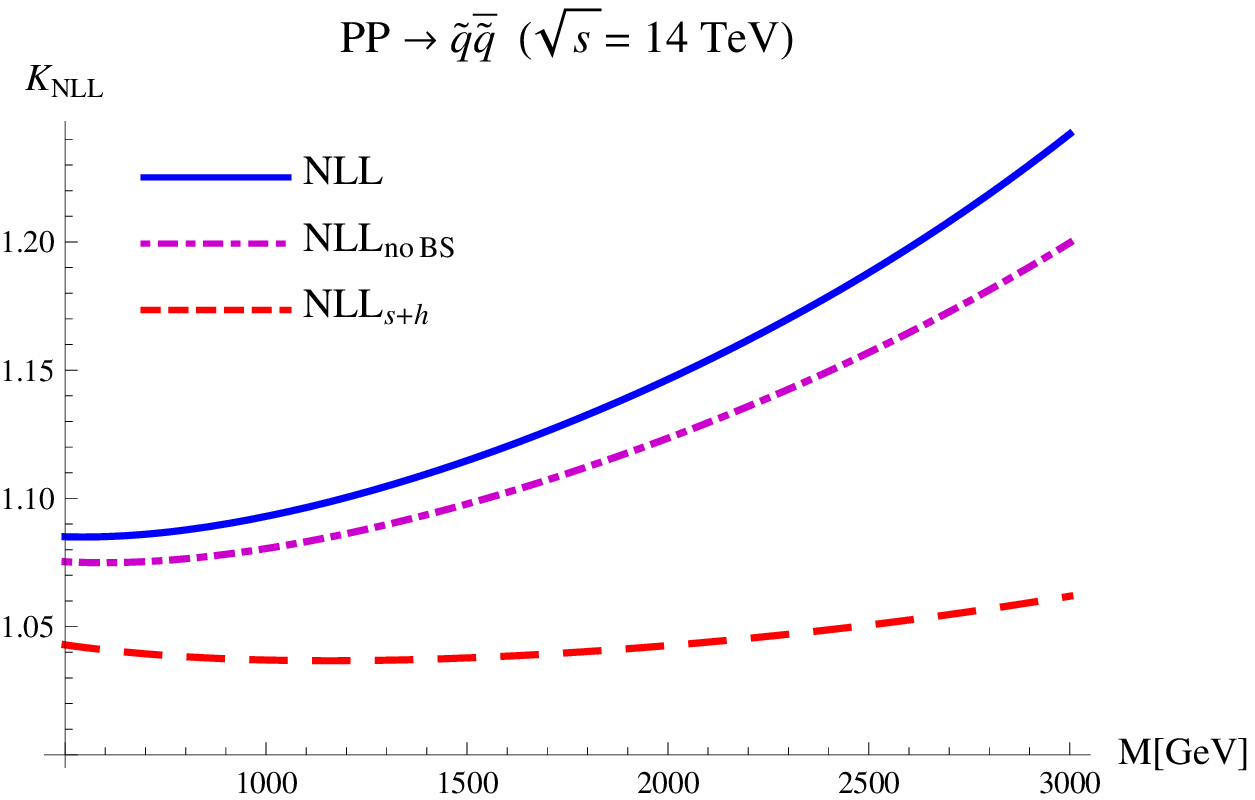}
\includegraphics[width=0.49 \linewidth]{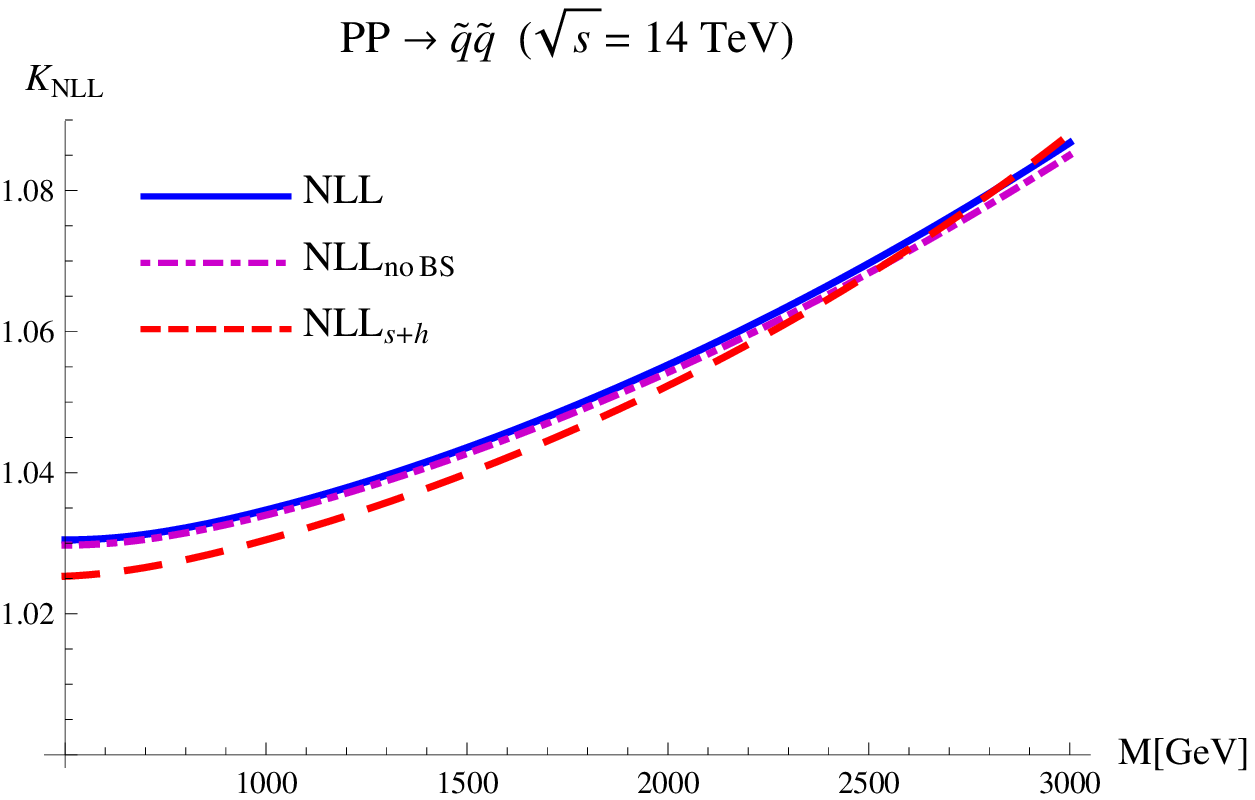}\\
\includegraphics[width=0.49 \linewidth]{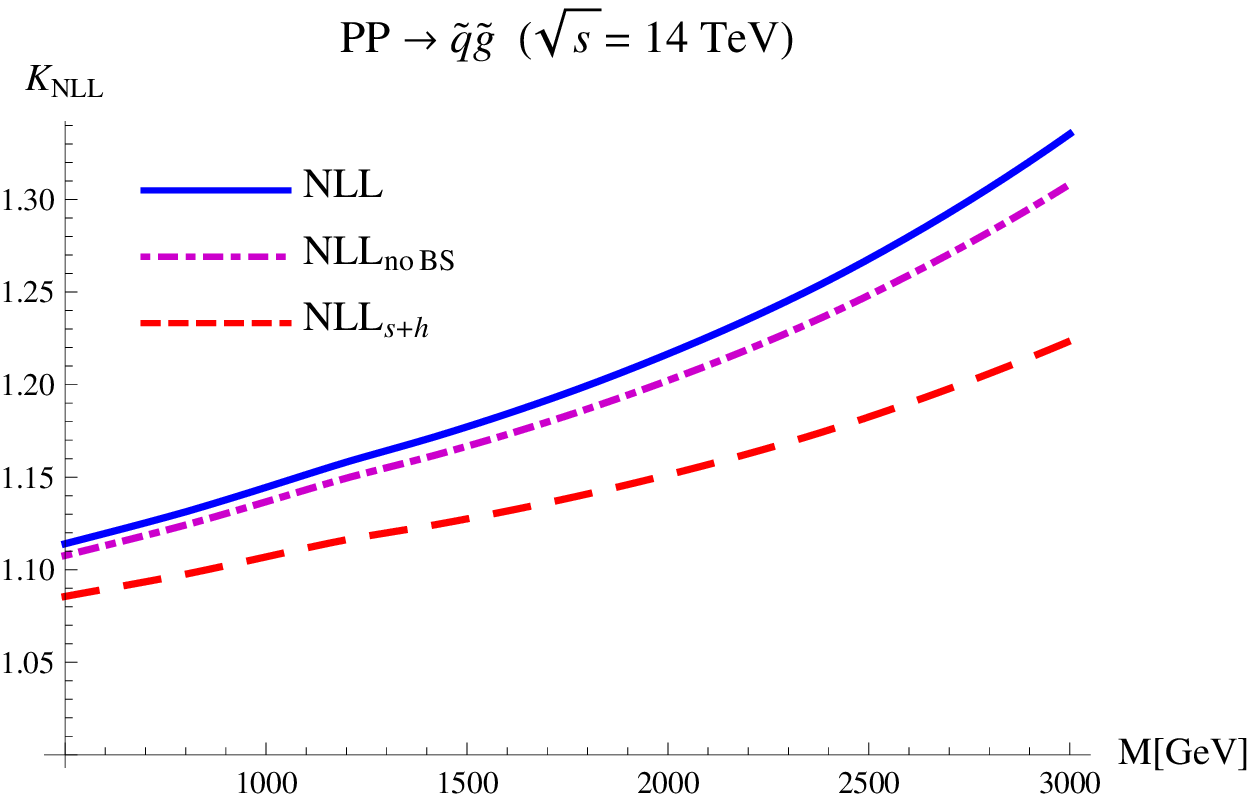}
\includegraphics[width=0.49 \linewidth]{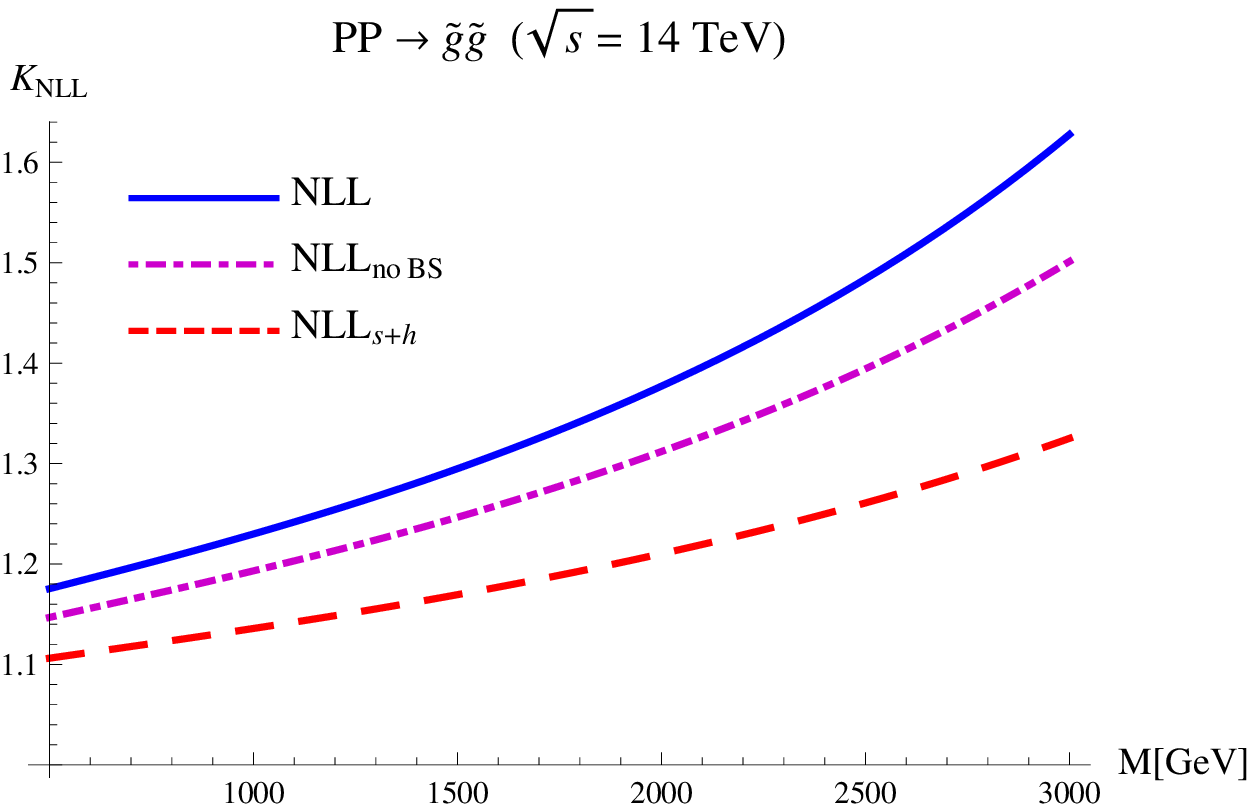}\\
\includegraphics[width=0.49 \linewidth]{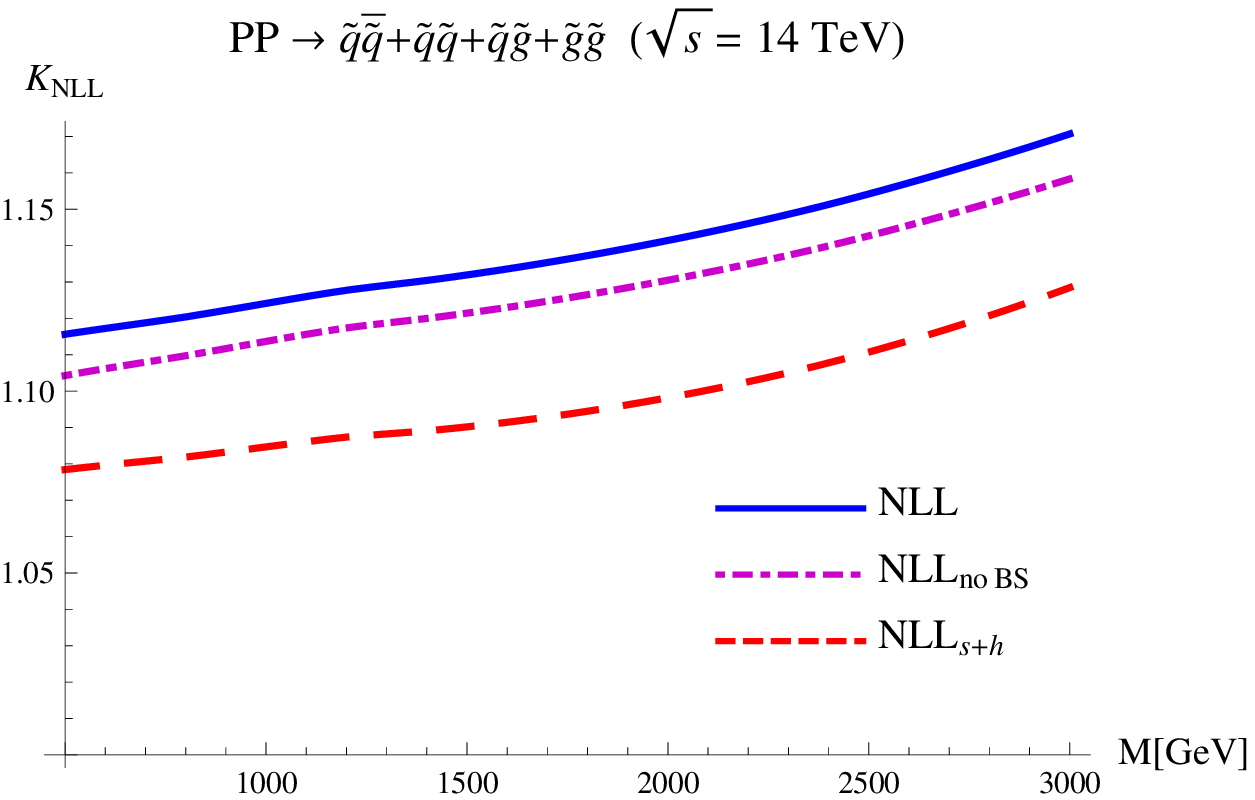}
\end{center}
\caption{NLL $K$-factor for squark-antisquark (top-left), squark-squark (top-right), squark-gluino (centre-left)
and gluino-gluino (centre-right) production at LHC with $\sqrt{s} = 14\,$ TeV, and for the sum of the four processes (bottom). The plots show $K_\text{NLL}$ as a function of $M$ for different NLL approximations: NLL (solid blue), NLL$_\text{no BS}$ (dot-dashed purple) and NLL$_{s+h}$ (dashed red). See the text for explanation.}
\label{fig:K14}
\end{figure}

We start presenting results for the NLL $K$-factor, defined as
\begin{equation} \label{eq:Kfact}
 K_\text{NLL}=\frac{\sigma^\text{matched}}{\sigma^\text{\text{NLO}}} \, ,
\end{equation}
where $\sigma^\text{matched}$ is our matched result for one of the NLL
implementations defined in the beginning of this Section and $\sigma^\text{NLO}$ the
fixed-order NLO result obtained using
\texttt{PROSPINO}. The NLL-$K$-factor for LHC with $7$ TeV centre-of-mass
energy is plotted in Figure \ref{fig:K7}, for the four
light-squark/gluino production processes and the mass range
$m_{\tilde{q}}=m_{\tilde{g}}=500$-$2000\,$GeV.  The results for
$\sqrt{s}=14\,$ TeV and the mass range
$m_{\tilde{q}}=m_{\tilde{g}}=500$-$3000\,$GeV are 
 given in Figure
\ref{fig:K14}. The NLL corrections for our default implementation (solid
blue lines) can be large, with corrections to the fixed-order NLO
results of up to $120 \%$ in the upper mass range for gluino-gluino
production at 7 TeV. The higher-order effects are smaller, but still sizeable,
for the other three processes, due to the smaller colour charges
involved in squark-antisquark, squark-squark and squark-gluino
production.  Furthermore, for a fixed SUSY mass the $K_{\text{NLL}}$-factor
decreases from 7 to 14 TeV, consistently with the expectation that at
lower centre-of-mass energies the threshold region plays a more
prominent role.

The effect of including Coulomb resummation and its interference with soft resummation is on average as large as (or
even larger than) the effect of pure soft and hard corrections, as can be seen comparing our default implementation NLL with 
NLL$_{s+h}$ (dashed red lines). Pure soft contributions beyond ${\cal O}(\alpha_s)$ 
amount to $5-60 \%$ of the fixed-order NLO result, depending on the mass and process considered, 
whereas pure Coulomb effects and interference of soft and Coulomb corrections can amount to up to $60 \%$. 
An exception to this is the squark-squark production process, where the effect of 
Coulomb corrections is small. This particular behaviour originates from cancellations between the 
cross sections for same-flavour squark production, where the repulsive colour-sextet channel is numerically dominant
and gives rise to negative ${\cal O}(\alpha_s^2 \ln^2 \beta/\beta)$ corrections, and different-flavour squark production, 
where the corresponding term is positive, due to the dominance of the attractive colour-triplet channel.

\begin{figure}[t!]
\begin{center}
  \includegraphics[width=0.65 \linewidth]{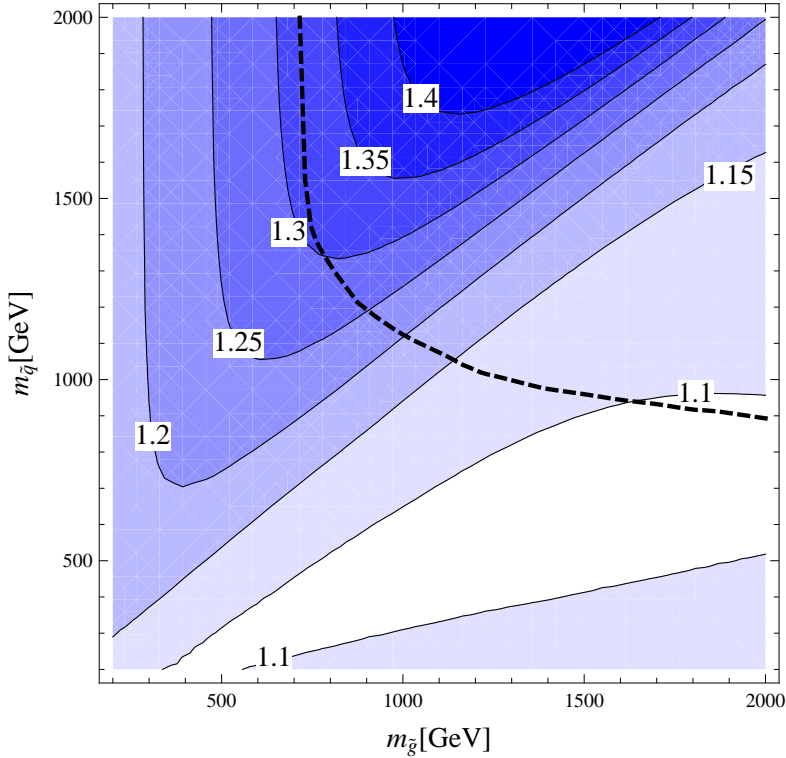}
\end{center}
\caption{NLL $K$-factor for the total SUSY production rate at LHC with $\sqrt{s}=7\,$TeV as a function of the
gluino mass $m_{\tilde{g}}$ and average squark mass $m_{\tilde{q}}$. The dashed line corresponds
to the most recent exclusion limit presented in \cite{Aad:2011ib}.}
    \label{fig:contour}
\end{figure}

\begin{figure}[t!]
\begin{center}
  \includegraphics[width=0.49 \linewidth]{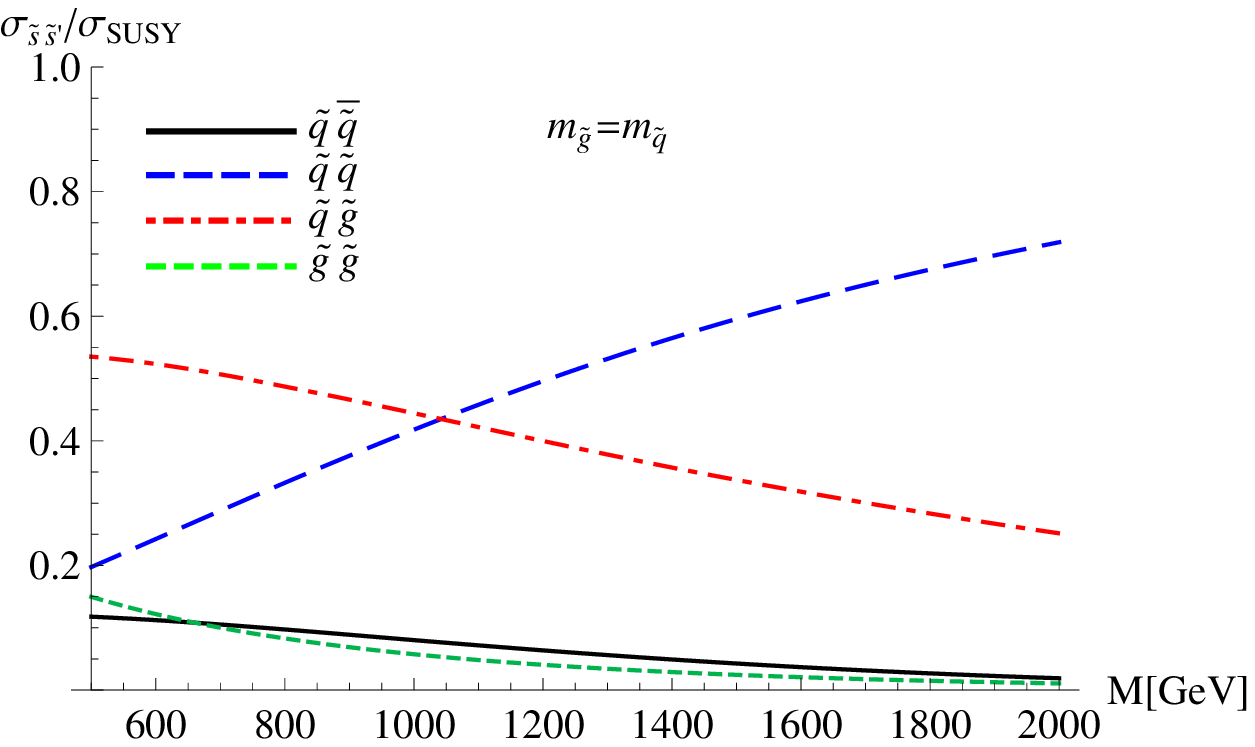}
  \includegraphics[width=0.49 \linewidth]{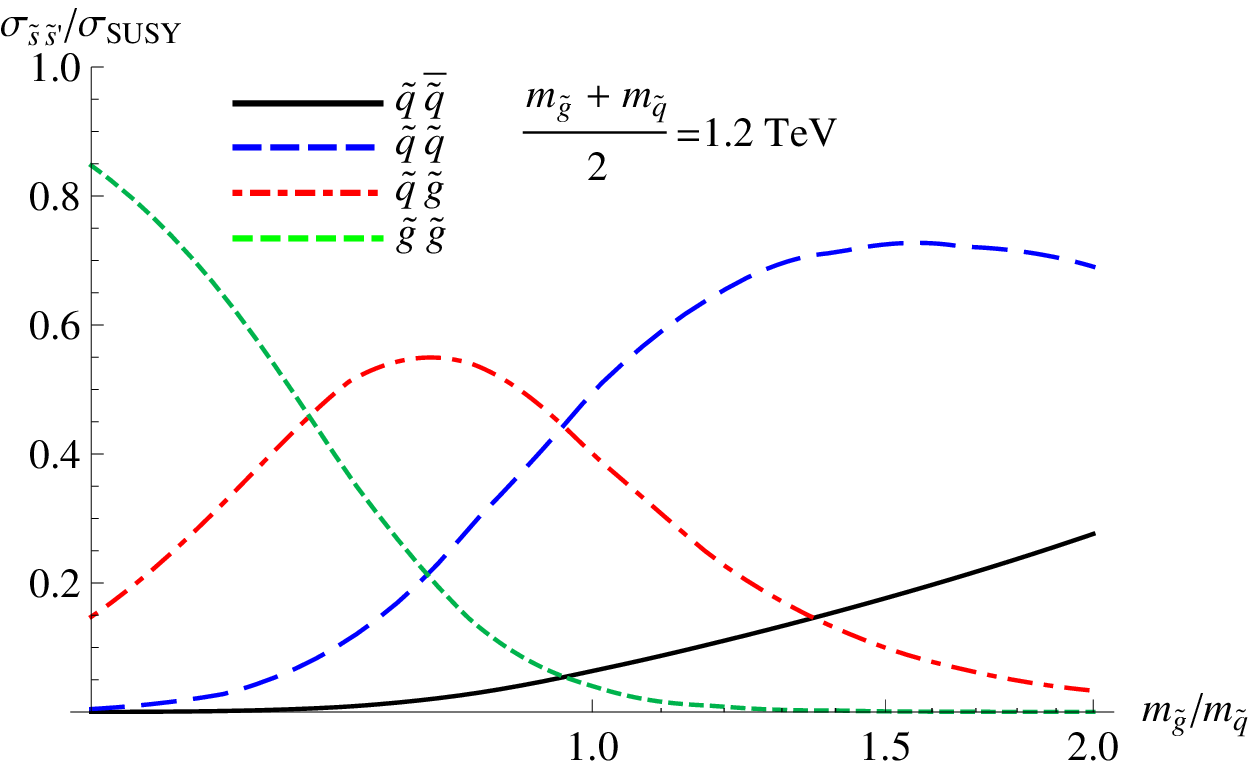}
\end{center}
\caption{Ratio of the NLL production-cross sections for the
    processes~\eqref{eq:processes} to the total NLL rate of coloured
    sparticle production $\sigma_{\text{SUSY}}$ for the LHC with $\sqrt s=7$
    TeV. Left:  Mass dependence for a fixed mass-ratio $m_{\tilde q}=m_{\tilde
      g}=M$ , Right:  Dependence on the ratio $m_{\tilde g}/m_{\tilde q}$ for
    a fixed average mass $(m_{\tilde{q}}+m_{\tilde{g}})/2=1.2$~TeV.}
    \label{fig:sigmaNLL}
\end{figure}

For squark-antisquark, squark-gluino and gluino-gluino production, a significant portion of the total Coulomb and
soft-Coulomb corrections originates from bound-state effects below threshold. These correspond to the difference
between the NLL and NLL$_\text{no BS}$ (dot-dashed purple) curves in the plots.  For squark-antisquark
and squark-gluino production bound-state corrections amount to $2-10 \%$ of the fixed-order NLO cross section, 
whereas for gluino-gluino production they can be as large as $30 \%$.

Figure \ref{fig:contour} shows the NLL $K$-factor for the total SUSY
production rate at the $7$ TeV LHC as a contour plot in the
$(m_{\tilde{g}},m_{\tilde{q}})$-plane. The $r$-dependence
of the total resummed cross section arises from an interplay of the 
$r$-dependence of the single-process cross sections and of the relative dominance
of the four subprocesses for a given $r$. The largest $K$-factor is obtained for $m_{\tilde{q}}\sim
2\,$TeV and $m_{\tilde{g}}\sim 1.4\,$TeV, with corrections of $50\%$
to the NLO cross section. The plot shows also the recent exclusion
limit published by the ATLAS collaboration in \cite{Aad:2011ib}
assuming a simplified model of a massless neutralino, a gluino
octet and degenerate squarks of the first two generations,
while all the other supersymmetric particles, including stops and
sbottoms, are decoupled by giving them a mass of $5\,$TeV.  The limits
are therefore not directly comparable to our results which treat the
sbottom as degenerate with the light-flavour squarks, but are shown
here as an indication of the current LHC reach.  We do not
attempt to estimate how resummation would affect the determination of
this limit. However, one can observe that in the large squark-mass
region the exclusion limit crosses regions with a $K$-factor bigger
than $1.3$, where resummation effects on the limit extraction might be
relevant.

Given the large effect of resummation, especially for squark-gluino and gluino-gluino production, it is interesting to
study how the relative contribution of the four production processes to the total SUSY production rate is modified by 
the inclusion of NLL corrections. This is shown in Figure \ref{fig:sigmaNLL}.
The qualitative behaviour of the relative contribution of the four different processes is very similar to the LO result
(Figure \ref{fig:sigmaLO}).
However at large masses one can notice an enhancement of the squark-gluino production rate compared to the 
squark-squark channel (left plot), as one would expect from the larger NLL $K$-factor for the first processes. For a fixed average
squark and gluino mass of 1.2 TeV (right plot) the relative ratios are basically unchanged for moderate values of $r=m_{\tilde{g}}/m_{\tilde{q}}$,
though one observes a significant enhancement of the squark-antisquark cross section for large gluino masses ($r=2$).

\begin{figure}[t!]
\begin{center}
\includegraphics[width=0.49 \linewidth]{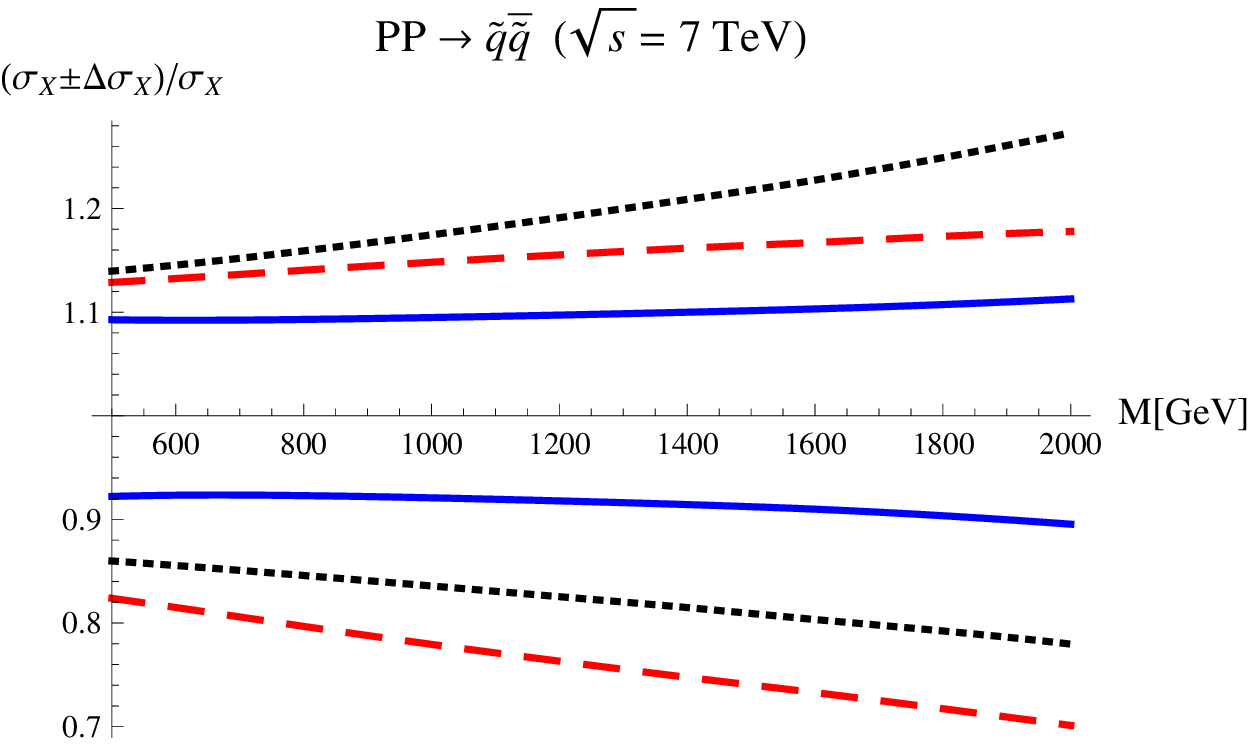}
\includegraphics[width=0.49 \linewidth]{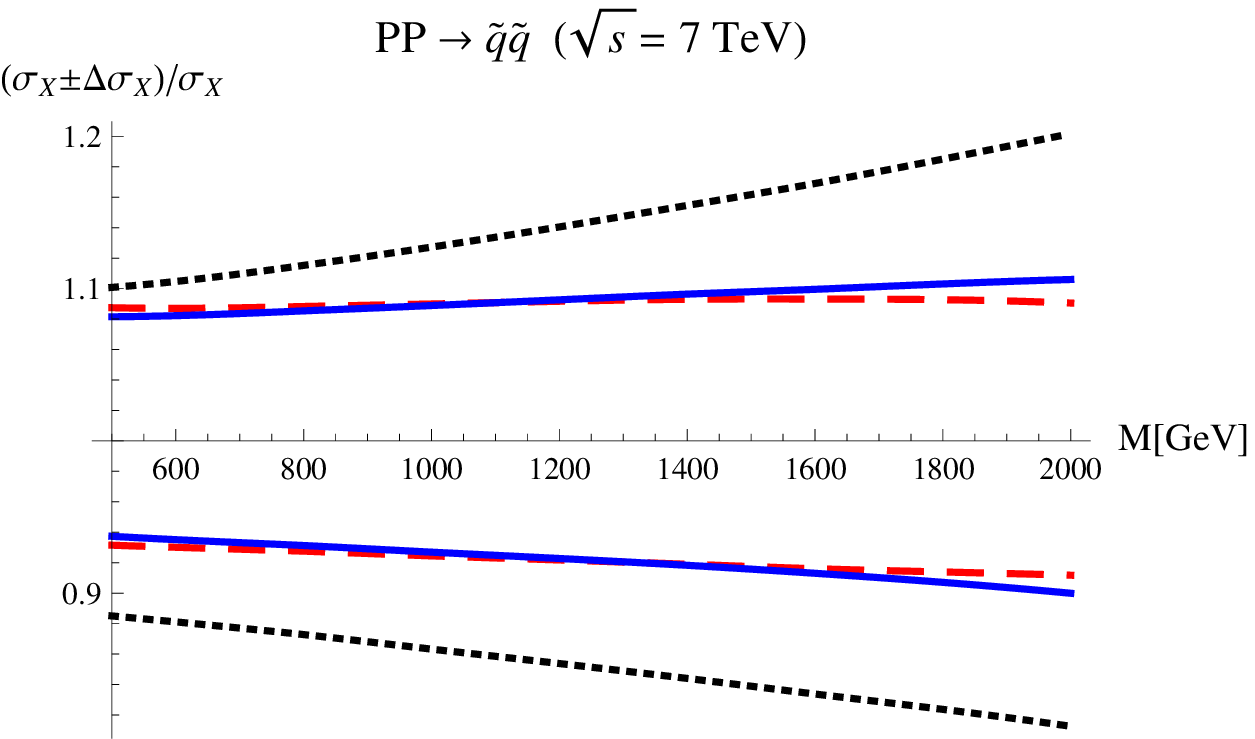}\\
\includegraphics[width=0.49 \linewidth]{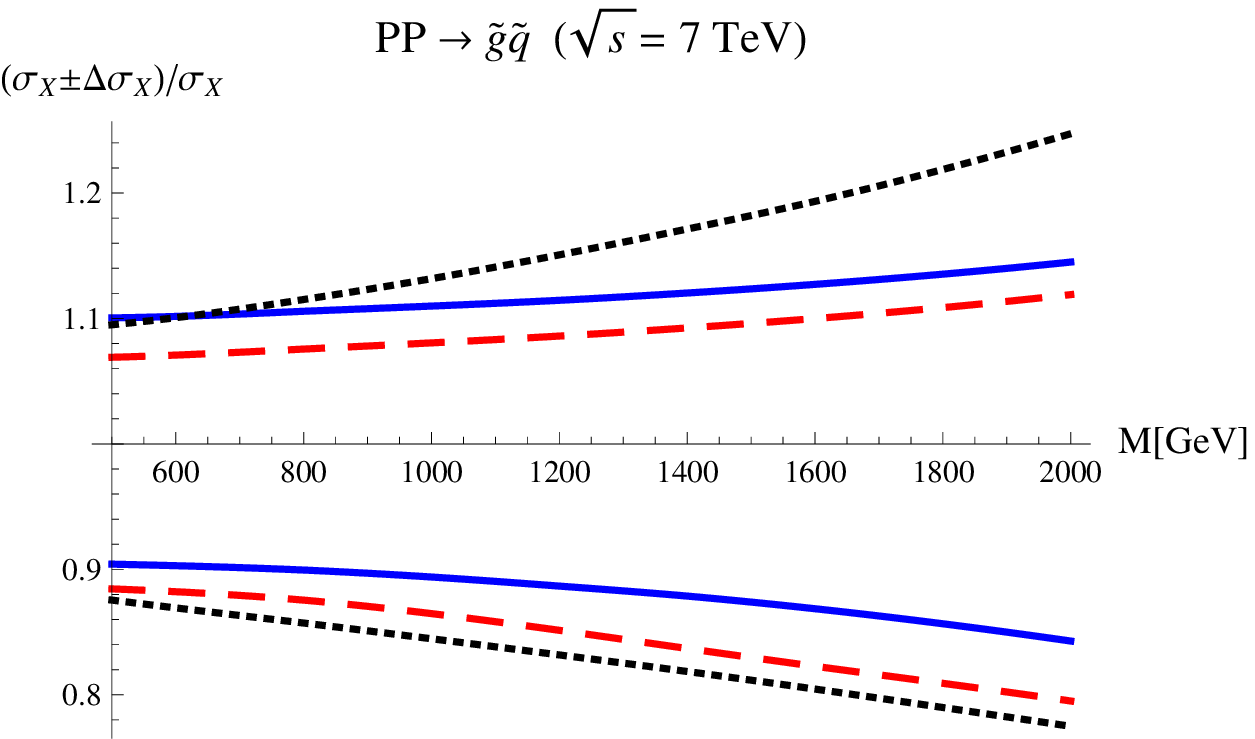}
\includegraphics[width=0.49 \linewidth]{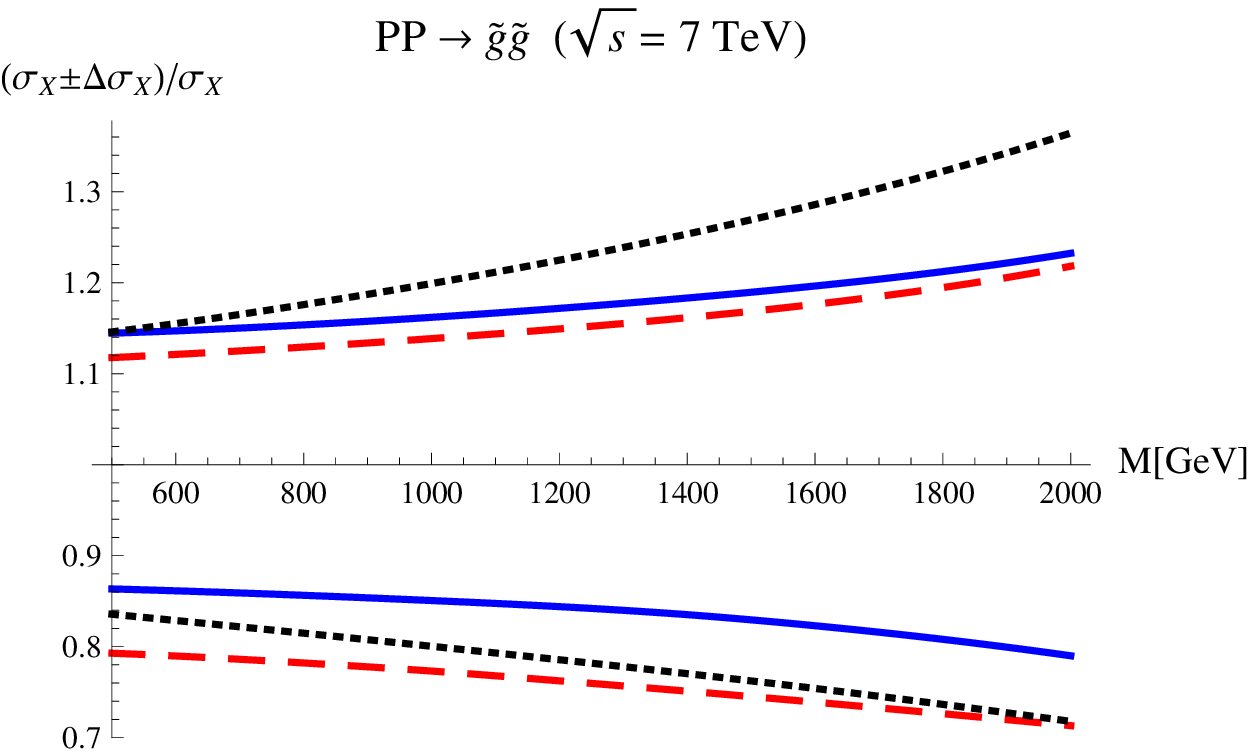}\\
\includegraphics[width=0.49 \linewidth]{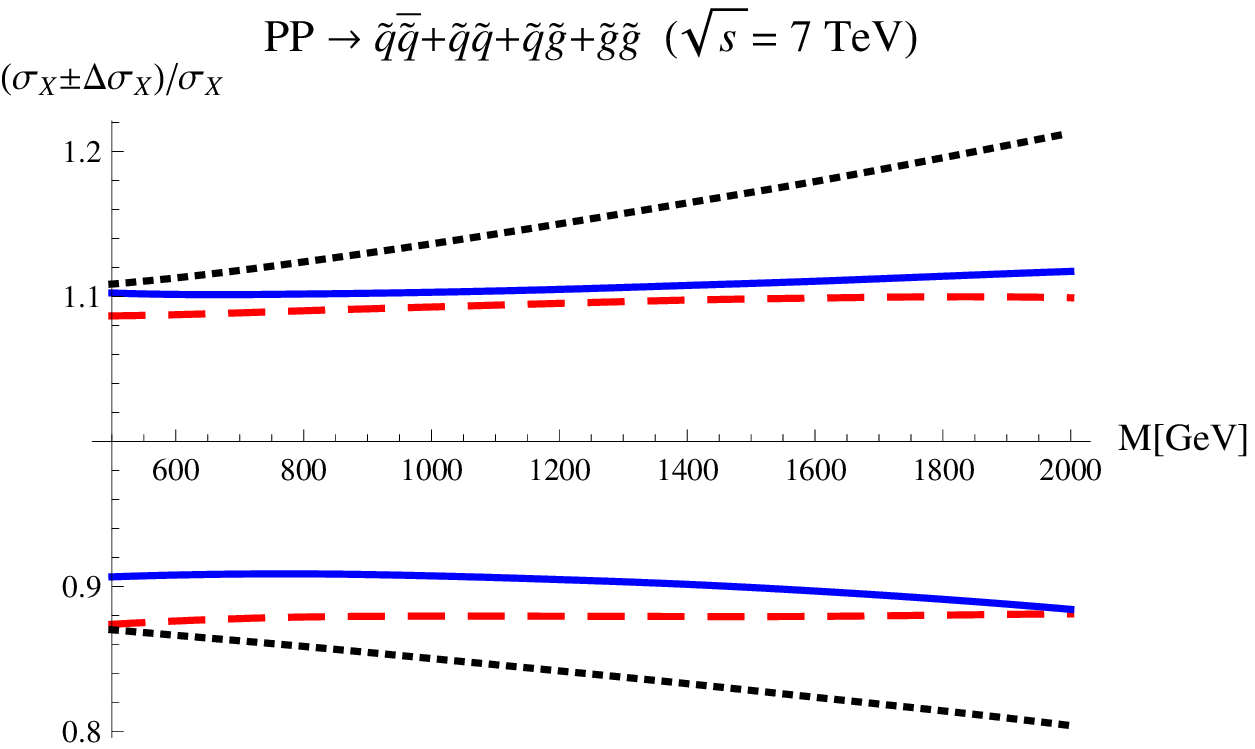}
\end{center}
\caption{Total theoretical uncertainty of  the NLO approximation (dotted black), full NLL resummed result (solid blue)
and NLL$_{s+h}$ (dashed red) at the LHC with $\sqrt{s}=\,$7 TeV. All cross sections are normalized to one at the central value of the scales.}
\label{fig:totaluncertainty7}
\end{figure}

\begin{figure}[t!]
\begin{center}
\includegraphics[width=0.49 \linewidth]{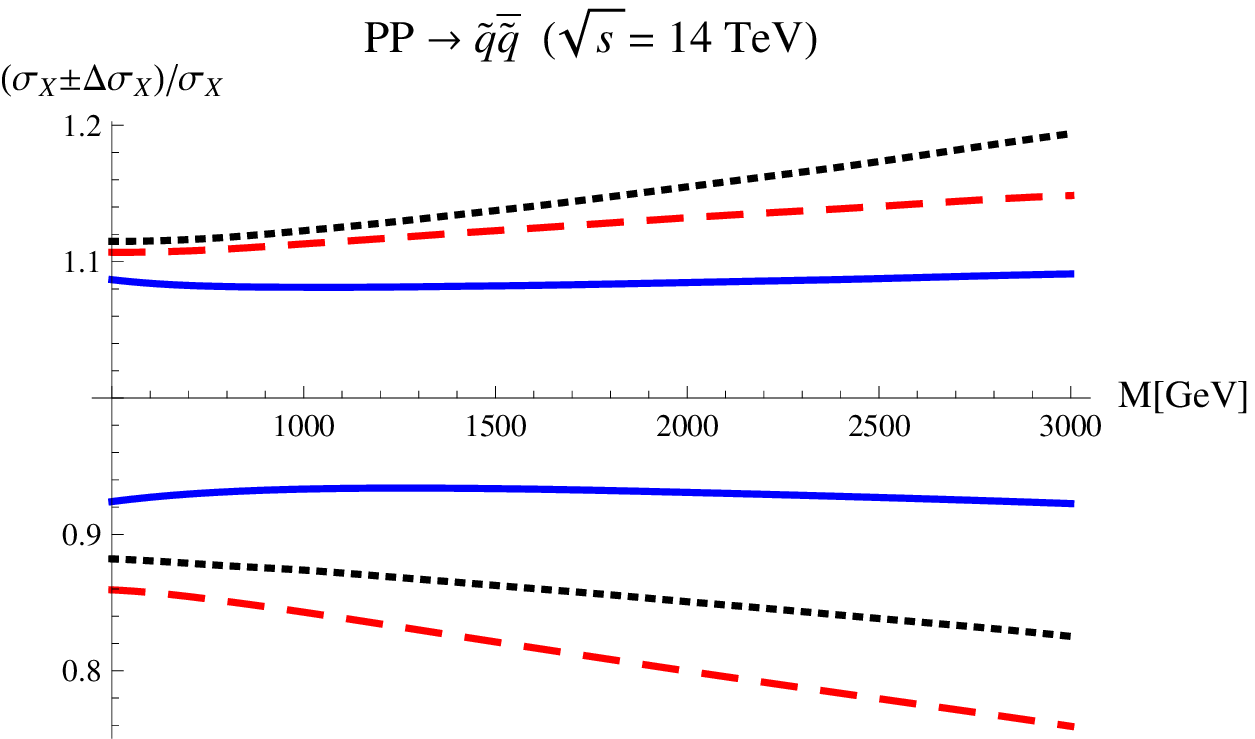}
\includegraphics[width=0.49 \linewidth]{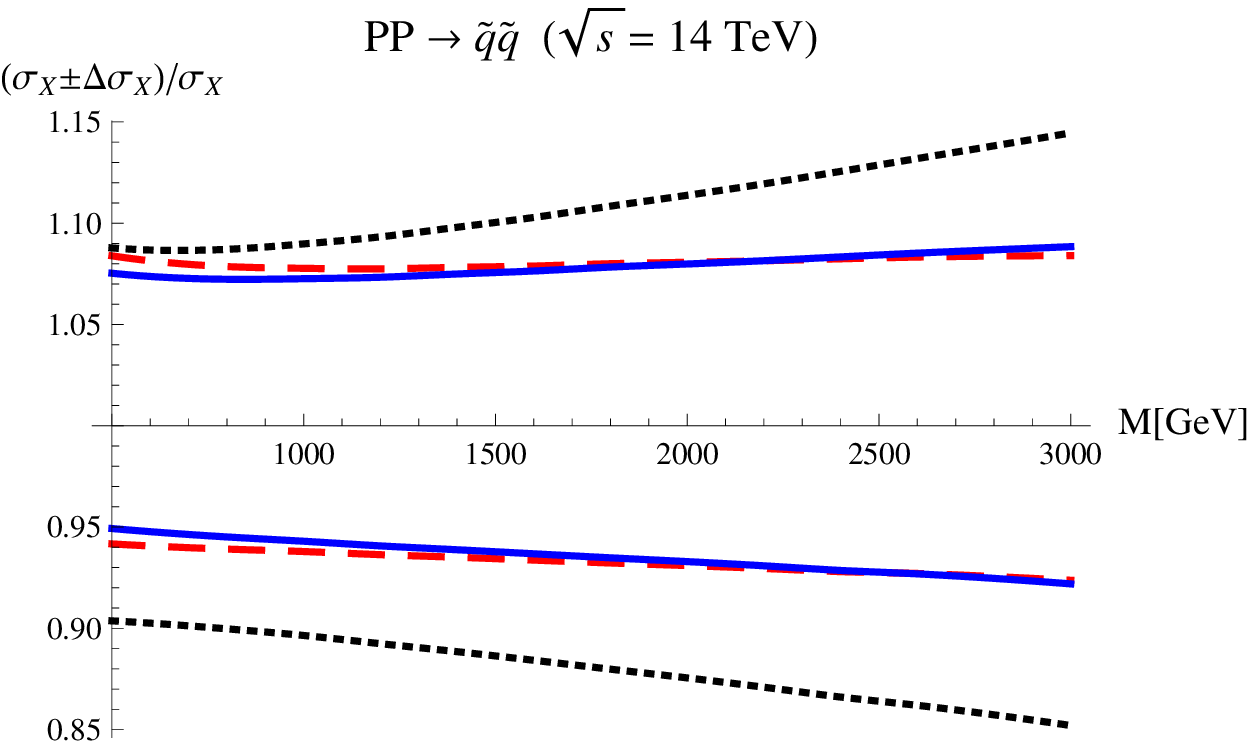}\\
\includegraphics[width=0.49 \linewidth]{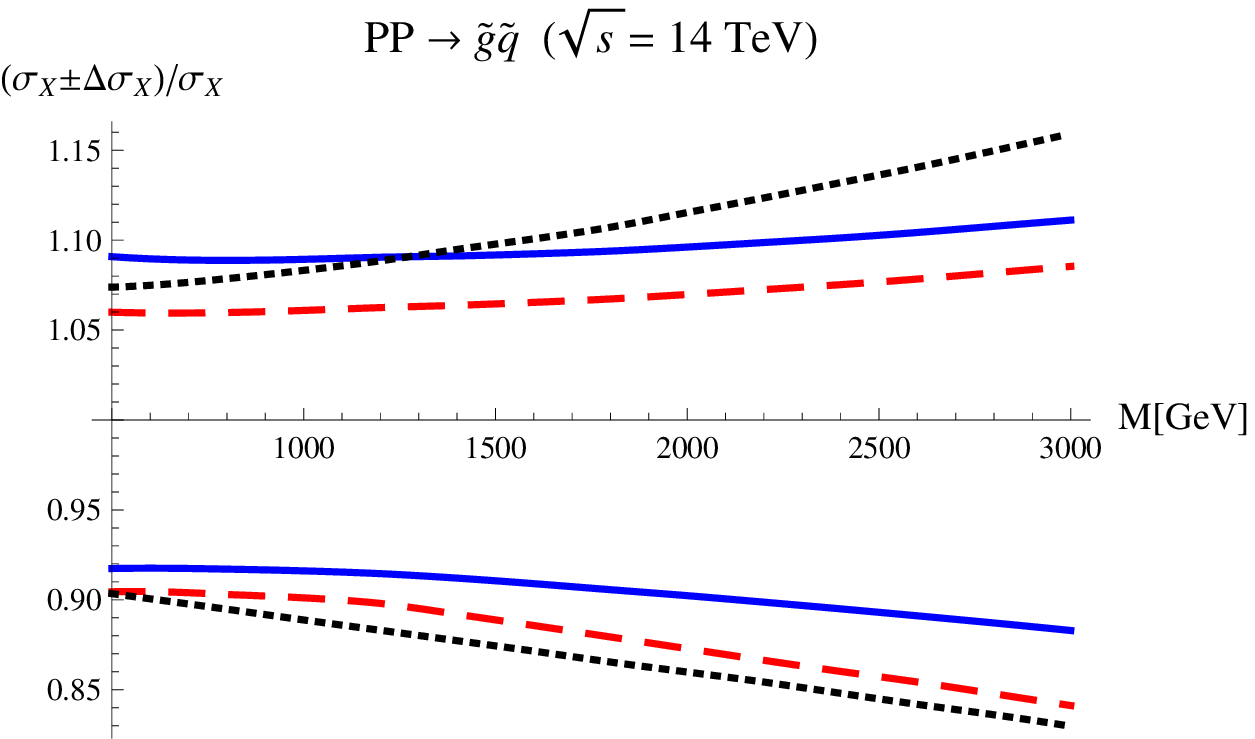}
\includegraphics[width=0.49 \linewidth]{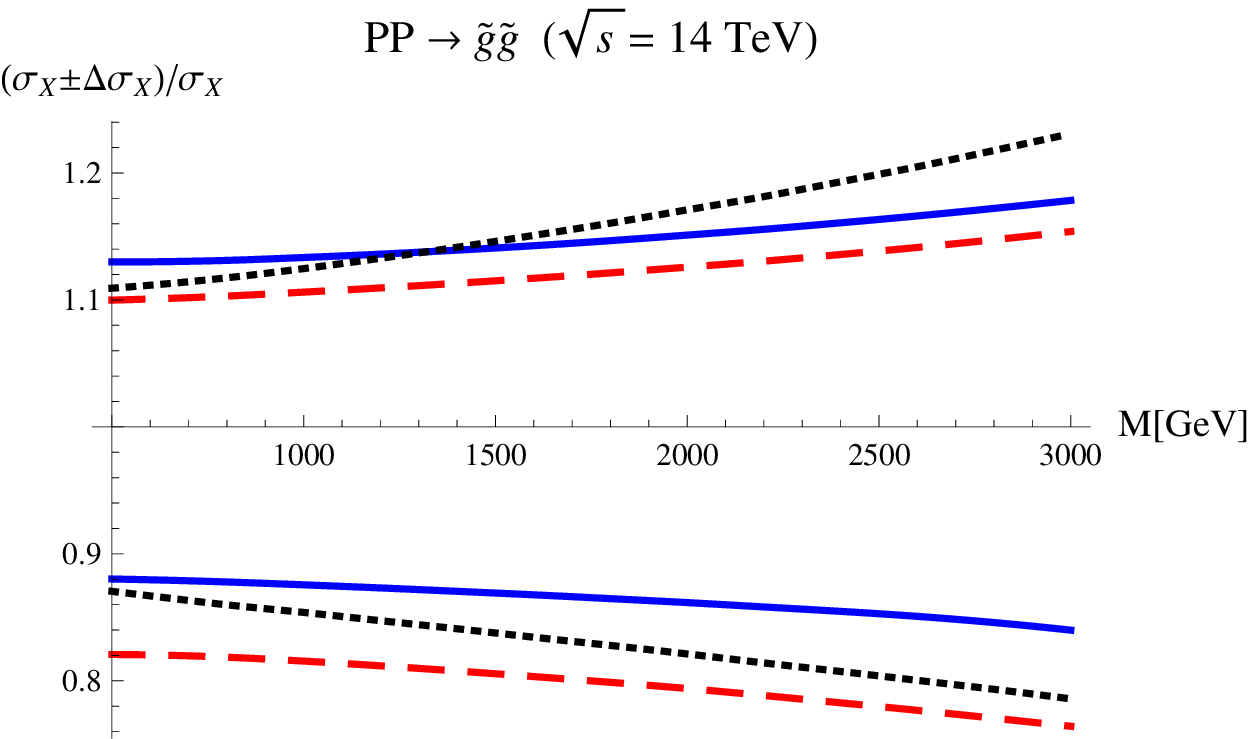}\\
\includegraphics[width=0.49 \linewidth]{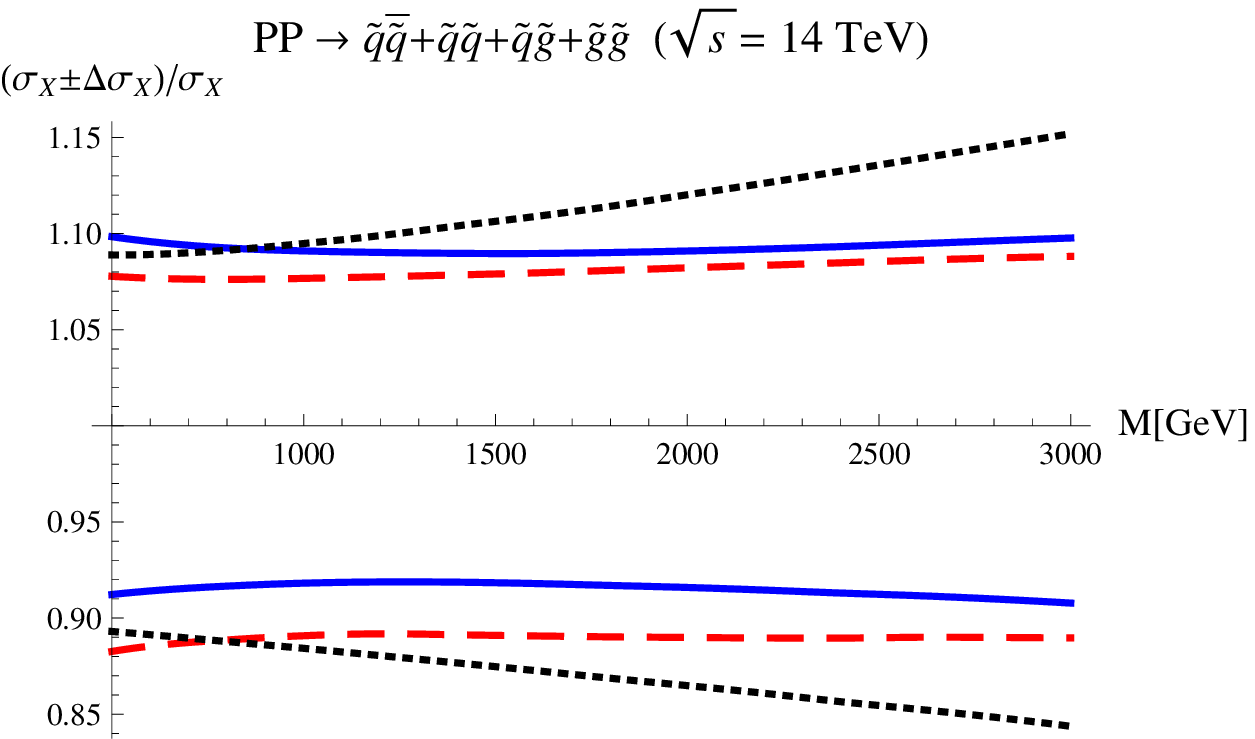}
\end{center}
\caption{Total theoretical uncertainty of  the NLO approximation (dotted black), full NLL resummed result (solid blue)
and NLL$_{s+h}$ (dashed red) at the LHC with $\sqrt{s}=\,$14 TeV. All cross sections are normalized to one at the central value of the scales.}
\label{fig:totaluncertainty14}
\end{figure}

As a result of including the threshold-enhanced higher-order corrections, 
one expects that the uncertainty due to missing perturbative corrections is reduced compared to the NLO results.
While for NLO the theoretical 
uncertainty arises from scale-variation only,  the total theoretical error of the NLL results is obtained by adding scale and resummation uncertainties in quadrature, as defined in Section \ref{sec:scales}. The uncertainty bands for NLO, NLL$_{s+h}$ and NLL approximations 
are shown in Figure \ref{fig:totaluncertainty7} for the LHC with $\sqrt{s} =\,$7 TeV and in Figure \ref{fig:totaluncertainty14} for the LHC with $\sqrt{s} =\,$14 TeV .
In all plots the cross sections are normalized to unity at the central values of the scales and other input parameters. 
It is evident that the combined resummation of soft and Coulomb effects (NLL, solid blue) generally leads to a significant reduction
of theoretical uncertainties compared to the NLO result (dotted black), especially in squark-antisquark and squark-squark
production, where the error is reduced by a factor 2 or more in the large-mass region. The behaviour of NLL$_{s+h}$ (dashed red) 
is more process-dependent, with basically no uncertainty reduction compared to the fixed-order NLO result for squark-antisquark
production, and moderate effects for squark-gluino and gluino-gluino production. For squark-squark production (and, as a consequence of
the dominance of squark-squark production, for the total SUSY production rate) the uncertainties of NLL$_{s+h}$ and NLL are very similar, 
due to the smallness of Coulomb effects in this particular production channel. 
The large reduction of the scale dependence for squark-antisquark production by soft-Coulomb interference effects is consistent with recent
NNLL studies in this channel~\cite{Beenakker:2011sf} that include the first Coulomb correction.

\subsection{Benchmark points for SUSY searches at LHC}
\label{sec:bench}

In addition to the grid files provided with the arXiv submission of this paper, we here present numerical predictions 
for some benchmark points at the LHC with $7$~TeV centre-of-mass
energy, in order to illustrate the effect of our NLL results on the
production cross sections.  We employ the sets of benchmark points
defined in~\cite{AbdusSalam:2011fc}, that are compatible with recent
LHC bounds and other data such as $b\to s\gamma$, but not necessarily
with constraints from the anomalous magnetic moment of the muon, or
from the dark matter relic abundance. We consider the seven lines in
the constrained MSSM~(CMSSM) parameter space defined
in~\cite{AbdusSalam:2011fc} and one line for the minimal gauge
mediated SUSY breaking~(mGMSB) scenario.  For each line we selected
one benchmark point expected to be relevant for $5\text{ fb}^{-1}$ of
data and a second point relevant for $10-30 \text{ fb}^{-1}$, where we
naively extrapolate the reach of the $1\text{ fb}^{-1}$ LHC
data~\cite{Aad:2011ib,Chatrchyan:2011zy} that exclude CMSSM benchmark
points and simplified models with total SUSY production cross sections
of the order of $\sigma_{\text{SUSY}}\gtrsim 0.04$ pb for $m_{\tilde
  q}\lesssim m_{\tilde g}$ and $\sigma_{\text{SUSY}}\gtrsim 0.1$ pb
for $m_{\tilde q}> m_{\tilde g}$.  The mGMSB scenario we have selected
has a quasi-stable neutralino as next-to-lightest SUSY
particle~(NLSP), so a similar reach as for CMSSM-type scenarios are
expected.  Since only the squark and gluino masses are relevant for
the production cross sections, we have chosen points with a reasonable
spread of masses and mass ratios, covering the range of average
sparticle masses, $1.3 \text{ TeV}\lesssim (m_{\tilde
  g}+m_{\tilde q})/2\lesssim 1.5 \text{ TeV}$, and the mass
ratios $0.75\lesssim \frac{m_{\tilde g}}{m_{\tilde q}}\lesssim 1.12$.  This mass range is also compatible with the
estimated discovery reach~\cite{Baer:2011aa} of $m_{\tilde g}\sim 1.3$
TeV ($1.6$ TeV) for $m_{\tilde q}\sim m_{\tilde g}$ at $5\text{
  fb}^{-1}$ ($30 \text{ fb}^{-1}$) and $m_{\tilde g}\sim 0.8$ TeV ($1$
TeV) for $m_{\tilde q}\gg m_{\tilde g}$ in a CMSSM scenario with
$\tan\beta=0.45$, $A_0=0$.  The remaining families of benchmark
scenarios introduced in~\cite{AbdusSalam:2011fc} tend to have very
similar mass ratios as our selected points. Therefore the relative
contributions of the different production channels and the effect of
the higher-order QCD corrections will be similar, although the decay
chains and the resulting collider signatures can be very
different. For some scenarios lighter squarks and gluinos than the
ones considered here might still be allowed, for instance in GMSB with
a stau NLSP. Predictions for such scenarios can be obtained by an
interpolation of the grid files provided with the arXiv submission of
this paper.

The SUSY breaking parameters and the resulting mass
spectrum of the coloured SUSY particles for the selected points is
shown in Tables~\ref{tab:bench1}, \ref{tab:bench2}
and~\ref{tab:bench3} together with our best NLL predictions for the
total cross section for light-flavour squark and gluino
production (including simultaneous soft-gluon and Coulomb resummation
as well as bound-state effects).  Here $m_{\tilde q}$ denotes
the average mass of all squarks except the stops, following the setup
of~\cite{Beenakker:2011sf,NLLfast}. The low-scale mass parameters 
 have been generated using
\texttt{SUSY-HIT}~\cite{Djouadi:2006bz} employing
\texttt{SuSpect2.41}~\cite{Djouadi:2002ze} with the standard model
input $m_t=172.5$~GeV, $ \alpha_s(m_Z)= 0.1172$, $ \overline
m_b(m_b)=4.25$~GeV.\footnote{ Note that in the mass-spectra quoted
  in~\cite{AbdusSalam:2011fc} the masses are rounded to $5$~GeV
  accuracy and only the light-flavour squark masses are
  averaged. Furthermore version 2.3 of \texttt{SuSpect} has been used
  for the CMSSM benchmark points and \texttt{SOFTSUSY} for the mGMSB
  benchmark points.
} The cross sections for the separate squark and gluino production
processes are shown in Tables~\ref{tab:bench1-nll}
and~\ref{tab:bench2-nll}.
The stops are always heavier than $m_{\tilde t_1}>750$ GeV for the considered benchmark points so direct stop-antistop production will be out of reach at the LHC with $\sqrt{s}=7$~TeV (for some of the benchmark points, it might be possible to discover them
 in the gluino-decay products).
Therefore we will give results for stop-antistop production separately in Section~\ref{sec:stop-nll}.

 \begin{table}[t]
\renewcommand{\tabcolsep}{0.45pc} 
 \renewcommand{\arraystretch}{1.3} 
  \centering
   \begin{tabular}{|c||c|c||c|c||c|c|c|c|} \hline 
& & &&&   \multicolumn{3}{c}{ $\sigma_{\text{SUSY}}$(pb),
$\sqrt{s}=7$ TeV}&\\
  Point& $m_0$& $m_{1/2}$ & $m_{\tilde g}$& $m_{\tilde q}$&
     NLO& NLL & $\Delta_{\text{PDF}}$& $K_{\text{NLL}}$\\\hline
    10.1.3& 150 &600& 1357  & 1209 & $0.91_{-0.14}^{+0.13}\times 10^{-2}$&
    $1.04_{-0.09}^{+0.10}\times 10^{-2} $ &  ${}_{-2.9\%}^{+4.2\%}$    & $1.15$  \\\hline
    10.1.4& 162.5 &650& 1461  & 1300 &$4.17_{-0.64}^{+0.60}\times 10^{-3}$&
  $4.79_{-0.42}^{+0.45}\times 10^{-3}$   & ${}_{-3.0\%}^{+4.2\%} $&$1.15$\\ \hline\hline
    10.2.2 &225& 550& 1255  &1130 & $1.83_{-0.27}^{+0.25}\times 10^{-2}$  &
$2.10_{-0.18}^{+0.20}\times 10^{-2}$  &${}_{-3.0\%}^{+4.2\%}$    & $1.15$
\\\hline
     10.2.5& 300&700&1569 & 1412 & $1.66_{-0.27}^{+0.26}\times 10^{-3}$ &
 $1.92_{-0.17}^{+0.18}\times 10^{-3}$ &${}_{-3.1\%}^{+4.4\%}$  & $1.16$\\\hline\hline  
 10.3.2& 350&525& 1209 &1115 &$2.21_{-0.33}^{+0.30}\times 10^{-2}$  & 
 $2.54_{-0.23}^{+0.24}\times 10^{-2}$ &
 ${}_{-3.2\%}^{+4.4\%} $  & $1.15$
 \\\hline
  10.3.3& 400&600& 1367 &1261& $6.32_{-0.98}^{+0.92}\times 10^{-3}$ &
$7.31_{-0.66}^{+0.71}\times 10^{-3}$ &
   ${}_{-3.3\%}^{+4.4\%} $& $1.16$
 \\\hline\hline 
      10.4.2& 850&400&983 &1160 &$3.48_{-0.59}^{+0.57}\times 10^{-2}$  & 
$4.24_{-0.43}^{+0.48} \times 10^{-2}$ & ${}_{-6.2\%}^{+7.0\%}$& $1.22$
\\\hline
       10.4.4& 1050&500&1207 &1427& $3.65_{-0.66}^{+0.68}\times 10^{-3}$  &
  $4.59_{-0.50}^{+0.54}\times 10^{-3}$     &
 ${}_{-7.4\%}^{+8.5\%}$& $1.26$
\\\hline
 \end{tabular}
   \caption{CMSSM benchmark points and total inclusive SUSY production  cross sections for $\tan\beta=10$, $A_0=0$. All masses are given in GeV.}
   \label{tab:bench1}
 \end{table}

\begin{table}[t]
  \renewcommand{\tabcolsep}{0.45pc} 
 \renewcommand{\arraystretch}{1.3} 
   \centering
   \begin{tabular}{|c||c|c||c|c||c|c|c|c|} \hline
  \centering
& & &&&   \multicolumn{3}{c}{ $\sigma_{\text{SUSY}}$(pb),
$\sqrt{s}=7$ TeV}&\\
  Point& $m_0$& $m_{1/2}$ & $m_{\tilde g}$& $m_{\tilde q}$&
     NLO& NLL & $\Delta_{\text{PDF}}$ &$K_{\text{NLL}}$\\\hline
     40.1.2& 345 &550& 1259  & 1144 & $1.66_{-0.25}^{+0.23}\times 10^{-2}$ & 
 $1.91_{-0.17}^{+0.18}\times 10^{-2}$ &$ {}_{-3.1\%}^{+4.3\%}$& $1.15$ \\\hline
  40.1.4& 375 &650& 1468  & 1325 & $3.48_{-0.55}^{+0.52}\times 10^{-3}$  & 
 $4.02_{-0.36}^{+0.38}\times 10^{-3} $ &
 ${}_{-3.1\%}^{+4.4\%}$ &  $1.15$\\ \hline\hline
 40.2.2& 600 &500& 1172  &1153  &$1.90_{-0.29}^{+0.28}\times 10^{-2}$
      &$2.22_{-0.21}^{+0.23}\times 10^{-2}$   &${}_{-3.8\%}^{+5.0\%} $& $1.17$\\ \hline
 40.2.5& 750 &650&1492  & 1460 &$1.34_{-0.23}^{+0.22}\times 10^{-3}$ &
 $1.59_{-0.16}^{+0.17}\times 10^{-3}$ &  
${}_{-4.0\%}^{+5.3\%} $   &$1.19$ \\\hline\hline
40.3.1& 1000 &350& 886  &1182  & $5.30_{-0.94}^{+0.96}\times 10^{-2}$   &
$6.63_{-0.71}^{+0.77}\times 10^{-2}$&
 ${}_{-7.8\%}^{+8.4\%}$ & $1.25$\\ \hline
40.3.5& 1200 &450&1111  & 1446 &  $5.29_{-1.00}^{+1.06}\times 10^{-3}$  &
$6.86_{-0.77}^{+0.83}\times 10^{-3} $&${}_{-9.3\%}^{+10\%}$  &$1.30$ \\ \hline
 \end{tabular}
   \caption{ CMSSM benchmark points and total inclusive SUSY production  cross sections for $\tan\beta=40$, $A_0=-500$ GeV.  All masses are given in GeV.}
   \label{tab:bench2}
 \end{table}
\begin{table}[ht]
\renewcommand{\tabcolsep}{0.45pc} 
 \renewcommand{\arraystretch}{1.3} 
 \begin{tabular}{|c||c||c|c||c|c|c|c|} \hline
  \centering
& &&&   \multicolumn{3}{c}{ $\sigma_{\text{SUSY}}$(pb),
$\sqrt{s}=7$ TeV}&\\
Point & $\Lambda_{\text{SUSY}}$ & $m_{\tilde g}$& $m_{\tilde q}$&
     NLO& NLL & $\Delta_{\text{PDF}}$ &$K_{\text{NLL}}$\\\hline
\small{mGMSB2.2.1}&
 $1.2\times 10^5$ &943  & 1142 & $4.55_{-0.77}^{+0.76}\times 10^{-2}$ & 
 $5.57_{-0.57}^{+0.63}\times 10^{-2} $ &$ {}_{-6.4\%}^{+7.3\%}$& $1.22$ \\\hline
\small{mGMSB2.2.4} &  $1.5\times 10^5$& 1154 &1408  & $5.04_{-0.92}^{+0.95}\times 10^{-3}$  & 
 $6.38_{-0.69}^{+0.76}\times 10^{-3}$ &${}_{-7.9\%}^{+8.9\%}$ &  $1.27$\\ \hline
 \end{tabular}
   \caption{ mGMSB benchmark points and total inclusive SUSY production  cross sections for $\tan\beta=15$, $N_{\text{mess}}=1$, $M_{\text{mess}}=10^9$ GeV. All masses and scales in GeV.}
   \label{tab:bench3}
 \end{table}

 For comparison, NLO results obtained using \texttt{PROSPINO} are also
 shown in the tables.  Our setup agrees with the one
 in~\cite{Beenakker:2011fu,NLLfast} and the NLO results agree at the
 expected one-percent level or better with the results obtained from
 an interpolated grid using \texttt{NLL-fast}~\cite{NLLfast}.  For the
 NLO results the scale uncertainty is estimated by varying the
 factorisation scale in the interval $M/2<\mu_f<M$, with the
 renormalization scale set equal to the factorization scale.  The
 tables also include the relative PDF uncertainties of the NLO cross
 sections. Based on experience with top-pair
 production~\cite{Beneke:2011mq}, we expect that these agree with the
 relative PDF uncertainties of the NLL results at the relevant
 accuracy.  Note that, due to correlations, the PDF uncertainty of the
 total SUSY production cross section is not equal to the sum of the
 uncertainties of the individual channels~\cite{Beenakker:2011fu}.
 For the NLL results we quote the total uncertainty including scale
 and resummation uncertainties, as discussed in
 Section~\ref{sec:scales}. In this case, the uncertainty of the total SUSY
 production cross section was obtained by neglecting correlations and
 adding the uncertainties of the individual production channels
 linearly, which gives a conservative upper bound.

 As can be seen from Tables~\ref{tab:bench1}, \ref{tab:bench2}
 and~\ref{tab:bench3}, for the benchmark points with $m_{\tilde
   q}\lesssim m_{\tilde g}$ the NLL correction to the inclusive squark
 and gluino production cross section is in the $15-19\%$-range, which is
 consistent with Figure~\ref{fig:contour}. For these points the
 theoretical uncertainty is reduced from $\pm 14$-$15\%$ at NLO to
 $\pm 9$-$10\%$ at NLL, in agreement with the behaviour seen in
 Figure~\ref{fig:totaluncertainty7}.  The PDF uncertainty is also
 small for the benchmark points with $m_{\tilde q}< m_{\tilde g}$,
 since only the relatively precisely known quark PDFs are relevant at leading order for the
 dominant squark-squark production channel.
 For scenarios where the gluinos
are lighter than the squarks, the NLL corrections grow to $20$-$30\%$
and the theoretical uncertainty at NLL is at the $11$-$12\%$ level.
These features can be understood by considering the relative
contributions of the different production processes in
Figure~\ref{fig:sigmaNLL} and Tables~\ref{tab:bench1-nll} and
~\ref{tab:bench2-nll}. For $m_{\tilde q}\lesssim m_{\tilde g}$,
squark-pair production with moderate NLL corrections in the $6$-$10\%$
range is the dominant production channel, with a non-negligible
contribution from squark-gluino production with larger NLL corrections
of the order of $30\%$. 
 For $m_{\tilde g}<m_{\tilde q}$ the roles are
reversed, resulting in larger NLL corrections to the total rate and a
slightly larger theoretical uncertainty, as can be seen from the
uncertainties of the different production channels in
Figure~\ref{fig:totaluncertainty7}. 
Due to the large uncertainties in the
current gluon PDF sets for large $x$, the PDF error for these points rises to the $10\%$-level. In
gluino-pair production, especially for large gluino masses, the PDF error can even become $\sim 30\%$.
 At a mass ratio $\frac{m_{\tilde
    g}}{m_{\tilde q}}\lesssim 0.75$, as realized for the $40.3.1$ and
$40.3.5$ points, gluino-pair production overtakes squark-pair
production as the second-most important channel.  For these points,
the gluinos are relatively light, and the NLL corrections to gluino-pair production are at most $45\%$. For benchmark points with heavier
gluinos, the NLL corrections can grow to $70$-$80\%$, but the
gluino-pair production rate is negligible for these points.  For the
moderate mass ratios considered here, squark-antisquark production is
always much suppressed, but would have the second-largest cross section
for $\frac{m_{\tilde g}}{m_{\tilde q}}\to 2$ .

\begin{table}[p]
\begin{center}
\begin{tabular}{|c|c|l|l|l|c|}
\hline& &
\multicolumn{3}{c}{ $\sigma(pp\to \tilde s \tilde s')$(pb),
$\sqrt{s}=7$ TeV}&
\\\hline
Point&$\tilde s\tilde s'$ &  NLO   &NLL &  $\Delta_{\text{PDF}}$& $ K_{\text{NLL}}$
\\[0.08cm]\hline
10.1.3&$ \tilde q\bar{\tilde q}$&$7.26_{-1.25}^{+1.34}\times 10^{-4}$ & 
$8.80_{-0.73}^{+0.84}\times 10^{-4} $
 & ${}^{+11\%}_{-11\%}$&  $1.21$ \\[0.08cm]
& $ \tilde q\tilde q$& 
$6.22_{-0.88}^{+0.81}\times 10^{-3}$ & $6.72_{-0.48}^{+0.55}\times 10^{-3}$
 & ${}^{+3.8\%}_{-2.8\%}$&  $1.08$\\[0.08cm]
& $ \tilde q\tilde g$& 
$2.06_{-0.35}^{+0.30}\times 10^{-3}$ & $2.69_{-0.32}^{+0.32}\times 10^{-3}$
 & ${}^{+11\%}_{-10\%}$&  $1.31$\\[0.08cm]
& $ \tilde g\tilde g$& 
$0.77_{-0.17}^{+0.17}\times 10^{-4}$ & $1.26_{-0.22}^{+0.24}\times 10^{-4}$
 & ${}^{+25\%}_{-23\%}$&  $1.64$
\\[0.08cm]\hline
10.1.4&$ \tilde q\bar{\tilde q}$&$2.91_{-0.52}^{+0.56} \times 10^{-4}$ &
 $3.59_{-0.31}^{+0.35}\times 10^{-4}$
 & ${}_{-12\%}^{+13\%}$&  $1.23$ \\[0.08cm]
& $ \tilde q\tilde q$& 
$3.02_{-0.44}^{+0.41}\times 10^{-3}$ & $3.28_{-0.24}^{+0.27}\times 10^{-3} $
 & ${}_{-2.8\%}^{+3.9\%} $&  $1.09$\\[0.08cm]
& $ \tilde q\tilde g$& 
$0.83_{-0.15}^{+0.13}\times 10^{-3}$ & $1.11_{-0.14}^{+0.13}\times 10^{-3}$
 & ${}_{-11\%}^{+12\%} $&  $1.33$\\[0.08cm]
& $ \tilde g\tilde g$& 
$2.54_{-0.57}^{+0.59}\times 10^{-5}$ & $4.36_{-0.81}^{+0.87}\times 10^{-5}$
 & ${}_{-25\%}^{+28\%}$&  $1.72$
\\[0.08cm]\hline\hline
10.2.2&$ \tilde q\bar{\tilde q}$&$1.60_{-0.27}^{+0.29}\times 10^{-3}$ & 
$1.92_{-0.16}^{+0.18}\times 10^{-3} $
 & ${}^{+9.9\%}_{-9.6\%}$&  $1.20$ \\[0.08cm]
& $ \tilde q\tilde q$& 
$1.17_{-0.16}^{+0.15}\times 10^{-2}$ & $1.25_{-0.09}^{+0.10}\times 10^{-2}$
 & ${}^{+3.7\%}_{-2.7\%}$&  $1.07$\\[0.08cm]
& $ \tilde q\tilde g$& 
$4.81_{-0.79}^{+0.67}\times 10^{-3}$ & $6.18_{-0.72}^{+0.71}\times 10^{-3}$
 & ${}^{+9.7\%}_{-9.2\%}$&  $1.28$\\[0.08cm]
& $ \tilde g\tilde g$& 
$2.27_{-0.48}^{+0.48}\times 10^{-4}$ & $3.56_{-0.61}^{+0.65}\times 10^{-4} $
 & ${}^{+23\%}_{-22\%}$&  $1.57$
\\[0.08cm]\hline
10.2.5&$ \tilde q\bar{\tilde q}$&$0.96_{-0.18}^{+0.19}\times 10^{-4}$ &
$1.21_{-0.10}^{+0.12}\times 10^{-4}$
 & ${}_{-14\%}^{+14\%}$&  $1.26$ \\[0.08cm]
& $ \tilde q\tilde q$& 
$1.26_{-0.19}^{+0.18}\times 10^{-3}$ & $ 1.38_{-0.11}^{+0.12}\times 10^{-3}$
 & ${}_{-3.0\%}^{+4.0\%}$&  $1.10$\\
& $ \tilde q\tilde g$& 
$3.00_{-0.55}^{+0.51}\times 10^{-4}$ & $4.11_{-0.53}^{+0.51}\times 10^{-4}$
 & ${}_{-12\%}^{+13\%}$&  $1.37$\\[0.08cm]
& $ \tilde g\tilde g$& 
$0.80_{-0.19}^{+0.20}\times 10^{-5}$ & $1.44_{-0.28}^{+0.30}\times 10^{-5} $
 & ${}_{-27\%}^{+31\%}$&  $1.81$
\\[0.08cm]\hline\hline
10.3.2&$ \tilde q\bar{\tilde q}$&$1.88_{-0.32}^{+0.34}\times 10^{-3} $
& $2.24_{-0.18}^{+0.21}\times 10^{-3}$
& ${}_{-9.2\%}^{+9.8\%}$&  $1.19$ \\[0.08cm]
& $ \tilde q\tilde q$& 
$1.33_{-0.19}^{+0.17}\times 10^{-2} $ & $1.43_{-0.10}^{+0.12}\times 10^{-2}$
 & ${}_{-2.7\%}^{+3.7\%}$&  $1.07$\\[0.08cm]
& $ \tilde q\tilde g$& 
$6.49_{-1.05}^{+0.89}\times 10^{-3}$ & $8.30_{-0.95}^{+0.94}\times 10^{-3} $
 & ${}_{-8.9\%}^{+9.4\%} $&  $1.28$\\[0.08cm]
& $ \tilde g\tilde g$& 
$3.70_{-0.77}^{+0.77}\times 10^{-4}$ & $ 5.69_{-0.96}^{+1.03}\times 10^{-4}$
 & ${}_{-21\%}^{+22\%}$&  $1.54$
\\[0.08cm]\hline
10.3.3&$ \tilde q\bar{\tilde q}$&$4.40_{-0.78}^{+0.85}\times 10^{-4}$ & 
$5.37_{-0.45}^{+0.52}\times 10^{-4}$
 & ${}_{-11\%}^{+12\%}$&  $1.22$ \\[0.08cm]
& $ \tilde q\tilde q$& 
$4.23_{-0.62}^{+0.58}\times 10^{-3}$ & $4.58_{-0.34}^{+0.39}\times 10^{-3} $
 & ${}_{-2.9\%}^{+3.8\%}$&  $1.08$\\[0.08cm]
& $ \tilde q\tilde g$& 
$1.58_{-0.27}^{+0.24}\times 10^{-3}$ & $ 2.08_{-0.25}^{+0.24}\times 10^{-3}$
 & ${}_{-10\%}^{+11\%}$&  $1.32$\\[0.08cm]
& $ \tilde g\tilde g$& 
$0.69_{-0.15}^{+0.16}\times 10^{-4}$ & $1.13_{-0.21}^{+0.22}\times 10^{-4}$
 & ${}_{-23\%}^{+26\%}$&  $1.64$
\\[0.08cm]\hline\hline
10.4.2&$ \tilde q\bar{\tilde q}$&$1.25_{-0.21}^{+0.22}\times 10^{-3}$ &
$1.48_{-0.12}^{+0.14}\times 10^{-3}$
 & ${}_{-9.9\%}^{+10\%}$&  $1.19$ \\[0.08cm]
& $ \tilde q\tilde q$& 
$1.09_{-0.16}^{+0.16}\times 10^{-2}$ & $1.16_{-0.10}^{+0.12}\times 10^{-2} $
 & ${}_{-2.7\%}^{+3.7\%}$&  $1.07$\\[0.08cm]
& $ \tilde q\tilde g$& 
$1.82_{-0.31}^{+0.29}\times 10^{-2}$ & $2.30_{-0.23}^{+0.24}\times 10^{-2} $
 & ${}_{-8.0\%}^{+8.3\%} $&  $1.27$\\[0.08cm]
& $ \tilde g\tilde g$& 
$4.55_{-0.94}^{+0.99}\times 10^{-3}$ & $6.37_{-0.88}^{+0.95}\times 10^{-3} $
 & ${}_{-17\%}^{+17\%}$&  $1.40$
\\[0.08cm]\hline
10.4.4&$ \tilde q\bar{\tilde q}$&$0.85_{-0.16}^{+0.17}\times 10^{-4}$ &
 $1.07_{-0.09}^{+0.10}\times 10^{-4}$ & ${}_{-15\%}^{+15\%} $&  $1.25$ \\[0.08cm]
& $ \tilde q\tilde q$& 
$1.29_{-0.21}^{+0.22}\times 10^{-3}$ & $1.40_{-0.13}^{+0.16}\times 10^{-3} $
 & ${}_{-3.0\%}^{+4.0\%}$&  $1.08$\\[0.08cm]
& $ \tilde q\tilde g$& 
$1.86_{-0.34}^{+0.35}\times 10^{-3}$ & $2.47_{-0.27}^{+0.28}\times 10^{-3} $
 & ${}_{-10\%}^{+11\%} $&  $1.33$\\[0.08cm]
& $ \tilde g\tilde g$& 
$4.12_{-0.91}^{+1.02}\times 10^{-4}$ & $6.19_{-0.90}^{+0.99}\times 10^{-4} $
 & ${}_{-20\%}^{+22\%}$&  $1.50$
\\[0.08cm]\hline
\end{tabular}
\end{center}\caption{NLL results for CMSSM benchmark points for $\tan\beta=10$, $A_0=0$.}
   \label{tab:bench1-nll}
\end{table}

\begin{table}[p]
\begin{center}
\begin{tabular}{|c|c|l|l|l|c|}
\hline& &
\multicolumn{3}{c}{ $\sigma(pp\to \tilde s \tilde s')$(pb),
$\sqrt{s}=7$ TeV}&
\\\hline
Point&$\tilde s\tilde s'$ &  NLO   &NLL &  $\Delta_{\text{PDF}}$& $ K_{\text{NLL}}$
\\[0.08cm]\hline
40.1.2&$ \tilde q\bar{\tilde q}$&$1.41_{-0.24}^{+0.26}\times 10^{-3}$ & 
$1.68_{-0.14}^{+0.16}\times 10^{-3}$
 & ${}_{-9.7\%}^{+10\%}$&  $1.20$ \\[0.08cm]
& $ \tilde q\tilde q$& 
$1.05_{-0.15}^{+0.13}\times 10^{-2}$ & $1.13_{-0.08}^{+0.09}\times 10^{-2} $
 & ${}_{-2.7\%}^{+3.7\%} $&  $1.07$\\[0.08cm]
& $ \tilde q\tilde g$& 
$4.46_{-0.73}^{+0.63}\times 10^{-3}$ & $5.74_{-0.67}^{+0.66}\times 10^{-3} $
 & ${}_{-9.2\%}^{+9.8\%}$&  $1.29$\\[0.08cm]
& $ \tilde g\tilde g$& $2.18_{-0.46}^{+0.46}\times 10^{-4}$
 & $3.42_{-0.58}^{+0.63}\times 10^{-4} $
 & ${}_{-22\%}^{+23\%}$&  $1.57$
\\[0.08cm]\hline
40.1.4&$ \tilde q\bar{\tilde q}$&$2.28_{-0.41}^{+0.45}\times 10^{-4}$ &
 $2.83_{-0.24}^{+0.27} \times 10^{-4}$
 & ${}_{-13\%}^{+13\%}$&  $1.24$ \\[0.08cm]
& $ \tilde q\tilde q$& 
$2.51_{-0.37}^{+0.35}\times 10^{-3}$ & $2.73_{-0.21}^{+0.23}\times 10^{-3} $
 & ${}_{-2.9\%}^{+3.9\%} $&  $1.09$\\[0.08cm]
& $ \tilde q\tilde g$& 
$7.25_{-1.28}^{+1.16}\times 10^{-4}$ & $9.71_{-1.21}^{+1.17}\times 10^{-4} $
 & ${}_{-11\%}^{+12\%} $&  $1.34$\\[0.08cm]
& $ \tilde g\tilde g$& 
$2.37_{-0.54}^{+0.57}\times 10^{-5}$ & $ 4.08_{-0.75}^{+0.81}\times 10^{-5}  $
 & ${}_{-25\%}^{+28\%}$&  $1.72$
\\[0.08cm]\hline\hline 
40.2.2&$ \tilde q\bar{\tilde q}$&$1.34_{-0.23}^{+0.26}\times 10^{-3}$ & $1.60_{-0.13}^{+0.16}\times 10^{-3}$
 & ${}_{-9.7\%}^{+10\%}$&  $1.19$ \\[0.08cm]
& $ \tilde q\tilde q$& 
$1.04_{-0.15}^{+0.14}\times 10^{-2}$ & $1.11_{-0.08}^{+0.10}\times 10^{-2}$
 & ${}_{-2.7\%}^{+3.7\%}$&  $1.07$\\[0.08cm]
& $ \tilde q\tilde g$& 
$6.74_{-1.11}^{+0.97}\times 10^{-3}$ &$ 8.65_{-0.98}^{+0.98}\times 10^{-3}$
 & ${}_{-8.9\%}^{+9.4\%}$&  $1.28$\\[0.08cm]
& $ \tilde g\tilde g$& 
$5.48_{-1.14}^{+1.15}\times 10^{-4}$ & $8.28_{-1.35}^{+1.46}\times 10^{-4}$
 & ${}_{-20\%}^{+22\%}$&  $1.51$
\\[0.08cm]\hline
40.2.5&$ \tilde q\bar{\tilde q}$&$6.14_{-1.17}^{+1.34}\times 10^{-5} $ & 
$7.75_{-0.67}^{+0.79}\times 10^{-5}$
 & ${}_{-15\%}^{+15\%}$&  $1.26$ \\[0.08cm]
& $ \tilde q\tilde q$& 
$9.01_{-1.43}^{+1.41}\times 10^{-4}$ & $9.87_{-0.82}^{+0.94}\times 10^{-4}  $
& ${}_{-3.1\%}^{+4.0\%} $&  $1.10$\\[0.08cm]
& $ \tilde q\tilde g$& 
$3.64_{-0.67}^{+0.64}\times 10^{-4}$ & $ 4.98_{-0.62}^{+0.61}\times 10^{-4} $
 & ${}_{-12\%}^{+13\%}$&  $1.37$\\[0.08cm]
& $ \tilde g\tilde g$& 
$1.82_{-0.42}^{+0.46}\times 10^{-5}$ & $3.14_{-0.57}^{+0.62}\times 10^{-5}$
 & ${}_{-26\%}^{+29\%}$&  $1.72$
\\[0.08cm]\hline\hline
40.3.1&$ \tilde q\bar{\tilde q}$&$1.00_{-0.17}^{+0.18}\times 10^{-3}$ & $1.19_{-0.10}^{+0.11}\times 10^{-3}$
 & ${}_{-10\%}^{+11\%}$&  $1.19$ \\[0.08cm]
& $ \tilde q\tilde q$& 
$0.95_{-0.15}^{+0.15}\times 10^{-2}$ & $1.01_{-0.09}^{+0.12}\times 10^{-2}$
 & ${}_{-2.7\%}^{+3.7\%}$&  $1.06$\\[0.08cm]
& $ \tilde q\tilde g$& 
$2.83_{-0.49}^{+0.48}\times 10^{-2}$ & $3.57_{-0.35}^{+0.37}\times 10^{-2} $
 & ${}_{-7.5\%}^{+7.9\%}$&  $1.26$\\[0.08cm]
& $ \tilde g\tilde g$& 
$1.42_{-0.29}^{+0.31}\times 10^{-2}$ & $1.93_{-0.25}^{+0.27}\times 10^{-2}$
 & ${}_{-15\%}^{+15\%}$&  $1.36$
\\[0.08cm]\hline
40.3.5&$ \tilde q\bar{\tilde q}$&$7.05_{-1.30}^{+1.45}\times 10^{-5}$ &
 $8.81_{-0.75}^{+0.87}\times 10^{-5}$
 & ${}_{-15\%}^{+15\%}$&  $1.25$ \\[0.08cm]
& $ \tilde q\tilde q$& 
$1.14_{-0.19}^{+0.20}\times 10^{-3}$ & $1.23_{-0.12}^{+0.15}\times 10^{-3}$
 & ${}_{-3.0\%}^{+4.2\%}$&  $1.08$\\[0.08cm]
& $ \tilde q\tilde g$& 
$2.87_{-0.53}^{+0.55}\times 10^{-3}$ & $3.78_{-0.40}^{+0.41}\times 10^{-3}$
 & ${}_{-9.8\%}^{+10\%} $&  $1.32$\\[0.08cm]
& $ \tilde g\tilde g$& 
$1.21_{-0.26}^{+0.29}\times 10^{-3}$ & $1.75_{-0.24}^{+0.26}\times 10^{-3}$
 & ${}_{-18\%}^{+19\%}$&  $1.45$
\\[0.08cm]\hline\hline
\small{mGMSB2.2.1}&$ \tilde q\bar{\tilde q}$&
$1.48_{-0.25}^{+0.26}\times 10^{-3}$ & $1.76_{-0.14}^{+0.17}\times 10^{-3}$
 & ${}_{-9.6\%}^{+10\%}$&  $1.19$ \\[0.08cm]
& $ \tilde q\tilde q$& 
$1.26_{-0.19}^{+0.19}\times 10^{-2}$ & $1.34_{-0.12}^{+0.15}\times 10^{-2}$
 & ${}_{-2.6\%}^{+3.8\%} $&  $1.06$\\[0.08cm]
& $ \tilde q\tilde g$& 
$2.43_{-0.41}^{+0.39}\times 10^{-2}$ & $3.07_{-0.31}^{+0.32}\times 10^{-2}$
 & ${}_{-7.6\%}^{+8.1\%}$&  $1.26$\\[0.08cm]
& $ \tilde g\tilde g$& 
$7.17_{-1.46}^{+1.54}\times 10^{-3}$ & $9.92_{-1.35}^{+1.45}\times 10^{-3} $
 & ${}_{-16\%}^{+17\%}$&  $1.38$
\\[0.08cm]\hline
\small{mGMSB2.2.4}&$ \tilde q\bar{\tilde q}$&
$1.04_{-0.19}^{+0.21}\times 10^{-4}$ & $1.29_{-0.11}^{+0.13}\times 10^{-4}$
 & ${}_{-14\%}^{+14\%} $&  $1.24$ \\[0.08cm]
& $ \tilde q\tilde q$& 
$1.52_{-0.25}^{+0.26}\times 10^{-3}$ & $1.65_{-0.15}^{+0.19}\times 10^{-3}$
 & ${}_{-2.9\%}^{+4.1\%}$&  $1.08$\\[0.08cm]
& $ \tilde q\tilde g$& 
$2.67_{-0.49}^{+0.49}\times 10^{-3}$ & $3.52_{-0.38}^{+0.39}\times 10^{-3}$
 & ${}_{-9.9\%}^{+10\%} $&  $1.32$\\[0.08cm]
& $ \tilde g\tilde g$& 
$0.74_{-0.16}^{+0.18}\times 10^{-3}$ & $1.09_{-0.15}^{+0.17}\times 10^{-3}$
 & ${}_{-19\%}^{+20\%}$&  $1.47$\\\hline
\end{tabular}
\end{center}\caption{NLL results for CMSSM benchmark points for $\tan\beta=40$, $A_0=-500$ and mGMSB benchmark points.}
   \label{tab:bench2-nll}
\end{table}

\subsection{Results for stop-antistop production}
\label{sec:stop-nll}

\begin{figure}[t!]
\begin{center}
\includegraphics[width=0.49 \linewidth]{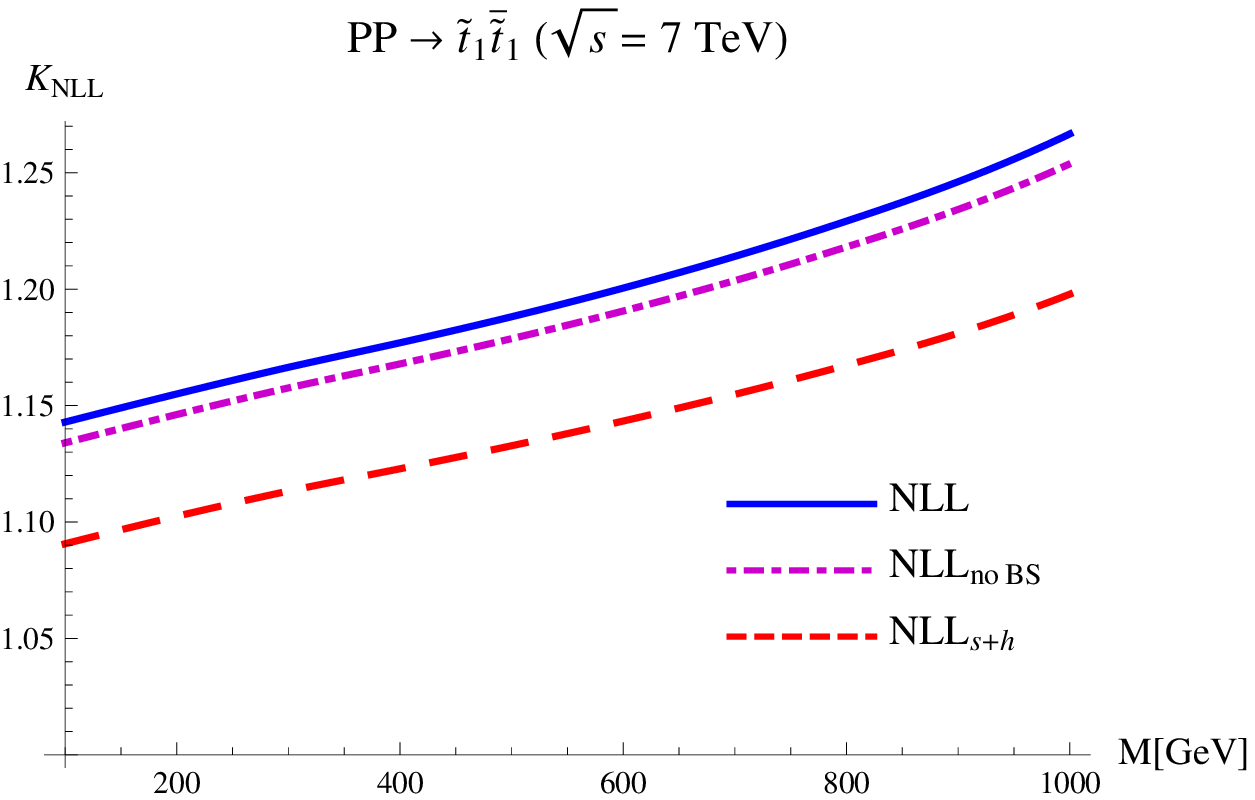}
\includegraphics[width=0.49 \linewidth]{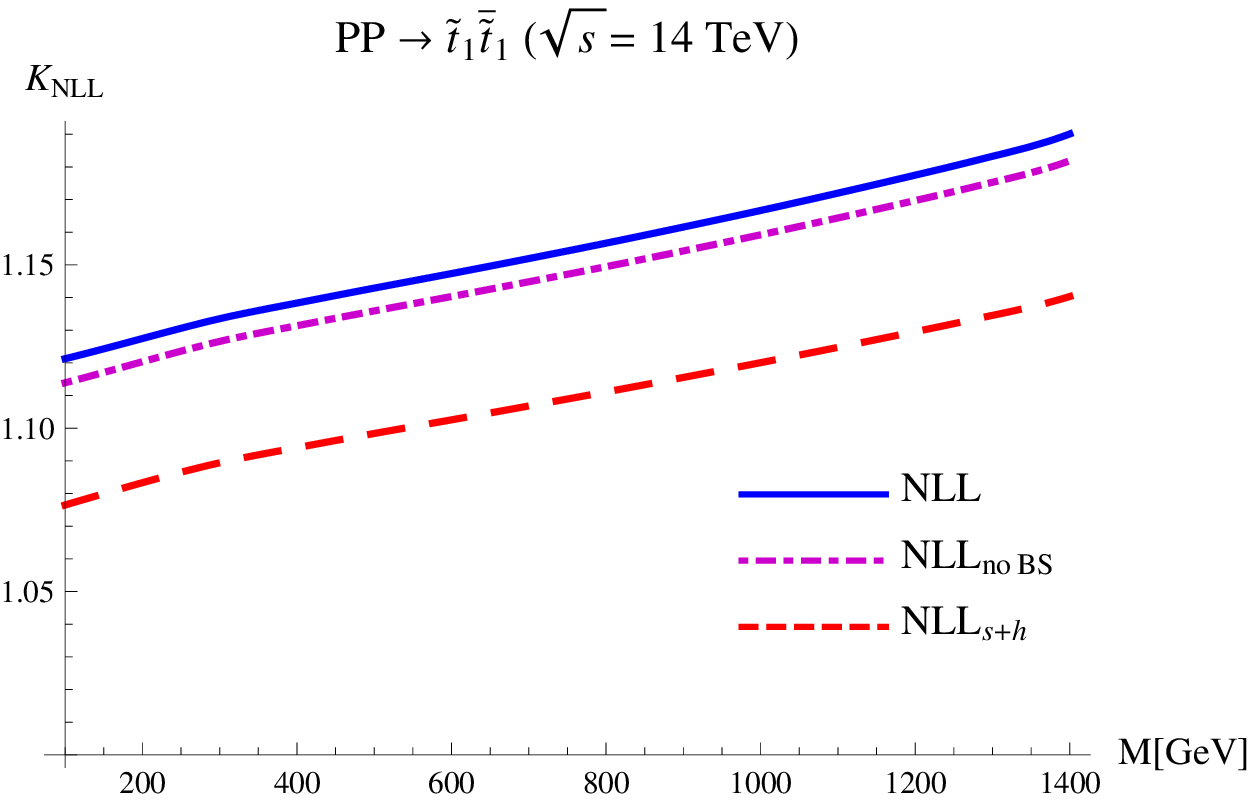}
\end{center}
\caption{$K_\text{NLL}$ for stop-antistop production at the LHC at $\sqrt{s}=7\,$TeV (left) and $\sqrt{s}=14\,$TeV (right)
for different NLL approximations: NLL (solid blue), NLL$_\text{no BS}$ (dot-dashed purple) and NLL$_{s+h}$ (dashed red).
} 
\label{fig:Kstop} 
\end{figure}

\begin{figure}[t!]
\begin{center}
\includegraphics[width=0.49 \linewidth]{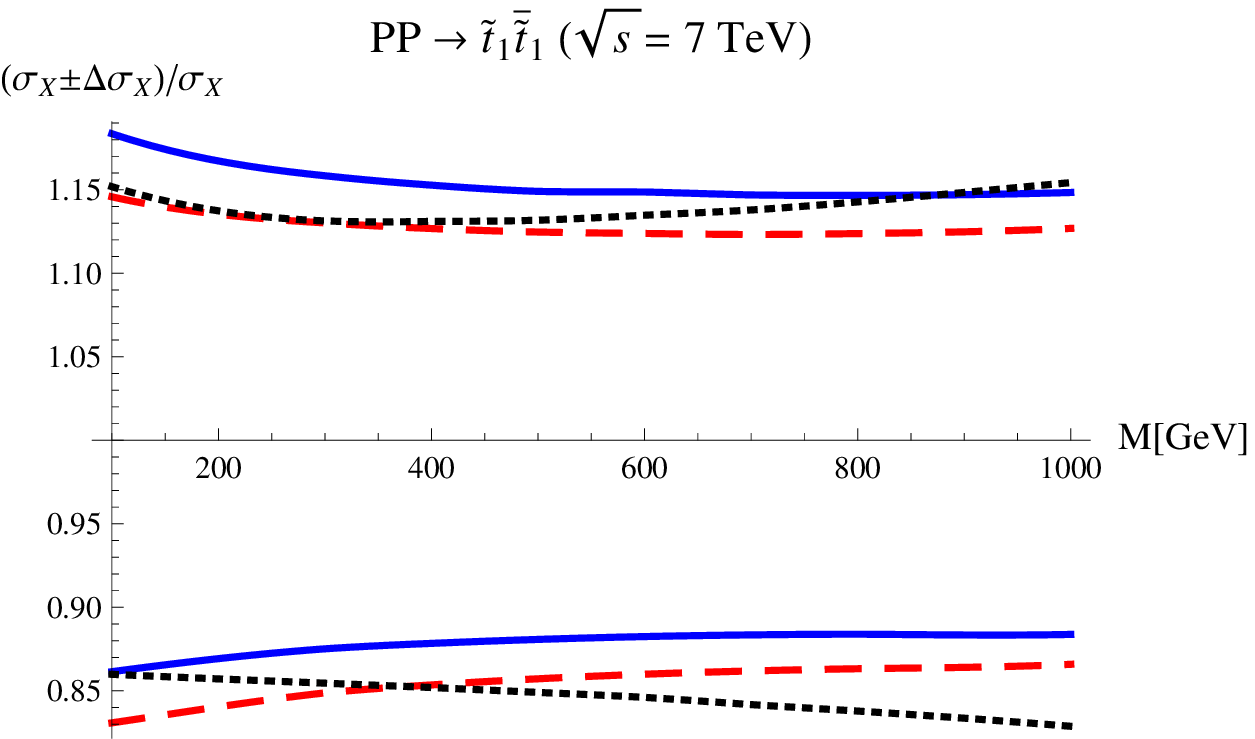}
\includegraphics[width=0.49 \linewidth]{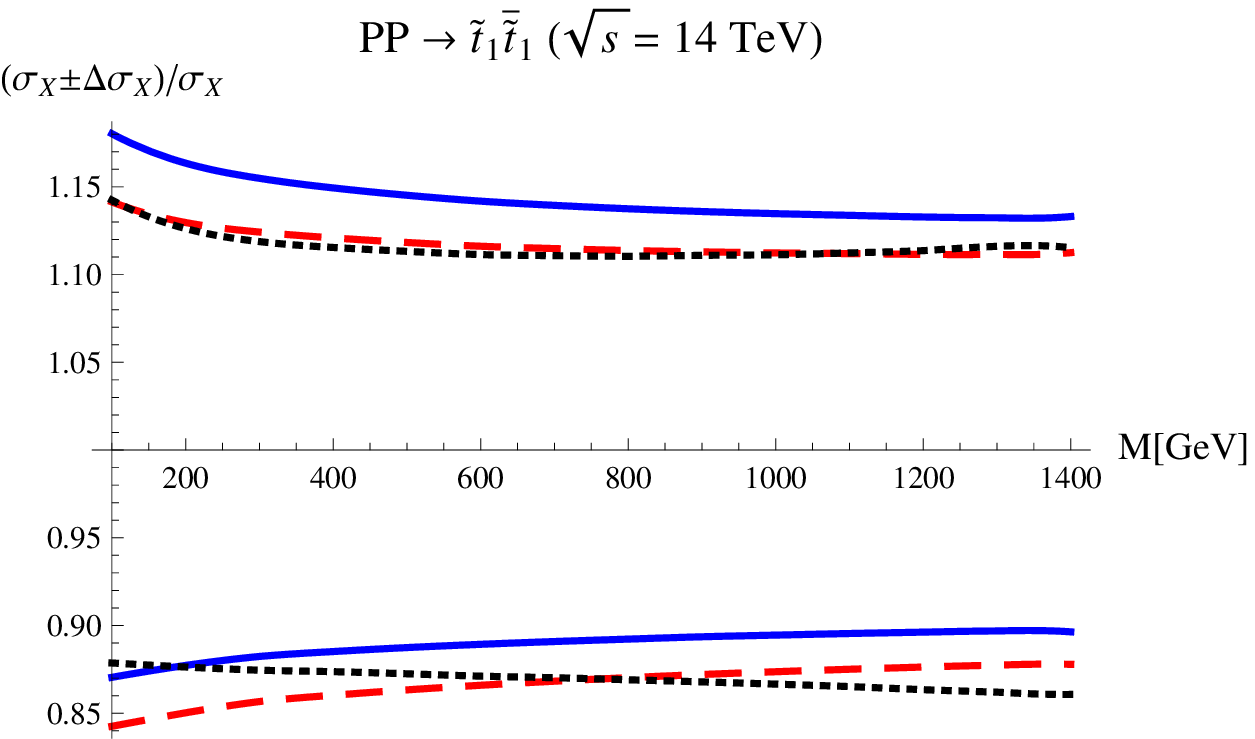}
\end{center}
\caption{Total theoretical uncertainty of  the NLO approximation (dotted black), full NLL resummed result (solid blue)
and NLL$_{s+h}$ (dashed red) for $\tilde{t}_1 \bar{\tilde{t}}_1$ 
at the LHC with $\sqrt{s}=\,$7 TeV (left) and $\sqrt{s}=\,$14 TeV (right). All cross sections are normalized to one at the central value of the scales.} 
\label{fig:unc_stop} 
\end{figure}

In this section we present results for resummation for stop-antistop production
$PP \rightarrow \tilde{t}_i \bar{\tilde{t}}_i$. At LO 
the cross section depends only on the stop mass, while at NLO it also presents a (much smaller) dependence on the mixing angle $\theta_{\tilde t}$ and the squark and gluino masses.
For definiteness, we follow~\cite{Beenakker:2011fu} and fix all the parameters, except for the stop mass, to 
the values corresponding to the CMSSM benchmark point 40.2.4 with SUSY breaking parameters $m_0=700\,$GeV, $m_{1/2}=600\,$GeV, $A_0=-500\,$GeV and
$\tan \beta=40$ \cite{AbdusSalam:2011fc}. Using \texttt{SUSY-HIT}~\cite{Djouadi:2006bz} we obtain the relevant input parameters $m_{\tilde g}= 1386$~GeV, $m_{\tilde q}=1358$ and $\cos\theta_{\tilde t}=0.39$. The numerical dependence  on these parameters has been studied in~\cite{Beenakker:2010nq} and found to be negligible. 
As discussed in Section~\ref{sec:stop} we will limit ourselves to results for the lighter mass-eigenstate $\tilde t_1$.
\begin{table}[t]
  \renewcommand{\tabcolsep}{0.45pc} 
 \renewcommand{\arraystretch}{1.3} 
\begin{center}
\begin{tabular}{|c|l|l|l|c|}
\hline& 
\multicolumn{3}{c}{ $\sigma(pp\to \tilde t_1 \bar{\tilde t}_1)$(pb),
$\sqrt{s}=7$ TeV}&
\\\hline
$m_{\tilde t_1}$ &  NLO   &NLL &  $\Delta_{\text{PDF}}$& $ K_{\text{NLL}}$
\\\hline
100 &$4.18_{-0.59}^{+0.63}\times 10^2$ & 
$4.77_{-0.66}^{+0.87}\times 10^2$ & ${}_{-2.4\%}^{+2.0\%}$&   1.14\\\hline
200& $1.28_{-0.18}^{+0.17}\times 10^1$ & $1.47_{-0.19}^{+0.25}\times 10^1$ &${}_{-4.1\%}^{+3.1\%} $ & 1.15\\\hline
300 & $1.28^{+0.17}_{-0.18}$  & $1.49^{+0.24}_{-0.19}$ & ${}_{-5.3\%}^{+4.4\%}$ & 1.17\\\hline
400 & $2.12_{-0.31}^{+0.28}\times 10^{-1}$  & $2.50_{-0.30}^{+0.38}\times 10^{-1}$ &${}_{-6.4\%}^{+5.7\%}$ & 1.18\\
\hline
\end{tabular}
\end{center}
 \caption{Stop-antistop production cross sections. All masses are given in GeV.}
   \label{tab:NLL-stop}
 \end{table}

Numerical results for the lightest stop mass eigenstates for masses in the $100$-$400$~GeV range are
presented in table~\ref{tab:NLL-stop}.  
The NLL $K$-factor is plotted as a
function of the mass in Figure \ref{fig:Kstop}.  For the stop-mass
range considered here the resummation effects are in the $15$-$25\%$
range at the LHC with $\sqrt{s}=7\,$TeV and at the $12$-$20\%$-level
at a centre-of-mass energy of 14 TeV.  The NLL-corrections are
therefore moderate for the mass-ranges accessible at the LHC, but
larger than for light-flavour squark-antisquark production for the
same masses (c.f. figure~\ref{fig:K7}).  Contrary to the latter, for stops the
most substantial contribution is given by pure soft resummation, with
Coulomb effects, including bound-state corrections, in the 5$\%$ range
of the NLO result.  As the partonic cross sections for the
gluon-fusion channel are identical for the light-flavour and top
squarks, the different behaviour can be attributed to the
quark-antiquark induced channel. For the light-flavour squarks it is
dominated by $S$-wave colour-singlet production with a large, attractive
Coulomb potential. For stop production the colour-singlet channel is
absent, and only the $P$-wave colour-octet channel with a smaller
repulsive Coulomb potential contributes. The increased relative size
of the soft corrections compared to the light-flavour squarks follows
from the $P$-wave suppression of the quark-antiquark channel and the
resulting dominance of the gluon-fusion channel with larger soft
corrections due to the colour factors $C_A=N_C$.

The total theoretical uncertainty of the fixed-order NLO result and the resummed cross section, with and without Coulomb resummation, is 
compared in Figure \ref{fig:unc_stop}. The width of the uncertainty band for NLL and NLL$_{s+h}$ is similar, consistent with the observed small 
effect of Coulomb resummation. One can also notice that the resummed results shows almost no uncertainty reduction compared 
to the NLO result, except for the high-mass range.

\subsection{Comparison with Mellin-space results}
\label{sec:compare}

Results for NLL resummation of soft logarithms for squark and gluino
production have been presented earlier in
\cite{Kulesza:2008jb,Kulesza:2009kq,Beenakker:2009ha,Beenakker:2010nq,Beenakker:2011fu}.
These works adopt the so-called Mellin-space formalism, in which the
threshold logarithms are exponentiated in Mellin-moment space, where
singular terms appear as logarithms of the Mellin-moment $N$, and the
resummed cross section is then numerically inverted back to
momentum-space.  In this section we compare these earlier predictions
to the momentum-space formalism adopted here, using the numerical code
\texttt{NLL-fast}~\cite{NLLfast} to compute the Mellin-space resummed
cross sections. Since \texttt{NLL-fast} provides results for
soft-resummation only (i.e. no Coulomb effects beyond ${\cal
  O}(\alpha_s)$ are included), for the comparison we introduce two
additional NLL implementations:
\begin{itemize}
\item {\bf NLL$_{s}$}: this implementation includes NLL resummation of soft logarithms, but no Coulomb or hard effects beyond ${\cal O}(\alpha_s)$. This is
achieved using Eqs. (\ref{eq:NLLs}) and (\ref{eq:NLLs-P}) and setting $\mu_h=\mu_f$. 
For the soft scale we adopt the running scale given in Eqs. (\ref{eq:runmus_a}), (\ref{eq:runmus_b}). 
\item {\bf NLL$_{s, \text{fixed}}$}: as above, but with the running scale replaced by the fixed soft scale $\mu_s$ determined from Eq. (\ref{eq:def-mus}). 
\end{itemize}

\begin{figure}[p]
\begin{center}
\includegraphics[width=0.49 \linewidth]{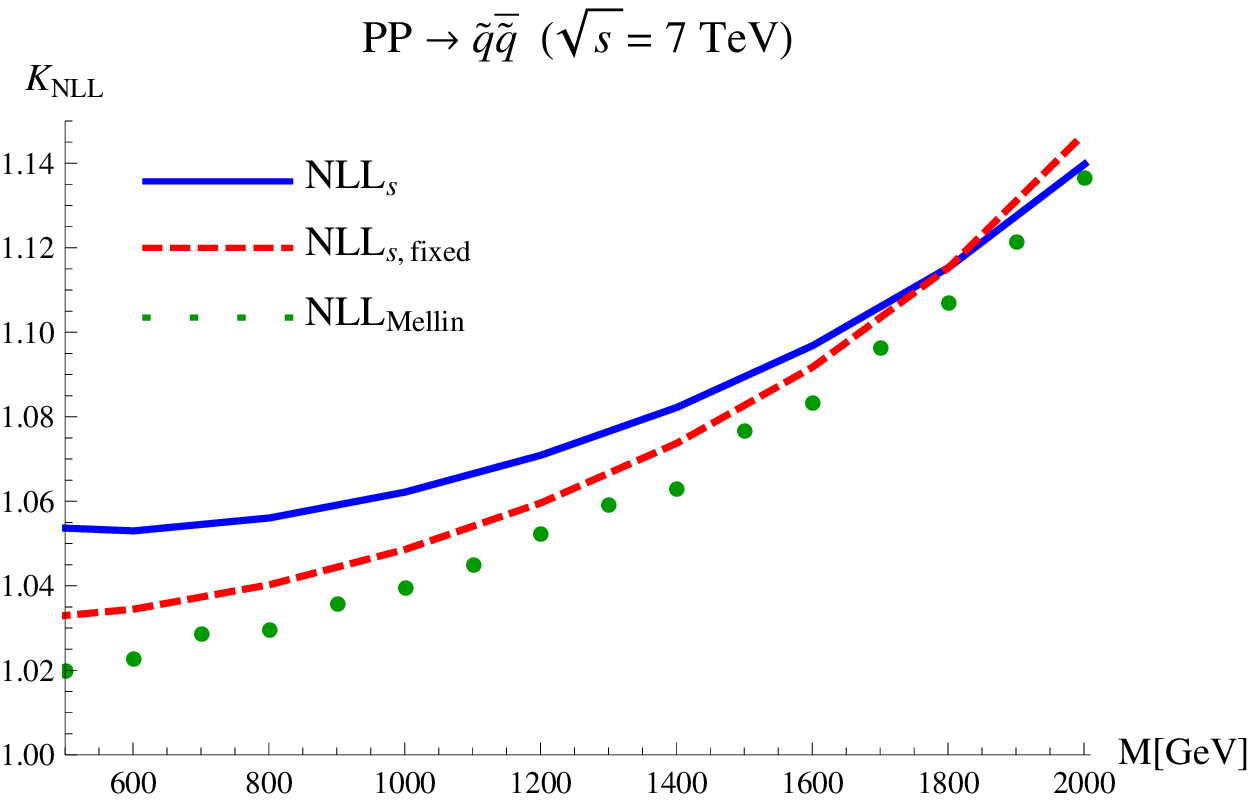}
\includegraphics[width=0.49 \linewidth]{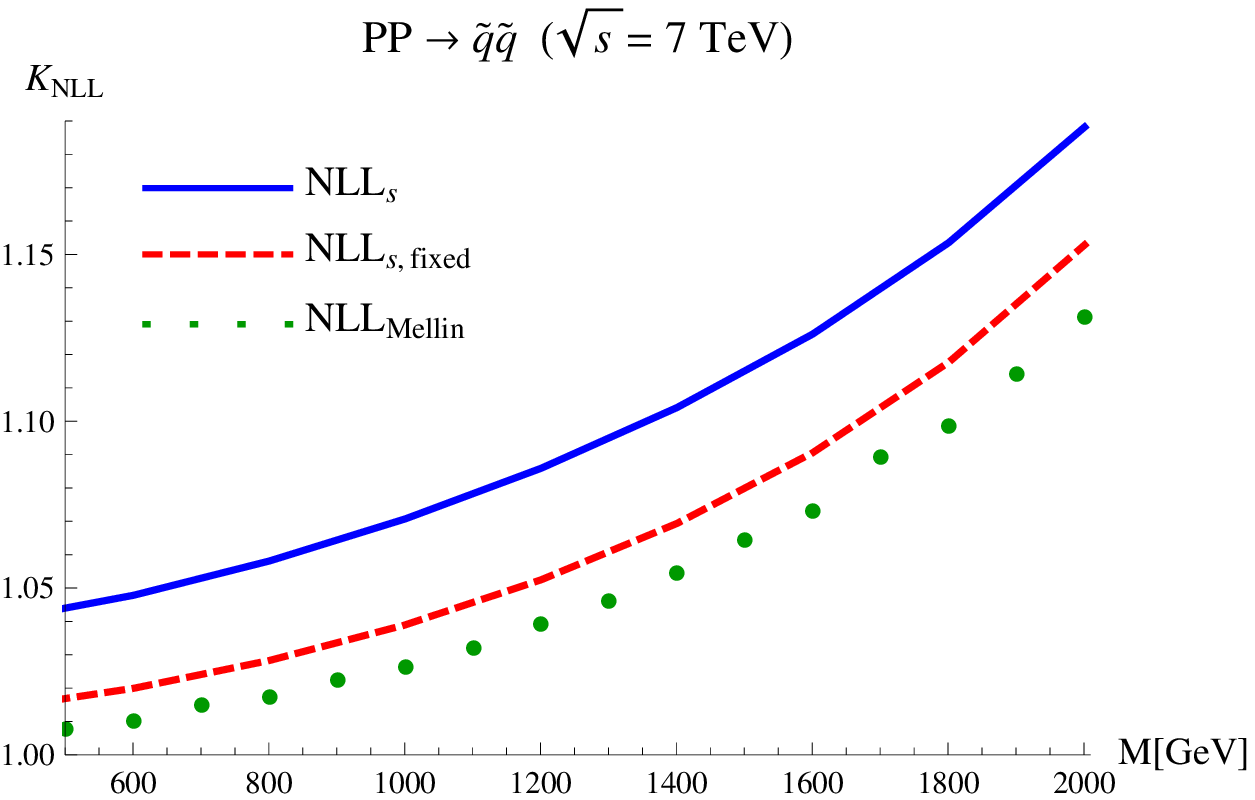}\\
\includegraphics[width=0.49 \linewidth]{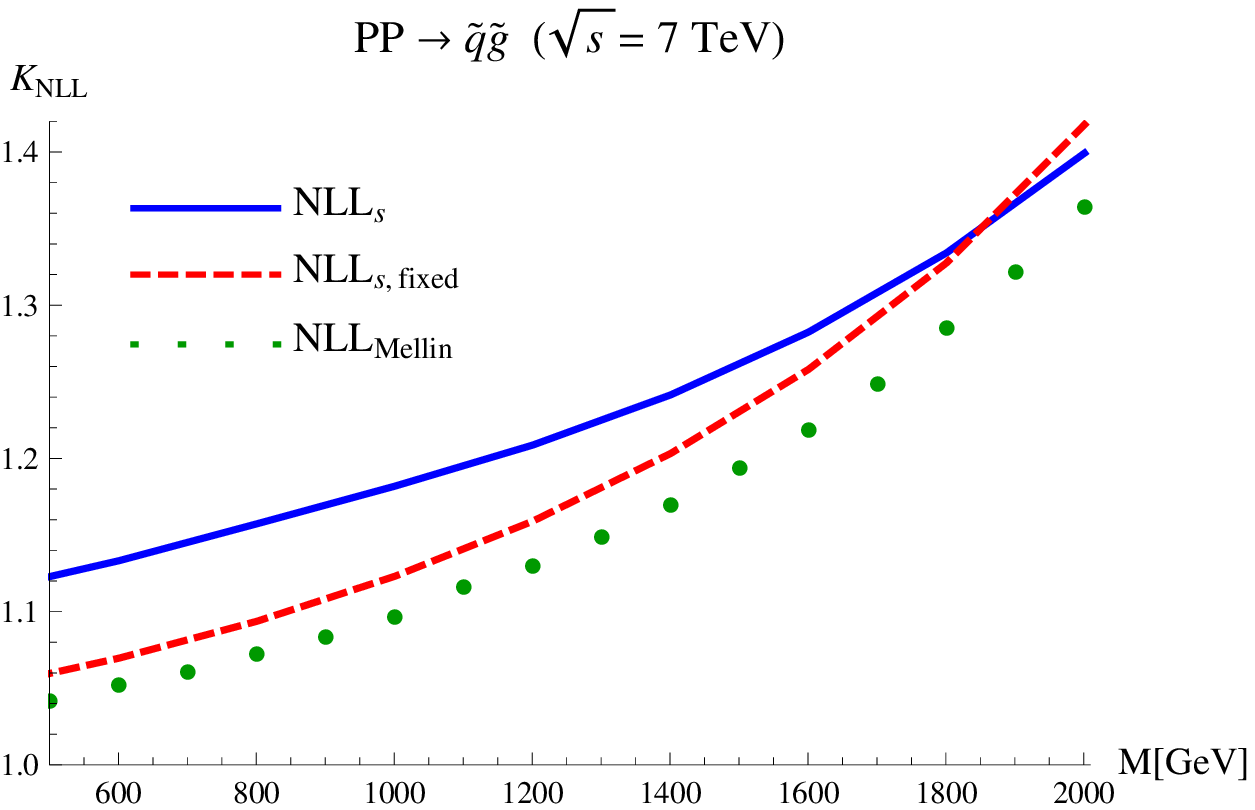}
\includegraphics[width=0.49 \linewidth]{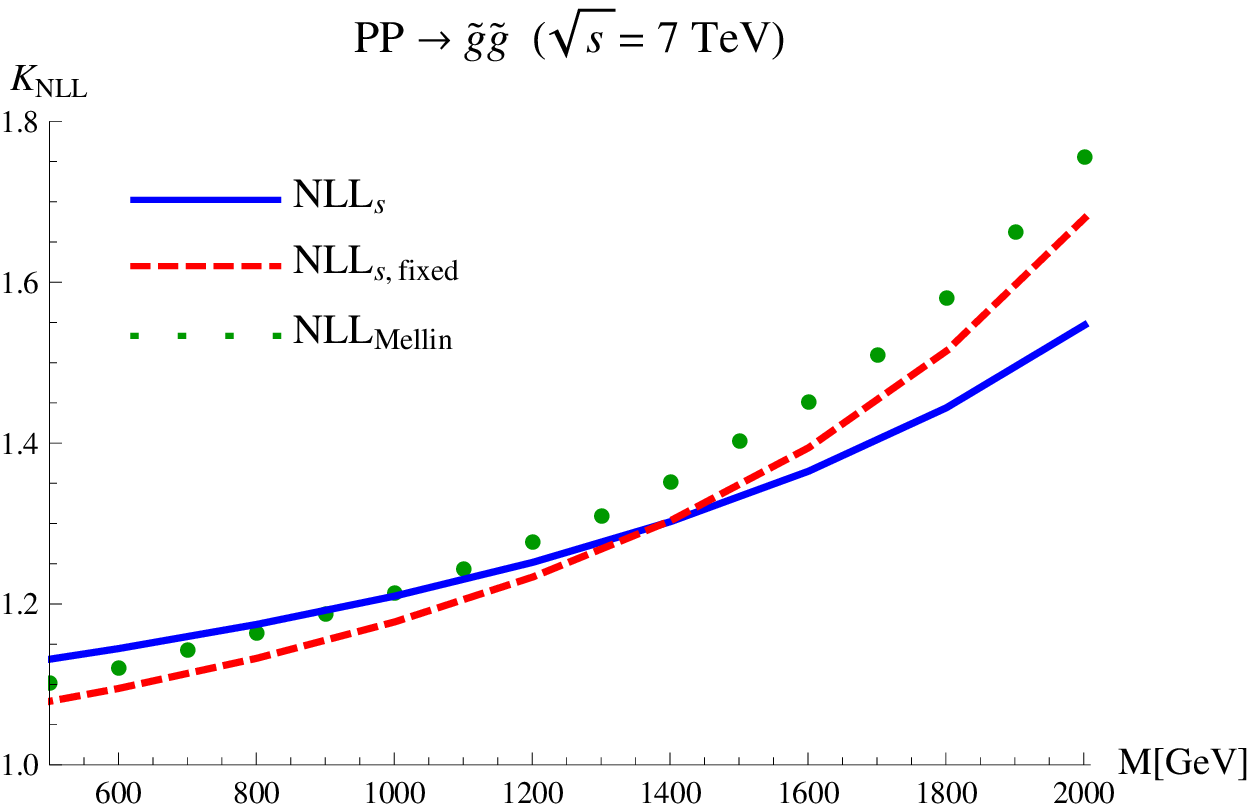}\\
\includegraphics[width=0.49 \linewidth]{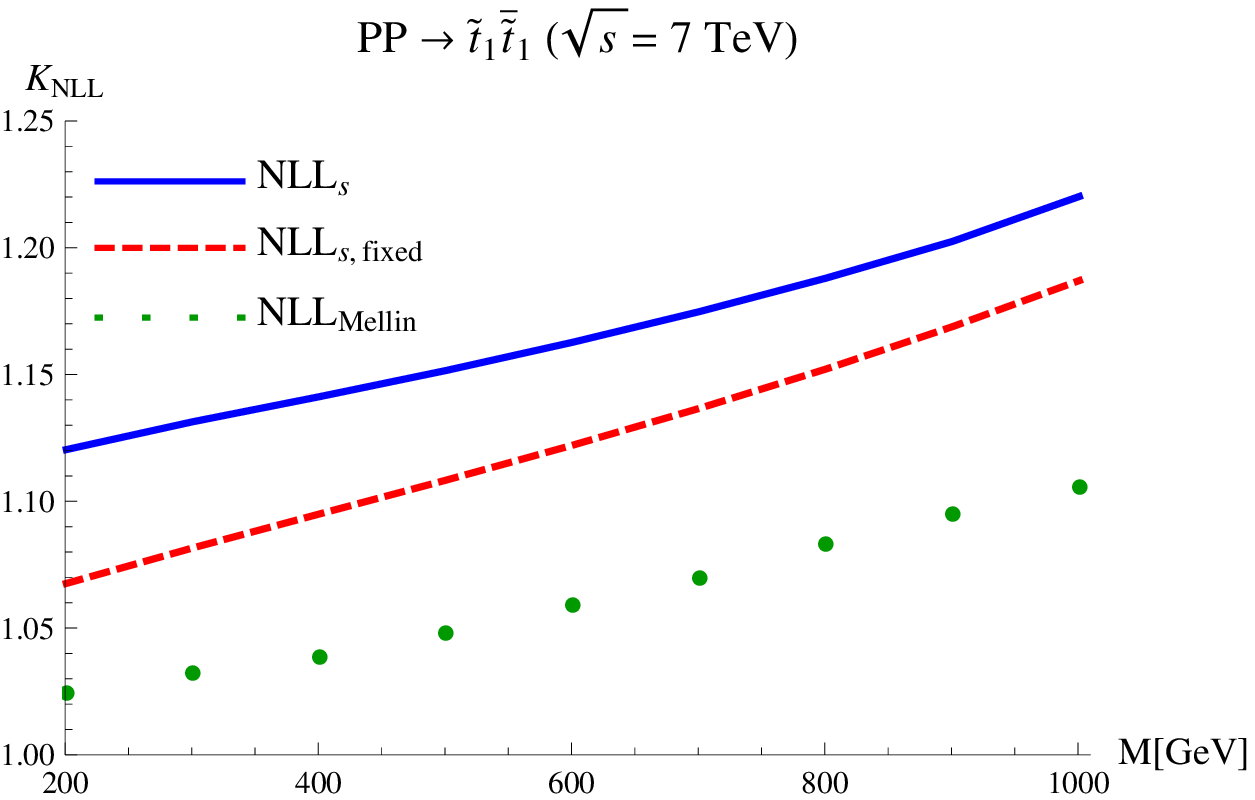}
\includegraphics[width=0.49 \linewidth]{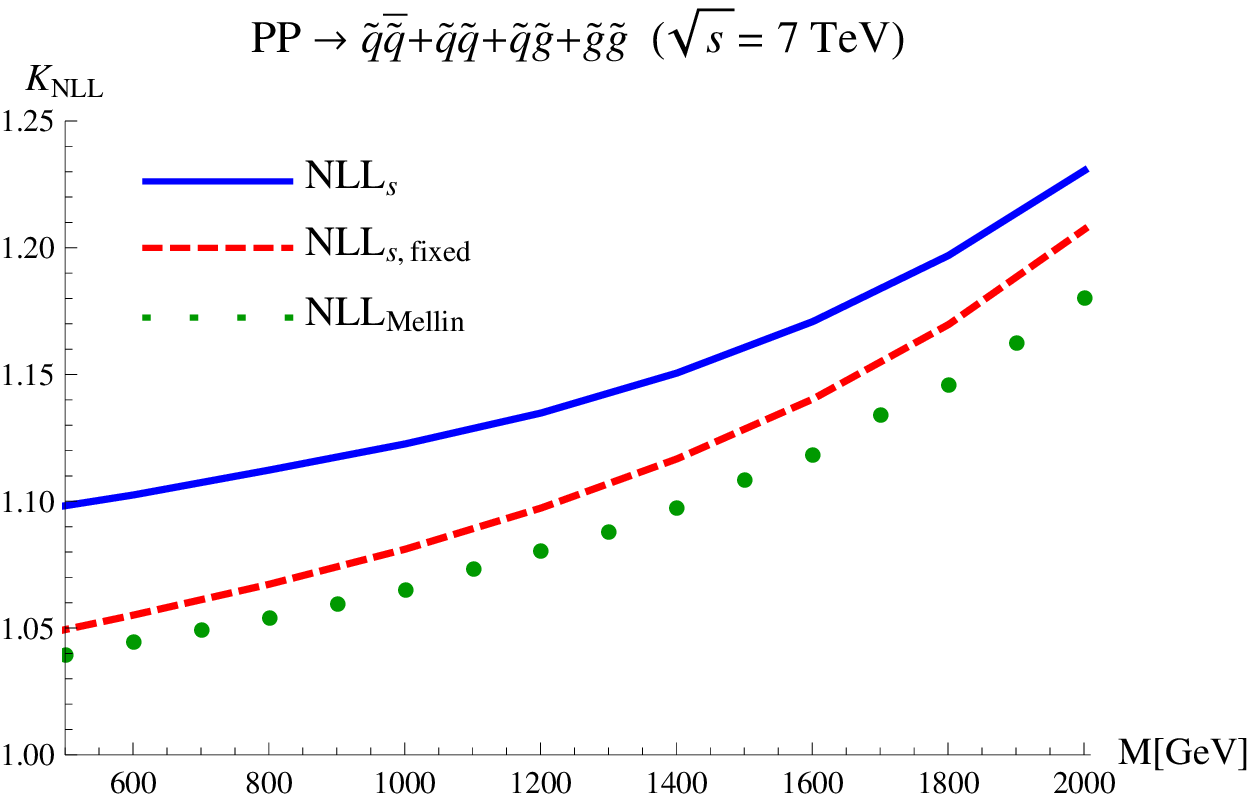}
\end{center}
\caption{Comparison of soft resummation in the momentum-space formalism
  adopted in this work and in the conventional Mellin-space approach
  for squark-antisquark (top-left), squark-squark (top-right),
  squark-gluino (centre-left), gluino-gluino (centre-right),
  stop-antistop (bottom-left) production and the inclusive gluino and light-flavour squark cross section (bottom-right) at LHC with $\sqrt{s} = 7\,$ TeV.
  The plots show the $K$-factor for our default running-scale
  implementation (NLL$_s$, solid blue), for the fixed-scale
  implementation (NLL$_{s, \text{fixed}}$, dashed red) and for
  the Mellin-space result (NLL$_{\text{Mellin}}$, green dots) obtained
  with \texttt{NLL-fast}. See the text for more details.}
\label{fig:comp_Mellin}
\end{figure}

The $K$-factor for NLL$_s$ (solid blue), NLL$_{s,\text{fixed}}$
(dashed red) and the Mellin-space result (green dots) from
\texttt{NLL-fast} are shown in Figure \ref{fig:comp_Mellin}.  For
clarity, we do not show the resummation uncertainties of the NLL$_s$
and NLL$_{s,\text{fixed}}$ results that are significant, especially
for smaller masses, as can be anticipated from the full NLL-results
including Coulomb resummation in Figure~\ref{fig:fix_VS_run}.  Even
though the difference between NLL$_s$ and NLL$_\text{Mellin}$ can be
sizable, the envelope of the three curves in the plots in Figure
\ref{fig:comp_Mellin} is well within the theoretical uncertainties of
the predictions, and is thus correctly accounted for by our estimate
of the intrinsic ambiguities of NLL resummation.

The comparison of our default running-scale implementation NLL$_s$
with NLL$_\text{Mellin}$ shows different features for the various
production channels.  For squark-antisquark and gluino-pair production
the agreement is overall reasonable, although the behaviour as a
function of the mass is different in both cases, leading to a better
agreement at large masses in the former case and at smaller masses in
the latter.  For squark-gluino production, the agreement also improves
at larger masses.  In contrast, for the squark-squark production
channel, there is a constant shift, and the differences are sizable
over the whole mass range. Since the NLL corrections are small for
this production channel, the discrepancies in the cross-section
predictions are however less important.  For all processes, the
fixed-scale momentum-space results NLL$_{s, \text{fixed}}$ are closer
to the Mellin-space results, with similar magnitude and
mass-dependence of the corrections. Since NLL$_s$ and
NLL$_\text{Mellin}$ both resum threshold logarithms ($\ln \beta$ and
$\ln N$ respectively) appearing in the partonic cross sections, while
NLL$_{s, \text{fixed}}$ resums logarithms in the hadronic cross
section, this behaviour is somewhat counter-intuitive and deserves
further studies in the future.\footnote{An analytic comparison of
  momentum-space resummation with fixed soft scale to $N$-space
  resummation has appeared recently for Drell-Yan
  production~\cite{Bonvini:2012yg}, but no such investigation for the
  running scale with a lower cutoff has been performed yet.}  The
spread of the three predictions for stop production is comparable to
the one observed for the other SUSY processes in the same mass
range. In this case, however, the two momentum-space predictions with
a fixed and running soft scale show a better agreement with each
other.

It is interesting to note that the NNLL results for squark-antisquark
production obtained in the $N$-space approach
in~\cite{Beenakker:2011sf} are very similar to our best prediction for
this channel in Figures~\ref{fig:K7} and~\ref{fig:K14}.  Part of this
agreement can be attributed to the fact that Ref.~\cite{Beenakker:2011sf}
includes the interference of the first Coulomb correction with
higher-order soft corrections, which give a dominant contribution to
our full NLL predictions (i.e. to the difference between the red and
blue curves in Figures~\ref{fig:K7} and~\ref{fig:K14}), and to the
difference between the NLL and NNLL results
in~\cite{Beenakker:2011sf}. However, our prediction includes
higher-order Coulomb corrections and bound-state effects not included
in the results of~\cite{Beenakker:2011sf}, while their results include
one-loop hard corrections and NNLL soft corrections not included in
ours. Therefore the good agreement is to some extent fortunate,
and cannot necessarily be expected for the other processes where the
Coulomb corrections are of less relative importance (in particular for
squark-squark production), and where an NNLL analysis remains to be performed. Furthermore, a combined NNLL soft/Coulomb resummation as performed for top-quark pair production in~\cite{Beneke:2011mq} would include higher-order Coulomb effects included neither in our present predictions nor in~\cite{Beenakker:2011sf}.

\section{Conclusions and outlook}
\label{sec:conclusion}

We have performed a combined resummation of soft-gluon and Coulomb
effects for all squark- and gluino-pair production channels at the LHC,
including bound-state effects, based on the formalism derived
in~\cite{Beneke:2009rj,Beneke:2010da}.  We have also extended the
derivation of the factorization and resummation formalism to the case
of stop-antistop production with a quark-antiquark initial state, that
proceeds through a $P$-wave.  The corrections from higher-order Coulomb
and bound-state effects and their interference with soft corrections
are not included in previous predictions and can be sizeable, in
particular for gluino-pair production, where they are as large as the
soft-gluon corrections alone, and for squark-antisquark production
where the effect is even bigger. For benchmark scenarios with moderate
squark-gluino mass splitting the effect on the total inclusive squark
and gluino production cross section is less pronounced but still
relevant, with the total NLL corrections of the order of $15-30\%$ of
the NLO cross section.  Therefore these predictions should be taken
into account in the analysis of the upcoming LHC results expected this
year. To facilitate the application of our results we provided
numerical predictions for some of the benchmark points defined
in~\cite{AbdusSalam:2011fc}, and include grid files with our results
for squark and gluino masses in the $200$-$2000$~GeV ($200$-$2500$~GeV) range at the $7$~TeV ($8$~TeV) LHC
with the arXiv submission of this paper.

Our results for soft-gluon resummation alone, obtained in the momentum-space resummation approach~\cite{Becher:2006nr,Becher:2006mr,Becher:2007ty} and with 
the scale-setting procedure introduced in~\cite{Beneke:2011mq}, agree within our estimate of resummation ambiguities with results obtained in the Mellin-space formalism~\cite{Beenakker:2009ha,Beenakker:2010nq}, although a more detailed study of the relation between the approaches and the different  scale choices within the momentum space framework would be desirable.  

In our analysis, the squarks and gluinos have been treated as stable, but contributions to the cross section from below the nominal production threshold have been included 
through bound-state corrections. This is expected to take the effect of small, but finite, widths to some extent into account. A more refined analysis is possible in our framework 
using a complex energy in the argument of the potential function in the factorization formula~\eqref{eq:fact}, as done in recent studies of the invariant mass spectrum of 
gluino-pair and squark-gluino production~\cite{Hagiwara:2009hq,Kauth:2011vg,Kauth:2011bz}. The investigation of the combined effect  with soft resummation, as well as the 
extension to NNLL accuracy following Ref.~\cite{Beneke:2011mq}, is left for future work.

\subsubsection*{Acknowledgements}
We would like to thank M.~Beneke, W.~Beenakker, A.~Kulesza and I.~Niessen for useful discussions
and G.~Watt for providing us with  MSTW PDFs  with improved accuracy in the evolution at large $x$.
The work of P.F. is supported by the ``Stichting voor Fundamenteel Onderzoek der Materie (FOM)", 
the work of C.W. by the research programme Mozaiek, which is partly financed by the Netherlands Organisation 
for Scientific Research (NWO).

\appendix
\section{NLL resummation for stop-antistop production}
\label{sec:Pwave}

In this appendix we derive the factorization
formula~\eqref{eq:fact} for stop-antistop production from
quark-antiquark annihilation that proceeds in a $P$-wave state. 
While the aim is to establish resummation at NLL accuracy, the arguments suggest that the combined soft-Coulomb resummation can  be performed at NNLL accuracy as well. Our reasoning applies to partonic subprocesses with a leading $P$-wave contribution, as relevant for stop-antistop production, but not to $P$-wave contributions to $S$-wave dominated subprocesses. 

Following~\cite{Beneke:2010da} the scattering process $q\bar q\to \tilde t\bar{\tilde t}+X$ is described in terms of an effective field theory using  soft-collinear effective theory
(SCET)~\cite{Bauer:2000yr,Bauer:2001yt,Beneke:2002ph,Beneke:2002ni} 
for the initial state quarks and  potential 
non-relativistic QCD (pNRQCD)~\cite{Pineda:1997bj,Beneke:1998jj,Brambilla:1999xf,Beneke:1999zr} for the stop and antistop. In this framework, the scattering amplitude in the full MSSM is expressed in terms of expectation values of EFT operators multiplied by short-distance coefficients $C$, that contain the dependence on hard modes not present in the EFT (e.g. gluinos, hard off-shell gluons): 
\begin{equation}
\label{eq:matrix}
\mathcal{A}(q\bar q\to \tilde t \bar{\tilde t}X) = \sum_{\ell,R_\alpha}
\;C^{(\ell),R_\alpha}(\mu)\;
\langle  \tilde t \bar{\tilde t}X_s |\mathcal{O}^{(\ell),R_\alpha}(\mu) |
q\bar q\rangle_{\rm EFT} \,.
\end{equation} 
The series 
in $\ell$  accounts for the threshold expansion 
in powers of $\beta$, while $R_\alpha=1,8$ are the irreducible colour representations of the stop-antistop pair.
 Due to the threshold kinematics, $(k_1+k_2)^2\sim 4M^2$, no
collinear modes can appear in the final state $X_s$, that only contains soft
modes. 
Both the short-distance coefficients $C^{(\ell),R_\alpha}$ and the operators $\mathcal{O}^{(\ell),R_\alpha}$ can  carry open Lorentz and spin indices, that have been suppressed.
For NLL resummation, only the colour-octet state is relevant.\footnote{
At NLO a finite, non-logarithmic
colour-singlet contribution is generated by real-gluon
emission~\cite{Younkin:2009zn}  that is suppressed by $\alpha_s\beta^2$ compared to the leading $P$-wave contribution, and therefore not even relevant at NNLL.
Since there is no leading  $S$-wave  contribution to $q\bar q\to \tilde t \bar{\tilde t}$, the situation is simpler than for quarkonium production, where singularities cancel  among 
 colour octet $S$-wave and singlet $P$-wave channels~\cite{Bodwin:1992ye}.
}
For the case of $P$-wave production, the leading $\ell=0$ term in the expansion of the scattering amplitude~\eqref{eq:matrix} vanishes and the first non-vanishing term occurs for $\ell=1$, i.e. it is suppressed by $\mathcal{O}(\beta)$ compared to the $S$-wave case.
 The $P$-wave production operator for a colour-octet state
reads
\begin{equation}
\label{eq:opdef}
\mathcal{O}^{(1),8}_\mu(\mu) = \frac{1}{\sqrt 2}
\left[(\bar{\xi}_{\bar{c}}W_{\bar{c}})T^a
  (W^\dagger_{c}  \xi_{c})\right]
\left[\psi^\dagger T^a\left(-\frac{i}{2} \overleftrightarrow D_{\mu,\top}\right)
\chi\right](\mu). 
\end{equation}
Here the field $\psi^\dagger$ ($\chi$) is a
non-relativistic  field that creates the
stop (antistop) while $\xi_c$ ($\bar\xi_{\bar
  c}$) is a collinear (anti-collinear) field that destroys the initial-state quark with momentum $\sim \sqrt{\hat s}/2\, n$ (antiquark with momentum $\sim \sqrt{\hat s}/2\, \bar n$), with light-cone vectors $n^2=\bar n^2=0$, $n\cdot\bar n=2$. The $W_c$ ($W_{\bar{c}}$) are collinear (anti-collinear) Wilson lines summing up collinear gluon emission to all orders~\cite{Bauer:2000yr,Bauer:2001yt}.
 $D_\top$ denotes the projection of the covariant derivative $D_\mu=(\partial_\mu -i g_s T^aA^a_{\mu}(x))$ in the direction orthogonal to the
timelike velocity $w^\mu$ of the stop-antistop pair.

 At this stage, the operator
~\eqref{eq:opdef} is written in NRQCD and contains the full gluon
field $A$. This  does not yet incorporate a systematic expansion in $\beta$. For this purpose we adopt potential 
non-relativistic QCD (pNRQCD)~\cite{Pineda:1997bj,Beneke:1998jj,Brambilla:1999xf,Beneke:1999zr}, where only soft gluons with momenta scaling as
\begin{equation}
  q\sim M \beta^2
\end{equation}
and the non-relativistic fields with so-called potential momenta scaling as
\begin{equation}
\label{eq:pot-def}
  k_0\sim M \beta^2\,,\quad \vec k \sim M\beta 
\end{equation}
 are retained in the EFT, while potential gluons with momenta scaling like~\eqref{eq:pot-def} and modes with momenta of order $M\beta$ are integrated out. 
After this step, the covariant derivative in~\eqref{eq:opdef} contains only the soft gluon field $A_s$ with the same scaling as the soft momentum, 
\begin{equation}
\label{eq:soft-scaling}
  A_s^\mu \sim M\beta^2.
\end{equation}
Therefore the term $g_s A_{s,\mu\top}$ in the covariant derivative is suppressed by a factor $g_s\beta$ compared to the derivative term $\partial_{\mu\top}\sim\beta$ and the production operator~\eqref{eq:opdef} at leading power in pNRQCD is simply given by
\begin{equation}
\label{eq:opdef-2}
\mathcal{O}^{(1),8}_\mu(\mu) = \frac{1}{\sqrt 2}
\left[(\bar{\xi}_{\bar{c}}W_{\bar{c}})T^a
  (W^\dagger_{c}  \xi_{c})\right]
\left[\psi^\dagger T^a\left(-\frac{i}{2} \overleftrightarrow \partial_{\mu,\top}\right) \chi\right](\mu) .
\end{equation}

The potential, collinear and anti-collinear fields interact only through the exchange of soft gluons. These interactions can be removed from the leading effective-theory Lagrangians by a decoupling transformation of the collinear~\cite{Bauer:2001yt,Bauer:2002nz} and potential~\cite{Beneke:2010da} fields involving soft Wilson lines of the schematic form 
$ W_c^\dagger\xi_{c} = S_{n} W_c^{(0)\dagger}\xi_{c}^{(0)}$ and $\chi=S_{w}\chi^{(0)}$, and similar transformations for the conjugate fields.
For  definitions of the Wilson lines $S_n$ and $S_w$ see~\cite{Beneke:2010da}.
Since the fields with superscript $(0)$ do not interact with each other in the leading effective Lagrangians, the
scattering amplitude assumes a factorized form:
\begin{equation}
\label{eq:matrix-P}
\begin{aligned}
\mathcal{A}(q\bar q\to \tilde t_i\bar{\tilde t}_iX)=&
  \left(-\frac{i}{2 \sqrt{2}}\right)
\; C^{(1),8}_{\mu}(\mu)\;\braket{0|(W_c^{(0)\dagger }\xi_{c}^{(0)})_{j_1}|q}
\braket{0|(\bar{\xi}_{\bar{c}}W_{\bar{c}})_{j_2} |\bar q}\\
&\times\braket{ X_s|(S^\dagger_{\bar n} T^aS_{n})_{j_2j_1} 
(S_{w}^\dagger T^a S_{w})_{j_3j_4} | 0}\,
 \langle \tilde t_i\bar{\tilde t}_i |\psi^{(0)\dagger}_{j_3}
\overleftrightarrow {\partial^\mu_\top} \chi^{(0)}_{j_4}|0\rangle \, .
\end{aligned}
\end{equation}
Note that the factorization~\eqref{eq:matrix-P} holds only at leading
power in the $\beta$-expansion, since the field
redefinition does not remove the interaction
with the spatial components of the soft gluons in the $P$-wave
production operators, or in the subleading effective Lagrangians.
We will argue below that these corrections are not relevant 
at NLL accuracy for
stop-antistop production from quark-antiquark annihilation.

Using the representation of the scattering
amplitude~\eqref{eq:matrix-P} and following the steps discussed in detail
in~\cite{Beneke:2010da} one derives the factorization formula
\begin{equation}
\label{eq:factor-P}
\hat \sigma_{q\bar q}(\hat s,\mu) = 
H_{q\bar q}^8(\mu)
\int d \omega\;
J^P_{8}(E-\frac{\omega}{2})\,
W^{8}(\omega,\mu) \, .
\end{equation}
Here the potential function for $P$-wave production in a colour octet state is given by
\begin{equation}
J_{8}^P(q)=
\frac{1}{2}
\int d^4z \,e^{iq\cdot z} 
\braket{0|[\chi^{(0)\dagger}\overleftrightarrow{\partial_\top^\mu}T^a \psi^{(0)}](z) 
[\psi^{(0)\dagger}T^a \overleftrightarrow{{\partial}_{\top}}_{\!\!,\mu} 
\chi^{(0)}](0)|0}\,,
\end{equation}
while the colour-octet soft function (defined in terms of  the squared matrix element of the soft Wilson lines in~\eqref{eq:matrix-P}) is identical to that appearing for $S$-wave production~\cite{Beneke:2009rj}.
The factorization formula resembles
that for $S$-wave production, up to the replacement of the potential
function by the appropriate $P$-wave expression. 

We now show that no corrections to~\eqref{eq:factor-P} appear at NNLL.
As far as {\it soft-gluon effects} are concerned, we have argued that
the corrections to the production operator~\eqref{eq:opdef-2} from the
subleading $A_{s,\top}$-terms in~\eqref{eq:opdef} are of order
$g_s\beta$. Since the potential function for $S$-$P$-wave interference
vanishes, there can be no contributions to the total cross section
where one $A_\top$-operator interferes with a leading
operator~\eqref{eq:opdef-2}.  The squared $A_\top$-term gives a
correction to the cross section of the order\footnote{We have allowed
  for logarithmic corrections in the generalized potential function
  from contractions of the two gluon insertions. A simple estimate
  indicates, however, that these are in fact absent.}  $\Delta\sigma
\sim \sigma^0\alpha_s\beta^2\{1,\ln\beta\} $.  These corrections are
at most of order N$^3$LL in the combined counting $\alpha_s/\beta\sim
1$ and $\alpha_s\ln\beta\sim 1$.

Following~\cite{Beneke:2010da} one sees that insertions of
higher-order {\it soft-collinear} interactions into the matrix
element~\eqref{eq:matrix} are at least suppressed by
$\mathcal{O}(\beta^2)$ and therefore are beyond NNLL.  For $P$-wave
production it has to be shown in addition that no contributions to the
$\ell=1$ operators in~\eqref{eq:matrix} are generated from partonic
channels with leading $S$-wave ($\ell=0$) contributions.  In the EFT
language they would correspond to operators where one or both of the
quark/antiquark fields are replaced by (anti)collinear gluon fields.
A non-vanishing matrix element of such an operator in the $|\bar
qq\rangle$ initial state requires splittings of a collinear quark into
a collinear gluon and a soft quark that are mediated by
$\mathcal{O}(\beta)$-suppressed interactions in the SCET
Lagrangian~\cite{Beneke:2002ni} (recall that no collinear final state
particles appear at partonic threshold).  Since there is no
$qg$-initiated production of stop-antistop pairs at leading order, two
subleading splittings are required so these contributions to the
amplitude are suppressed by $g_s^2\beta$ compared to the $\ell=1$
term.  The resulting real corrections to the total cross section are
therefore even further suppressed compared to the subleading soft
effects.

Finally, subleading {\it soft-potential effects} due to chromoelectric
$\vec x\cdot \vec E_s$ interactions~\cite{Beneke:1999zr} have been
shown to be beyond NNLL in the $S$-wave case~\cite{Beneke:2010da}
since the potential function with a single insertion of an operator
$\sim \vec x$ vanishes by rotational invariance.  The same argument
holds for the $P$-wave potential function.\footnote{A possible
  interference of a single $A_\top$-term in the production operator
  with a chromoelectric vertex might lead to a non-vanishing potential
  function but would be of the same order $\alpha_s\beta^2
  \{1,\ln\beta\}$ as the soft contributions considered above.}
Furthermore, as mentioned before, due to the absence of a leading
$S$-wave contribution to the quark-antiquark channel, no mixing of
$S$-wave and $P$-wave states by subleading interactions appears.
Therefore, while we didn't perform an exhaustive study of effects
beyond NLL, we do not encounter an obstruction for NNLL resummation
for the quark-antiquark initial-state contribution to stop-antistop
production. Note, however, that Coulomb resummation at this accuracy
requires the computation of the NLO $P$-wave Coulomb Green's function
for scalar particles.


\section{Resummation functions}
\label{sec:resfunc}

The resummation functions appearing in~\eqref{eq:resum-NLL} are given by
 \begin{eqnarray}
U_i(M,\mu_h,\mu_f,\mu_s )&=&
\left(\frac{4M^2}{\mu_h^2}\right)^{-2a_\Gamma(\mu_h,\mu_s)}\,
\left(\frac{\mu_h^2}{\mu_s^2}\right)^{\eta}
\times \,\exp\Big[4  (S(\mu_h,\mu_f)-S(\mu_s,\mu_f))
\nonumber\\[0.1cm]
&&
 -\,2a_i^{V}(\mu_h,\mu_s) +2 a^{\phi,r}(\mu_s,\mu_f)+
2 a^{\phi,r'}(\mu_s,\mu_f)\Big]
\label{eq:def-u}
\end{eqnarray}
and $\eta = 2 a_{\Gamma}(\mu_s,\mu_f)$.  The labels $r, r'$ denote the
colour representation of the initial-state partons $p,p'$. 
At NLL the functions $S$, $a_i^{V}$ and $a_\Gamma$ are given
by~\cite{Becher:2007ty,Beneke:2010da}
\begin{eqnarray} 
\label{eq:res_funct_def}
S(\mu_i,\mu_j) &=&
\frac{C_r+C_{r'}}{2\beta_0^2}
\left[\frac{4\pi}{\alpha_s(\mu_i)}
\left(1-\frac{1}{r}-\ln r\right)
+\left( 2K-\frac{\beta_1}{\beta_0}\right)
\left(1-r+\ln r\right)
+\frac{\beta_1}{2\beta_0}\ln^2r \right],\quad
\nonumber\\
a_{\Gamma}(\mu_i,\mu_j) &=&
\frac{C_r+C_r}{\beta_0} \ln r
\, ,
\quad a^{V}_i (\mu_i,\mu_j) = 
\frac{\gamma_i^{(0),V} }{2\beta_0}
\ln r ,\;
\qquad a^{\phi,r} (\mu_i,\mu_j) = 
\frac{ \gamma^{(0)\phi,r} }{2\beta_0}
\ln r ,
\end{eqnarray}
where $r=\alpha_s(\mu_j)/\alpha_s(\mu_i)$ and  $\gamma_i^{V}=\gamma^{r}+\gamma^{r' }+\gamma_{H,s}^{R_\alpha}$.
The  one-loop anomalous-dimension coefficients appearing here are given by
\begin{align}
\gamma_{H,s}^{(0),R_\alpha}&=-2C_{R_\alpha},& 
  \gamma^{(0)3}&=-3C_F=-\gamma^{(0)\phi,3} ,&
  \gamma^{(0)8}&=-\beta_0=-\gamma^{(0)\phi,8}
\end{align}
and the coefficients of the beta function are
\begin{align}
\label{eq:beta}
\beta_0&=\frac{11}{3}C_A-\frac{2}{3}n_f ,&
 \beta_1&=\frac{34}{3}C_A^2-\frac{10}{3}C_An_f-2C_F n_f .
\end{align}
 We also used the factor $K=\left(\frac{67}{18}-\frac{\pi^2}{6}\right)C_A-\frac{10}{9}T_Fn_f$ appearing in the ratio of one- and two-loop cusp anomalous dimensions.
 The explicit values of the Casimir invariants for the $SU(3)$ representations relevant for squark and gluino production are:
 \begin{equation}
 C_1=0\,, \hspace{0.5 cm} C_3=\frac{4}{3}\,, \hspace{0.5 cm} C_6 =\frac{10}{3}\,, \hspace{0.5 cm} C_8 =3\,,\hspace{0.5 cm} C_{10}=6\,, \hspace{0.5 cm} C_{15}=\frac{16}{3}\,, \hspace{0.5 cm} C_{27}=8 \,. 
 \end{equation} 

\section{Determination of $\mu_s$ and $\beta_\text{cut}$}
\label{sec:bcutdeterm}

The soft scale $\mu_s$ used in the fixed-scale implementation, 
NLL$_\text{fixed}$, is defined by Eq. (\ref{eq:def-mus}).  For equal squark and gluino masses, $m_{\tilde{q}}=m_{\tilde{g}}$, the minimization procedure yields the following results: 
\begin{equation}
  \label{eq:muS}
  \begin{aligned}
    \tilde{q} \bar{\tilde{q}}&:& \mu_s&=127-170\,\text{GeV ($\sqrt{s}=7$ TeV)},&   \mu_s&=146-376\,\text{GeV ($\sqrt{s}=14$ TeV)},\\
\tilde{q} \tilde{q}&:& \mu_s&=122-135\,\text{GeV ($\sqrt{s}=7$ TeV)} ,&
\mu_s&=146-312\,\text{GeV ($\sqrt{s}=14$ TeV)},\\
\tilde{g} \tilde{q}&:&   \mu_s&=109-141\,\text{GeV ($\sqrt{s}=7$ TeV)},&
\mu_s&=131-310\,\text{GeV ($\sqrt{s}=14$ TeV)},\\
\tilde{g} \tilde{g}&:&  \mu_s&=106-141\,\text{GeV ($\sqrt{s}=7$ TeV)},&
 \mu_s&=127-308\,\text{GeV ($\sqrt{s}=14$ TeV)} \,.
  \end{aligned}
\end{equation}
The scales for a centre-of-mass energy of 7 TeV refers to a mass range $M=500-2000\,$GeV, while
for a 14 TeV LHC the mass interval $M=500-3000\,$GeV was considered. 
For stop-antistop production we obtain the scales
\begin{align} \label{eq:muSt}
  \tilde{t} \bar{\tilde{t}}&:& \mu_s&=29-138\,\text{GeV ($\sqrt{s}=7$ TeV)},& 
  \mu_s&=31-230\,\text{GeV ($\sqrt{s}=14$ TeV)}
\end{align}
 for the mass range $m_{\tilde t}=100-1000\,$GeV for a centre-of-mass energy of 7 TeV and  $m_{\tilde t}=100-1400\,$GeV at 14 TeV .
In \cite{Becher:2007ty} it was suggested to fit the mass and energy dependence of the soft scales (\ref{eq:muS}) 
 by a function of the form
\begin{equation} \label{eq:muSfit}
\mu_s = \frac{M (1-\rho)}{\sqrt{a+b \rho}} \, ,
\end{equation}  
with $\rho=4 M^2/s$. The coefficients $a$ and $b$ for the different processes are given in Table \ref{tab:coeffmuS},
and provide a fit to (\ref{eq:muS}) with accuracy better than $5\%$ for the mass range considered.
\begin{table}[t!]
\begin{center}
\renewcommand{\tabcolsep}{0.5pc} 
\renewcommand{\arraystretch}{1.3} 
\begin{tabular}{|l|c|c|}
\hline
& LHC(7 TeV) & LHC(14 TeV) \\
\hline
\hline
$\tilde{q} \bar{\tilde{q}}$ & $a=12.1328$ & $a=11.5289$ \\[-.1cm]
 & $b=159.828$ & $b=171.856$ \\
 \hline 
 $\tilde{q} \tilde{q}$ & $a=10.9627$ & $a=10.9981$ \\[-.1cm]
 & $b=265.414$ & $b=278.062$ \\
 \hline 
 $\tilde{g} \tilde{q}$ & $a=16.5459$ & $a=15.3779$ \\[-.1cm]
 & $b=235.227$ & $b=267.763$ \\
 \hline 
$\tilde{g} \tilde{g}$ & $a=17.95$ & $a=16.7483$ \\[-.1cm]
 & $b=231.545$ & $b=264.702$ \\
 \hline 
$\tilde{t} \bar{\tilde{t}}$ & $a=16.8468 $ & $a=15.7282$ \\[-.1cm]
 & $b=365.735$ & $b=481.045
$ \\
 \hline 
\end{tabular} 
\end{center}
\caption{Coefficients of the fit (\ref{eq:muSfit}) for the squark-gluino production processes at centre-of-mass energies of 7 and 14 TeV.}
\label{tab:coeffmuS}
\end{table}

The procedure used to determine $\beta_{\text{cut}}$, which enters the definition
of the soft scale used in our default NLL implementation, Eqs. (\ref{eq:runmus_a}) and
(\ref{eq:runmus_b}), was explained in detail in~\cite{Beneke:2011mq}. Following~\cite{Beneke:2011mq} 
we introduce eight different cross sections 
\begin{equation}
\label{eq:sigma-up-down}
  \hat \sigma_{pp'}(A_<,B_>,\beta_\text{cut})= 
\hat \sigma_{pp'}^{A_<}\;\theta(\beta_\text{cut}-\beta)
+\hat \sigma_{pp'}^{B_>}\;\theta(\beta-\beta_\text{cut}) \, ,
\end{equation}
defined using one of two possible matching prescriptions for the lower interval 
($A_<\in\{\text{NLL}_1,\text{NLL}_2\}$) and one of four possible 
approximations for the upper interval ($B_>\in\{\text{NLL}_2,\text{NLO}_{\text{app}},
\text{NNLO}_{\text{A}},\text{NNLO}_{\text{B}}\}$).
Here NLL$_2$ denotes our default approximation, Eq. (\ref{eq:cross_matched}), while
NLL$_1$ is the resummed result matched to the Born instead of the NLO cross section, 
\begin{equation}
  \hat\sigma^{\text{NLL}_1}_{pp'}(\hat s)
  =\left[\hat\sigma^{\text{NLL}}_{pp'}(\hat s)-
    \hat\sigma^{\text{NLL}(0)}_{pp'}(\hat s)\right]
 + \hat\sigma^{\text{LO}}_{pp'}(\hat s)
\label{eq:cross_matched_NLL1} \, .
\end{equation}
NLO$_\text{app}$ represents the sum of the full Born cross section and the 
approximated NLO corrections given in (\ref{eq:NLOapprox}), while the two
NNLO approximations contain in addition the $O(\alpha_s^4)$ terms arising
from the expansion of the NLL resummed result, including all of them 
(NNLO$_\text{A}$) or only the subset which is completely determined at
NLL (NNLO$_\text{B}$). $\beta_\text{cut}$ is then determined such that the width of the envelope of the
eight different implementations $\hat \sigma_{pp'}(A_<,B_>,\beta_\text{cut})$ is minimal.\footnote{The values (\ref{eq:bcut}) are also used for the NLL$_{s+h}$
approximation~\eqref{eq:NLLs}, 
instead of recalculating $\beta_{\text{cut}}$ using that approximation
in~\eqref{eq:sigma-up-down}. Since the difference of NLL and
NLL$_{s+h}$ is used in order to assess the effect of Coulomb
resummation, we consider it more meaningful to use the same
$\beta_{\text{cut}}$ for both in order not to
obscure the genuine Coulomb effects by different scale choices.}
 
\begin{figure}[t!]
\begin{center}
\includegraphics[width=0.48 \linewidth]{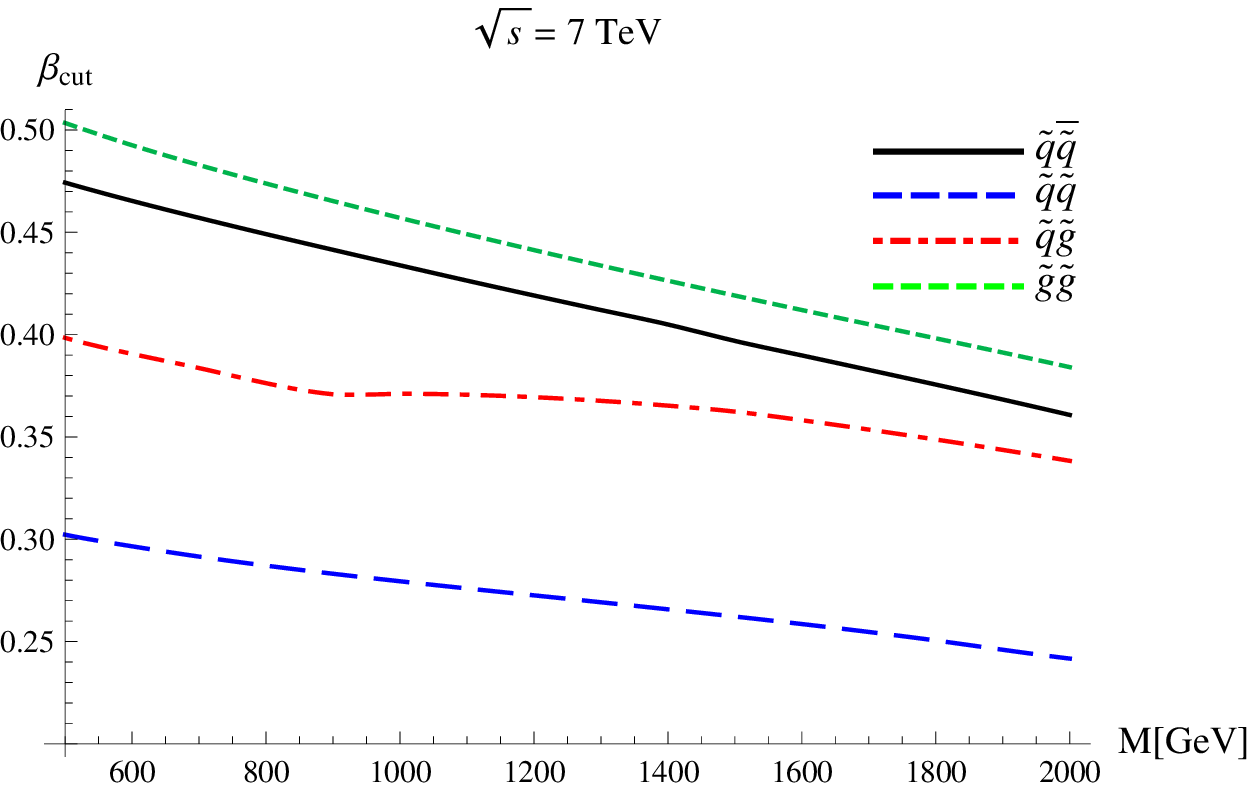}
\includegraphics[width=0.48 \linewidth]{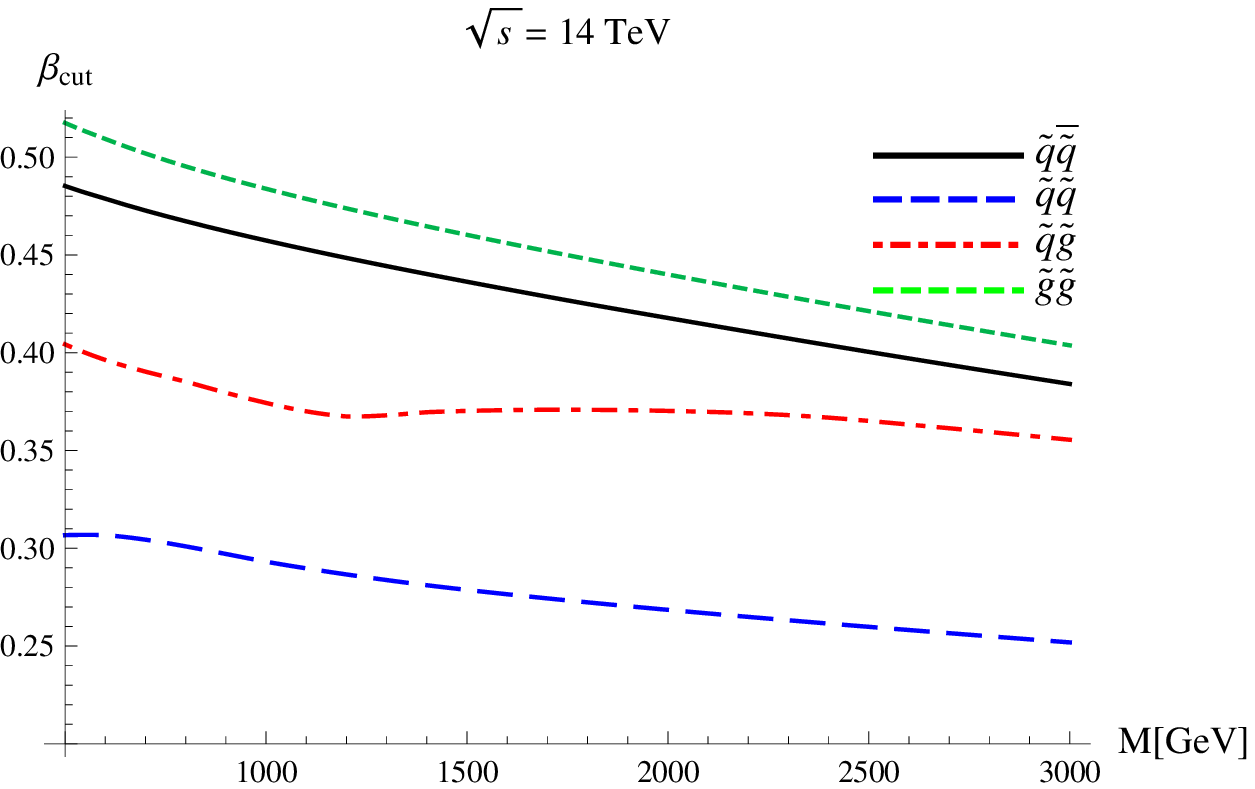}
\end{center}
\caption{Mass dependence of the parameter $\beta_\text{cut}$ for the four SUSY production processes.}
\label{fig:bcut} 
\end{figure} 
The procedure described provides, for the default choice $k_s=1$ and equal
squark and gluino masses, the following values for $\beta_\text{cut}$: 

\begin{equation}\label{eq:bcut}
  \begin{aligned}
  \tilde{q} \bar{\tilde{q}}&:& 
  \beta_{\text{cut}}^{(7\,\text{TeV})}&=0.47-0.36\, , &
\beta_{\text{cut}}^{(8\,  \text{TeV})}&=0.48-0.34\, ,&
  \beta_{\text{cut}}^{ (14\,\text{TeV})}&=0.48-0.38\,,\\
  \tilde{q} \tilde{q}&:& \beta_{\text{cut}}^{(7\,\text{TeV})}&=0.30-0.24\, , &
\beta_{\text{cut}}^{(8\,  \text{TeV})}&=0.31-0.24\, ,&
  \beta_{\text{cut}}^{ (14\,\text{TeV})}&=0.31-0.25\,,\\
\tilde{g} \tilde{q}&:&   \beta_{\text{cut}}^{(7\,\text{TeV})}&=0.40-0.34\,,&
 \beta_{\text{cut}}^{(8\,\text{TeV})}&=0.40-0.33\,,&
\beta_{\text{cut}}^{ (14\,\text{TeV})}&=0.40-0.36\,,  \\
\tilde{g} \tilde{g}&:&  \beta_{\text{cut}}^{(7\,\text{TeV})}&=0.50-0.39\,, &
\beta_{\text{cut}}^{(8\,  \text{TeV})}&=0.51-0.37\,,&
\beta_{\text{cut}}^{ (14\,\text{TeV})}&=0.52-0.40\,  \, .
  \end{aligned}
\end{equation}  
The numbers refer to the usual mass range of $500-2000\,$GeV at $\sqrt{s}=7\,$TeV and $500-3000\,$GeV at $\sqrt{s}=14\,$TeV, and the exact mass dependence is 
plotted in Figure \ref{fig:bcut}. We have also included the results for the
mass range  of $500-2500\,$GeV at $\sqrt{s}=8\,$TeV. 
For stop-antistop production the result is
\begin{align} \label{eq:bcutt}
  \tilde{t} \bar{\tilde{t}}&:& 
  \beta_{\text{cut}}^{(7\,\text{TeV})}&=0.53-0.40\, , &
\beta_{\text{cut}}^{(8\,  \text{TeV})}&=0.53-0.39\, ,&
  \beta_{\text{cut}}^{ (14\,\text{TeV})}&=0.54-0.41\,,
\end{align}
for the mass range $m_{\tilde t}=100-1000\,$GeV at 7 TeV, $m_{\tilde
  t}=100-1200\,$GeV at 8 TeV and $m_{\tilde t}=100-1400\,$GeV at 14 TeV.  The
theoretical ambiguities of the resummed result with the running scale are
estimated as follows: i) the default implementation NLL$_2$ is parametrized in
terms of $\hat E$ instead of $\beta$, ii) the NLL$_2$ result is computed for
the values $k_s=1/2,2 $, recomputing $\beta_{\text{cut}}$ anew for each choice
and iii), the envelope of the eight cross sections~\eqref{eq:sigma-up-down} is
taken while $\beta_{\text{cut}}$ is varied by $\pm 20\%$.  The uncertainties
from the three sources are added in quadrature, and iii) gives the dominant
contribution, in the $3-13\%$ range depending on the process, while i) and ii)
are usually below $2\%$.


\begin{thebibliography}{10}

\bibitem{Bueso:2011aa}
X.~P. Bueso,
\href{http://arxiv.org/abs/1112.1723}{{\tt arXiv:1112.1723 [hep-ex]}}.

\bibitem{Aad:2011ib}
{\bf ATLAS} Collaboration, G.~Aad {\em et al.}, {\em Phys.Lett.}
  {\bf B710} (2012)  67--85,
\href{http://arxiv.org/abs/1109.6572}{{\tt arXiv:1109.6572 [hep-ex]}}.

\bibitem{Chatrchyan:2011zy}
{\bf CMS} Collaboration, S.~Chatrchyan {\em et al.}, {\em
  Phys.Rev.Lett.} {\bf 107} (2011)  221804,
\href{http://arxiv.org/abs/1109.2352}{{\tt arXiv:1109.2352 [hep-ex]}}.

\bibitem{Baer:2009dn}
H.~Baer, V.~Barger, A.~Lessa, and X.~Tata,
  \href{http://dx.doi.org/10.1088/1126-6708/2009/09/063}{{\em JHEP} {\bf 09}
  (2009)  063},
\href{http://arxiv.org/abs/0907.1922}{{\tt arXiv:0907.1922 [hep-ph]}}.

\bibitem{Beenakker:1996ch}
W.~Beenakker, R.~H{\"o}pker, M.~Spira, and P.~M. Zerwas,
  \href{http://dx.doi.org/10.1016/S0550-3213(97)00084-9}{{\em Nucl. Phys.} {\bf
  B492} (1997)  51--103},
\href{http://arxiv.org/abs/hep-ph/9610490}{{\tt arXiv:hep-ph/9610490}}.

\bibitem{Beenakker:1996ed}
W.~Beenakker, R.~Hopker, and M.~Spira,
\href{http://arxiv.org/abs/hep-ph/9611232}{{\tt arXiv:hep-ph/9611232}}.

\bibitem{Beenakker:1997ut}
W.~Beenakker, M.~Kr{\"a}mer, T.~Plehn, M.~Spira, and P.~M. Zerwas,
  \href{http://dx.doi.org/10.1016/S0550-3213(98)00014-5}{{\em Nucl. Phys.} {\bf
  B515} (1998)  3--14},
\href{http://arxiv.org/abs/hep-ph/9710451}{{\tt arXiv:hep-ph/9710451}}.

\bibitem{Bornhauser:2007bf}
S.~Bornhauser, M.~Drees, H.~K. Dreiner, and J.~S. Kim,
  \href{http://dx.doi.org/10.1103/PhysRevD.76.095020}{{\em Phys.Rev.} {\bf D76}
  (2007)  095020},
\href{http://arxiv.org/abs/0709.2544}{{\tt arXiv:0709.2544 [hep-ph]}}.

\bibitem{Hollik:2007wf}
W.~Hollik, M.~Kollar, and M.~K. Trenkel,
  \href{http://dx.doi.org/10.1088/1126-6708/2008/02/018}{{\em JHEP} {\bf 0802}
  (2008)  018},
\href{http://arxiv.org/abs/0712.0287}{{\tt arXiv:0712.0287 [hep-ph]}}.

\bibitem{Hollik:2008yi}
W.~Hollik and E.~Mirabella,
  \href{http://dx.doi.org/10.1088/1126-6708/2008/12/087}{{\em JHEP} {\bf 0812}
  (2008)  087}, \href{http://arxiv.org/abs/0806.1433}{{\tt arXiv:0806.1433
  [hep-ph]}}.

\bibitem{Hollik:2008vm}
W.~Hollik, E.~Mirabella, and M.~K. Trenkel,
  \href{http://dx.doi.org/10.1088/1126-6708/2009/02/002}{{\em JHEP} {\bf 0902}
  (2009)  002}, \href{http://arxiv.org/abs/0810.1044}{{\tt arXiv:0810.1044
  [hep-ph]}}.

\bibitem{Mirabella:2009ap}
E.~Mirabella, \href{http://dx.doi.org/10.1088/1126-6708/2009/12/012}{{\em JHEP}
  {\bf 0912} (2009)  012}, \href{http://arxiv.org/abs/0908.3318}{{\tt
  arXiv:0908.3318 [hep-ph]}}.

\bibitem{Germer:2010vn}
J.~Germer, W.~Hollik, E.~Mirabella, and M.~K. Trenkel,
  \href{http://dx.doi.org/10.1007/JHEP08(2010)023}{{\em JHEP} {\bf 1008} (2010)
   023}, \href{http://arxiv.org/abs/1004.2621}{{\tt arXiv:1004.2621 [hep-ph]}}.

\bibitem{Kulesza:2008jb}
A.~Kulesza and L.~Motyka,
  \href{http://dx.doi.org/10.1103/PhysRevLett.102.111802}{{\em Phys. Rev.
  Lett.} {\bf 102} (2009)  111802},
\href{http://arxiv.org/abs/0807.2405}{{\tt arXiv:0807.2405 [hep-ph]}}.

\bibitem{Kulesza:2009kq}
A.~Kulesza and L.~Motyka,
  \href{http://dx.doi.org/10.1103/PhysRevD.80.095004}{{\em Phys. Rev.} {\bf
  D80} (2009)  095004},
\href{http://arxiv.org/abs/0905.4749}{{\tt arXiv:0905.4749 [hep-ph]}}.

\bibitem{Beenakker:2009ha}
W.~Beenakker {\em et al.},
  \href{http://dx.doi.org/10.1088/1126-6708/2009/12/041}{{\em JHEP} {\bf 12}
  (2009)  041},
\href{http://arxiv.org/abs/0909.4418}{{\tt arXiv:0909.4418 [hep-ph]}}.

\bibitem{Beenakker:2010nq}
W.~Beenakker, S.~Brensing, M.~Kr{\"a}mer, A.~Kulesza, E.~Laenen, {\em et al.},
  \href{http://dx.doi.org/10.1007/JHEP08(2010)098}{{\em JHEP} {\bf 1008} (2010)
   098}, \href{http://arxiv.org/abs/1006.4771}{{\tt arXiv:1006.4771 [hep-ph]}}.

\bibitem{Beenakker:2011fu}
W.~Beenakker, S.~Brensing, M.~Kr{\"a}mer, A.~Kulesza, E.~Laenen, {\em et al.},
  \href{http://dx.doi.org/10.1142/S0217751X11053560}{{\em Int.J.Mod.Phys.} {\bf
  A26} (2011)  2637--2664},
\href{http://arxiv.org/abs/1105.1110}{{\tt arXiv:1105.1110 [hep-ph]}}.

\bibitem{Sterman:1986aj}
G.~Sterman,
{\em Nucl. Phys.} {\bf B281} (1987)  310.

\bibitem{Catani:1989ne}
S.~Catani and L.~Trentadue,
{\em Nucl. Phys.} {\bf B327} (1989)  323.

\bibitem{Kidonakis:1997gm}
N.~Kidonakis and G.~Sterman,
  \href{http://dx.doi.org/10.1016/S0550-3213(97)00506-3}{{\em Nucl. Phys.} {\bf
  B505} (1997)  321--348},
\href{http://arxiv.org/abs/hep-ph/9705234}{{\tt arXiv:hep-ph/9705234}}.

\bibitem{Bonciani:1998vc}
R.~Bonciani, S.~Catani, M.~L. Mangano, and P.~Nason,
  \href{http://dx.doi.org/10.1016/S0550-3213(98)00335-6}{{\em Nucl. Phys.} {\bf
  B529} (1998)  424--450},
\href{http://arxiv.org/abs/hep-ph/9801375}{{\tt arXiv:hep-ph/9801375}}.

\bibitem{Czakon:2009zw}
M.~Czakon, A.~Mitov, and G.~Sterman,
  \href{http://dx.doi.org/10.1103/PhysRevD.80.074017}{{\em Phys. Rev.} {\bf
  D80} (2009)  074017},
\href{http://arxiv.org/abs/0907.1790}{{\tt arXiv:0907.1790 [hep-ph]}}.

\bibitem{Beenakker:2011sf}
W.~Beenakker, S.~Brensing, M.~Kr{\"a}mer, A.~Kulesza, E.~Laenen, {\em et al.}, {\em
  JHEP} {\bf 1201} (2012)  076,
\href{http://arxiv.org/abs/1110.2446}{{\tt arXiv:1110.2446 [hep-ph]}}.

\bibitem{Hagiwara:2009hq}
K.~Hagiwara and H.~Yokoya,
  \href{http://dx.doi.org/10.1088/1126-6708/2009/10/049}{{\em JHEP} {\bf 10}
  (2009)  049},
\href{http://arxiv.org/abs/0909.3204}{{\tt arXiv:0909.3204 [hep-ph]}}.

\bibitem{Kauth:2011vg}
M.~R. Kauth, J.~H. K{\"u}hn, P.~Marquard, and M.~Steinhauser,
  \href{http://dx.doi.org/10.1016/j.nuclphysb.2011.11.024}{{\em Nucl.Phys.}
  {\bf B857} (2012)  28--64},
\href{http://arxiv.org/abs/1108.0361}{{\tt arXiv:1108.0361 [hep-ph]}}.

\bibitem{Kauth:2011bz}
M.~R. Kauth, A.~Kress, and J.~H. K{\"u}hn,
  \href{http://dx.doi.org/10.1007/JHEP12(2011)104}{{\em JHEP} {\bf 1112} (2011)
   104},
\href{http://arxiv.org/abs/1108.0542}{{\tt arXiv:1108.0542 [hep-ph]}}.

\bibitem{Langenfeld:2009eg}
U.~Langenfeld and S.-O. Moch, {\em Phys. Lett.} {\bf B675} (2009)  210--221,
\href{http://arxiv.org/abs/0901.0802}{{\tt arXiv:0901.0802 [hep-ph]}}.

\bibitem{Langenfeld:2010vu}
U.~Langenfeld, \href{http://dx.doi.org/10.1007/JHEP07(2011)052}{{\em JHEP} {\bf
  1107} (2011)  052},
\href{http://arxiv.org/abs/1011.3341}{{\tt arXiv:1011.3341 [hep-ph]}}.

\bibitem{Beneke:2009rj}
M.~Beneke, P.~Falgari, and C.~Schwinn,
  \href{http://dx.doi.org/10.1016/j.nuclphysb.2009.11.004}{{\em Nucl. Phys.}
  {\bf B828} (2010)  69--101},
\href{http://arxiv.org/abs/0907.1443}{{\tt arXiv:0907.1443 [hep-ph]}}.

\bibitem{Beneke:2010da}
M.~Beneke, P.~Falgari, and C.~Schwinn,
  \href{http://dx.doi.org/10.1016/j.nuclphysb.2010.09.009}{{\em Nucl. Phys.}
  {\bf B842} (2011)  },
\href{http://arxiv.org/abs/1007.5414}{{\tt arXiv:1007.5414 [hep-ph]}}.

\bibitem{Beneke:2011mq}
M.~Beneke, P.~Falgari, S.~Klein, and C.~Schwinn,
  \href{http://dx.doi.org/10.1016/j.nuclphysb.2011.10.021}{{\em Nucl.Phys.}
  {\bf B855} (2012)  695--741},
\href{http://arxiv.org/abs/1109.1536}{{\tt arXiv:1109.1536 [hep-ph]}}.

\bibitem{Becher:2006nr}
T.~Becher and M.~Neubert, {\em Phys. Rev. Lett.} {\bf 97} (2006)  082001,
\href{http://arxiv.org/abs/hep-ph/0605050}{{\tt hep-ph/0605050}}.

\bibitem{Becher:2006mr}
T.~Becher, M.~Neubert, and B.~D. Pecjak, {\em JHEP} {\bf 01} (2007)  076,
\href{http://arxiv.org/abs/hep-ph/0607228}{{\tt hep-ph/0607228}}.

\bibitem{Becher:2007ty}
T.~Becher, M.~Neubert, and G.~Xu,
  \href{http://dx.doi.org/10.1088/1126-6708/2008/07/030}{{\em JHEP} {\bf 07}
  (2008)  030},
\href{http://arxiv.org/abs/0710.0680}{{\tt arXiv:0710.0680 [hep-ph]}}.

\bibitem{AbdusSalam:2011fc}
S.~AbdusSalam, B.~Allanach, H.~Dreiner, J.~Ellis, U.~Ellwanger, {\em et al.},
  \href{http://dx.doi.org/10.1140/epjc/s10052-011-1835-7}{{\em Eur.Phys.J.}
  {\bf C71} (2011)  1835},
\href{http://arxiv.org/abs/1109.3859}{{\tt arXiv:1109.3859 [hep-ph]}}.

\bibitem{Kane:1982hw}
G.~L. Kane and J.~Leveille,
  \href{http://dx.doi.org/10.1016/0370-2693(82)90968-6}{{\em Phys.Lett.} {\bf
  B112} (1982)  227}.

\bibitem{Harrison:1982yi}
P.~Harrison and C.~Llewellyn~Smith,
  \href{http://dx.doi.org/10.1016/0550-3213(83)90510-2,
  10.1016/0550-3213(83)90510-2}{{\em Nucl.Phys.} {\bf B213} (1983)  223}.

\bibitem{Dawson:1983fw}
S.~Dawson, E.~Eichten, and C.~Quigg,
  \href{http://dx.doi.org/10.1103/PhysRevD.31.1581}{{\em Phys.Rev.} {\bf D31}
  (1985)  1581}.

\bibitem{Martin:2009iq}
A.~D. Martin, W.~J. Stirling, R.~S. Thorne, and G.~Watt,
  \href{http://dx.doi.org/10.1140/epjc/s10052-009-1072-5}{{\em Eur. Phys. J.}
  {\bf C63} (2009)  189--285},
\href{http://arxiv.org/abs/0901.0002}{{\tt arXiv:0901.0002 [hep-ph]}}.

\bibitem{Beneke:2009ye}
M.~Beneke, M.~Czakon, P.~Falgari, A.~Mitov, and C.~Schwinn,
  \href{http://dx.doi.org/10.1016/j.physletb.2010.05.038}{{\em Phys. Lett.}
  {\bf B690} (2010)  483--490},
\href{http://arxiv.org/abs/0911.5166}{{\tt arXiv:0911.5166 [hep-ph]}}.

\bibitem{Kats:2009bv}
Y.~Kats and M.~D. Schwartz,
  \href{http://dx.doi.org/10.1007/JHEP04(2010)016}{{\em JHEP} {\bf 04} (2010)
  016},
\href{http://arxiv.org/abs/0912.0526}{{\tt arXiv:0912.0526 [hep-ph]}}.

\bibitem{Fadin:1987wz}
V.~S. Fadin and V.~A. Khoze,
{\em JETP Lett.} {\bf 46} (1987)  525--529.

\bibitem{Hoang:2000yr}
A.~H. Hoang {\em et al.}, {\em Eur. Phys. J. direct} {\bf C2} (2000)  1,
\href{http://arxiv.org/abs/hep-ph/0001286}{{\tt arXiv:hep-ph/0001286}}.

\bibitem{Beneke:1999qg}
M.~Beneke, A.~Signer, and V.~A. Smirnov,
  \href{http://dx.doi.org/10.1016/S0370-2693(99)00343-3}{{\em Phys. Lett.} {\bf
  B454} (1999)  137--146},
\href{http://arxiv.org/abs/hep-ph/9903260}{{\tt arXiv:hep-ph/9903260}}.

\bibitem{Bigi:1991mi}
I.~I.~Y. Bigi, V.~S. Fadin, and V.~A. Khoze,
\href{http://dx.doi.org/10.1016/0550-3213(92)90297-O}{{\em Nucl. Phys.} {\bf
  B377} (1992)  461--479}.

\bibitem{Kauth:2009ud}
M.~R. Kauth, J.~H. K{\"u}hn, P.~Marquard, and M.~Steinhauser,
  \href{http://dx.doi.org/10.1016/j.nuclphysb.2010.01.019}{{\em Nucl.Phys.}
  {\bf B831} (2010)  285--305},
\href{http://arxiv.org/abs/0910.2612}{{\tt arXiv:0910.2612 [hep-ph]}}.

\bibitem{Younkin:2009zn}
J.~E. Younkin and S.~P. Martin,
  \href{http://dx.doi.org/10.1103/PhysRevD.81.055006}{{\em Phys.Rev.} {\bf D81}
  (2010)  055006}, \href{http://arxiv.org/abs/0912.4813}{{\tt arXiv:0912.4813
  [hep-ph]}}.

\bibitem{Becher:2009kw}
T.~Becher and M.~Neubert,
  \href{http://dx.doi.org/10.1103/PhysRevD.79.125004}{{\em Phys. Rev.} {\bf
  D79} (2009)  125004},
\href{http://arxiv.org/abs/0904.1021}{{\tt arXiv:0904.1021 [hep-ph]}}.

\bibitem{Bozzi:2005sy}
G.~Bozzi, B.~Fuks, and M.~Klasen,
  \href{http://dx.doi.org/10.1103/PhysRevD.72.035016}{{\em Phys.Rev.} {\bf D72}
  (2005)  035016}, \href{http://arxiv.org/abs/hep-ph/0507073}{{\tt
  arXiv:hep-ph/0507073 [hep-ph]}}.

\bibitem{Cassel:2009wt}
S.~Cassel, \href{http://dx.doi.org/10.1088/0954-3899/37/10/105009}{{\em
  J.Phys.G} {\bf G37} (2010)  105009},
  \href{http://arxiv.org/abs/0903.5307}{{\tt arXiv:0903.5307 [hep-ph]}}.

\bibitem{Baer:2011aa}
H.~Baer, V.~Barger, A.~Lessa, and X.~Tata,
  \href{http://dx.doi.org/10.1103/PhysRevD.85.051701}{{\em Phys.Rev.} {\bf D85}
  (2012)  051701},
\href{http://arxiv.org/abs/1112.3044}{{\tt arXiv:1112.3044 [hep-ph]}}.

\bibitem{NLLfast}
\texttt{NLL-fast}, available at
  \url{http://web.physik.rwth-aachen.de/service/wiki/bin/view/Kraemer/SquarksandGluinos}.

\bibitem{Djouadi:2006bz}
A.~Djouadi, M.~Muhlleitner, and M.~Spira, {\em Acta Phys.Polon.} {\bf B38}
  (2007)  635--644,
\href{http://arxiv.org/abs/hep-ph/0609292}{{\tt arXiv:hep-ph/0609292
  [hep-ph]}}.

\bibitem{Djouadi:2002ze}
A.~Djouadi, J.-L. Kneur, and G.~Moultaka,
  \href{http://dx.doi.org/10.1016/j.cpc.2006.11.009}{{\em Comput.Phys.Commun.}
  {\bf 176} (2007)  426--455},
\href{http://arxiv.org/abs/hep-ph/0211331}{{\tt arXiv:hep-ph/0211331
  [hep-ph]}}.

\bibitem{Bonvini:2012yg}
M.~Bonvini, S.~Forte, M.~Ghezzi, and G.~Ridolfi,
\href{http://arxiv.org/abs/1201.6364}{{\tt arXiv:1201.6364 [hep-ph]}}.

\bibitem{Bauer:2000yr}
C.~W. Bauer, S.~Fleming, D.~Pirjol, and I.~W. Stewart, {\em Phys. Rev.} {\bf
  D63} (2001)  114020,
\href{http://arxiv.org/abs/hep-ph/0011336}{{\tt hep-ph/0011336}}.

\bibitem{Bauer:2001yt}
C.~W. Bauer, D.~Pirjol, and I.~W. Stewart, {\em Phys. Rev.} {\bf D65} (2002)
  054022,
\href{http://arxiv.org/abs/hep-ph/0109045}{{\tt hep-ph/0109045}}.

\bibitem{Beneke:2002ph}
M.~Beneke, A.~P. Chapovsky, M.~Diehl, and T.~Feldmann, {\em Nucl. Phys.} {\bf
  B643} (2002)  431--476,
\href{http://arxiv.org/abs/hep-ph/0206152}{{\tt hep-ph/0206152}}.

\bibitem{Beneke:2002ni}
M.~Beneke and T.~Feldmann,
  \href{http://dx.doi.org/10.1016/S0370-2693(02)03204-5}{{\em Phys. Lett.} {\bf
  B553} (2003)  267--276},
\href{http://arxiv.org/abs/hep-ph/0211358}{{\tt arXiv:hep-ph/0211358}}.

\bibitem{Pineda:1997bj}
A.~Pineda and J.~Soto,
  \href{http://dx.doi.org/10.1016/S0920-5632(97)01102-X}{{\em Nucl. Phys. Proc.
  Suppl.} {\bf 64} (1998)  428--432},
\href{http://arxiv.org/abs/hep-ph/9707481}{{\tt arXiv:hep-ph/9707481}}.

\bibitem{Beneke:1998jj}
M.~Beneke,
\href{http://arxiv.org/abs/hep-ph/9806429}{{\tt arXiv:hep-ph/9806429}}.

\bibitem{Brambilla:1999xf}
N.~Brambilla, A.~Pineda, J.~Soto, and A.~Vairo,
  \href{http://dx.doi.org/10.1016/S0550-3213(99)00693-8}{{\em Nucl. Phys.} {\bf
  B566} (2000)  275},
\href{http://arxiv.org/abs/hep-ph/9907240}{{\tt arXiv:hep-ph/9907240}}.

\bibitem{Beneke:1999zr}
M.~Beneke, \href{http://arxiv.org/abs/hep-ph/9911490}{{\tt
  arXiv:hep-ph/9911490}}.
Proceedings of the 8th International Symposium on Heavy Flavor Physics (Heavy
  Flavors 8), Southampton, England, 25-29 Jul 1999.

\bibitem{Bodwin:1992ye}
G.~T. Bodwin, E.~Braaten, and G.~P. Lepage,
  \href{http://dx.doi.org/10.1103/PhysRevD.46.R1914}{{\em Phys. Rev.} {\bf D46}
  (1992)  1914--1918},
\href{http://arxiv.org/abs/hep-lat/9205006}{{\tt arXiv:hep-lat/9205006}}.

\bibitem{Bauer:2002nz}
C.~W. Bauer, S.~Fleming, D.~Pirjol, I.~Z. Rothstein, and I.~W. Stewart, {\em
  Phys. Rev.} {\bf D66} (2002)  014017,
\href{http://arxiv.org/abs/hep-ph/0202088}{{\tt hep-ph/0202088}}.

\end{thebibliography}

\providecommand{\href}[2]{#2}\begingroup\raggedright\endgroup


\end{document}